\documentclass[journal=jctc,
manuscript=article
]{achemso}

\makeatletter
\let\l@addto@macro\relax
\makeatother
\usepackage[fontsize=10pt]{scrextend}
\usepackage[version=3]{mhchem} 
\usepackage{csvsimple}
\usepackage{adjustbox}
\usepackage{siunitx}
\usepackage{amsmath}
\usepackage{appendix}
\usepackage{xr}
\usepackage{booktabs}

\SectionNumbersOn
\DeclareUnicodeCharacter{2212}{-}
\makeatletter
\newcommand*{\addFileDependency}[1]{
  \typeout{(#1)}
  \@addtofilelist{#1}
  \IfFileExists{#1}{}{\typeout{No file #1.}}
}
\makeatother

\newcommand*{\myexternaldocument}[1]{%
    \externaldocument{#1}%
    \addFileDependency{#1.tex}%
    \addFileDependency{#1.aux}%
}
\myexternaldocument{DFTBML-supplement}

\newcolumntype{H}{>{\setbox0=\hbox\bgroup}c<{\egroup}@{}}
\DeclareUnicodeCharacter{2729}{}
\author{Frank Hu}
\email{frankhu@stanford.edu}
\author{Francis He}
\author{David J. Yaron}
\email{yaron@cmu.edu}
\phone{(412)-951-4516}
\affiliation[Carnegie Mellon University]
{Department of Chemistry, Carnegie Mellon University, Pittsburgh, PA}

\title[Learned Semiempirical Hamiltonians]
  {Semiempirical Hamiltonians learned from data can have accuracy comparable to Density Functional Theory}
\abbreviations{IR,NMR,UV}
\keywords{American Chemical Society, \LaTeX}

\begin{document}
\begin{abstract}
Quantum chemistry provides chemists with invaluable information, but the high computational cost limits the size and type of systems that can be studied. Machine learning (ML) has emerged as a means to dramatically lower cost while maintaining high accuracy. However, ML models often sacrifice interpretability by using components, such as the artificial neural networks of deep learning, that function as black boxes. These components impart the flexibility needed to learn from large volumes of data but make it difficult to gain insight into the physical or chemical basis for the predictions. Here, we demonstrate that semiempirical quantum chemical (SEQC) models can learn from large volumes of data without sacrificing interpretability. The SEQC model is that of Density Functional based Tight Binding (DFTB) with fixed atomic orbital energies and interactions that are one-dimensional functions of interatomic distance. This model is trained to \textit{ab initio} data in a manner that is analogous to that used to train deep learning models. Using benchmarks that reflect the accuracy of the training data, we show that the resulting model maintains a physically reasonable functional form while achieving an accuracy, relative to coupled cluster energies with a complete basis set extrapolation (CCSD(T)*/CBS), that is comparable to that of density functional theory (DFT). This suggests that trained SEQC models can achieve low computational cost and high accuracy without sacrificing interpretability. Use of a physically-motivated model form also substantially reduces the amount of \textit{ab initio} data needed to train the model compared to that required for deep learning models. 
\end{abstract}

\section{Introduction}
A substantial challenge for quantum chemistry is lowering the computational cost\cite{Whitfield2012ComputationalStructureb,Koppl2016ParallelFitting, Scuseria1992ComparisonTheory,Mardirossian2018LoweringRepresentation,Gruber2018ApplyingLimit,Gulania2021LimitationsResources} to enable accurate predictions on large systems such as those of interest in biological and material applications. Molecular systems have two properties that provide the basis for approximations that lower computational costs: nearsightedness and molecular similarity. Nearsightedness provides the chemical basis for methods that have, over the past few decades, substantially reduced computational cost without large sacrifices in accuracy. In particular, large reductions in cost can be achieved by replacing detailed Coulomb interactions, required at short range, with increasingly coarse-grained multi-polar interactions at long range\cite{Gordon2009AccurateSystems,Shang2010LinearSet,Riplinger2016SparseTheory,Prentice2020TheProgram}. Methods have also been developed that use molecular similarity to achieve dramatic reductions in computational cost, including molecular mechanics\cite{Jo2008CHARMM-GUI:CHARMM,amber22} and semiempirical quantum chemistry (SEQC)\cite{Thiel2014SemiempiricalMethods}. Unfortunately, these cost reductions have typically come with a substantial decrease in accuracy. More recently, machine learning (ML) has emerged as a means to leverage molecular similarity to develop models that are both low-cost and accurate\cite{vonLilienfeld2020ExploringLearning,Unke2021MachineFields,Poltavsky2021MachineChallenges,Kulik2022RoadmapStructure,Dral2023QuantumLearning}. However, current applications of ML in chemistry often incorporate little physics and function as black boxes that are difficult to interpret. Here, we combine ML with SEQC to create physics-based models that achieve high accuracy and computational efficiency without sacrificing interpretability.

The ability of ML to leverage molecular similarity stems from the use of highly flexible model forms such as the artificial neural networks (NNs)\cite{Behler2007GeneralizedSurfaces,Smith2017,Schutt2018SchNetMaterials,Schutt2019UnifyingWavefunctions,Smith2020TheMolecules,Qiao2020OrbNet:Features,Christensen2021OrbNetAccuracy,Zheng2021ArtificialApplicability} of deep learning. This flexibility enables ML models to learn from large volumes of training data. For example, the accuracy of the ANI-1 neural network potential\cite{Smith2017} improves as it is shown more training data, approaching chemical accuracy\cite{Dral2020HierarchicalSurfaces,Dral2021MLatomLearning,Afzal2019AMolecules,Haghighatlari2020ChemML:Data,Haghighatlari2019AdvancesSimulation} of ~1 kcal/mol when trained to \textit{ab initio} results on millions of molecular configurations. However, this flexibility of ML models is a double-edged sword. It leads to high accuracy, but it also makes it difficult to gain insight into the physical or chemical basis for the predictions.

SEQC provides alternative model forms that are capable of learning from data. Traditional SEQC model forms such as PM3\cite{Stewart1989OptimizationMethod} only have a handful of parameters and this limits their ability to take advantage of large volumes of data\cite{Dral2015MachineCalculations}. Replacing these single parameters with NNs imparts the flexibility to learn from large volumes of data\cite{Zhou2022DeepMechanics}, however the NNs function as black boxes and so decrease interpretability. Here, we increase the flexibility of SEQC models so that they can take advantage of larger volumes of data while retaining a purely physics-based form. This is operationalized using the Density Functional based Tight Binding (DFTB)\cite{Elstner1998Self-consistent-chargeProperties,Seifert2007Tight-bindingScheme} Hamiltonian with model parameters that can be expressed in the Slater-Koster File (SKF) format\cite{Hourahine2020DFTB+Simulations}. DFTB includes only valence electrons and uses a minimal atomic orbital basis. The atomic orbital energies are constants that can be adjusted during training, and the interactions and overlaps between atomic orbitals are one-dimensional functions of interatomic distance. We will refer to this as the SKF-DFTB model form and to our resulting trained models as DFTBML.

The flexibility of DFTBML lies primarily in the one-dimensional functions. Over the distances present in typical molecules, the interactions described by these functions vary by hundreds of kcal/mol. Because the molecular energy arises from many such interactions, changes of a few tenths of a kcal/mol can have significant effects on the total energy. For the model to learn effectively from data, we need a functional form with the sensitivity to fine tune these interactions while preventing oscillations and other non-physical behaviors. Here, the flexibility and sensitivity is provided through splines, i.e. piecewise polynomials, with a high polynomial order of five and a large number of 100 knots. To prevent oscillations and other non-physical behaviors, a strong regularization scheme is developed and implemented in our training of DFTBML.

\begin{figure}[ht]
    \centering
    \includegraphics[width=0.75\columnwidth]{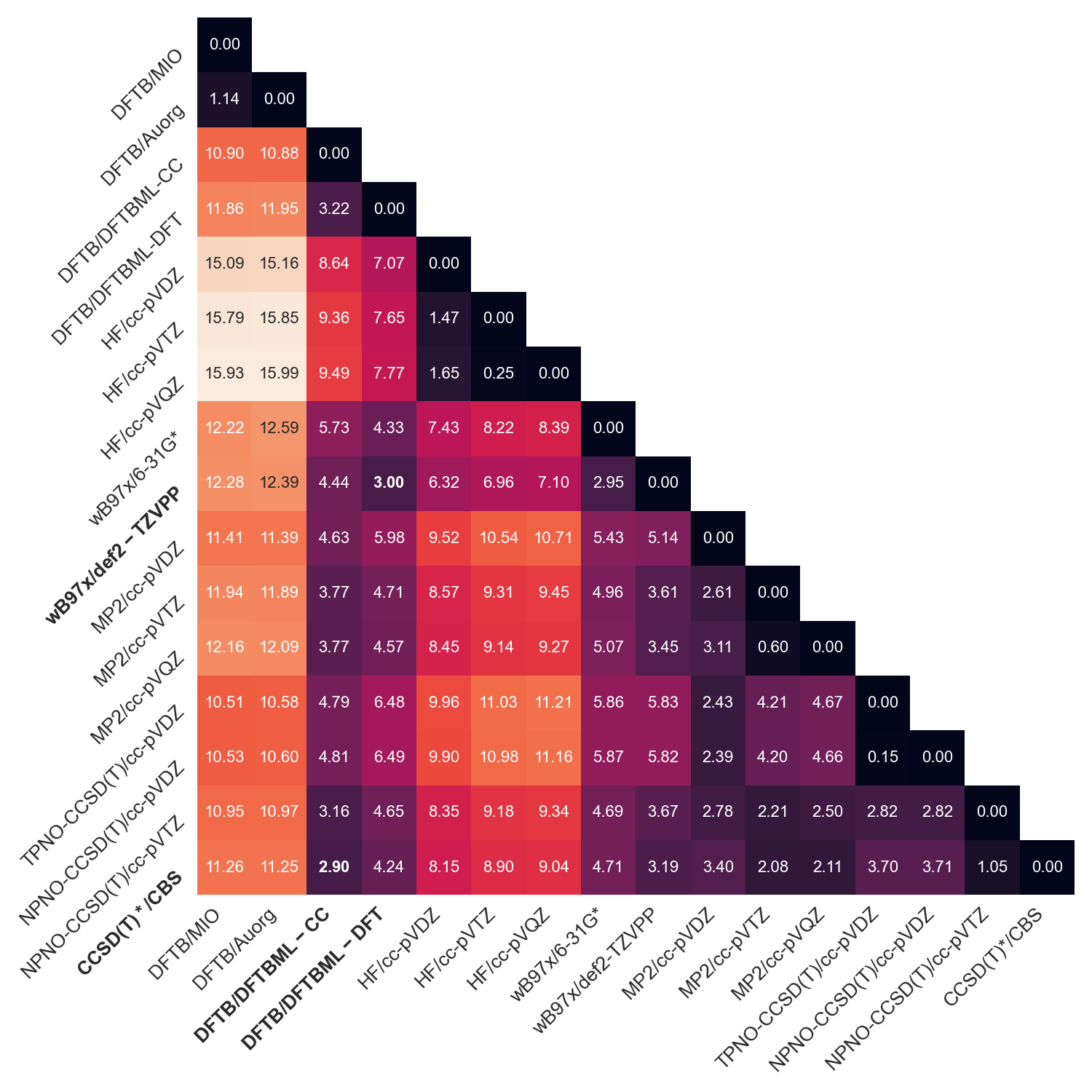}
    \caption{Comparison of different quantum chemistry methods on atomization energies (see Equation~\ref{eqn:ref_energy_main} in Section \ref{sec:experimental_details}). The heatmap is generated from the $\sim$230k molecular configurations in the ANI-1CCX dataset with up to eight heavy atoms, after removing configurations with incomplete entries. The DFTBML-CC/DFT parameterizations were trained to CCSD(T)*/CBS or wB97x/def2-TZVPP energies, respectively, on 20000 molecules with up to eight heavy atoms. DFTBML improves substantially on currently published DFTB parameters (MIO\cite{Elstner1998Self-consistent-chargeProperties} and Auorg\cite{Fihey2015SCC-DFTBCompounds}), with the agreement between DFTBML-CC and CCSD(T)*/CBS being somewhat better than that between DFT (wB97x/def2-TZVPP) and CCSD(T)*/CBS.}
    \label{fig:quant_chem_comparison}
\end{figure}

The DFTBML models explored here are trained to the ANI-1CCX dataset\cite{Smith2020TheMolecules}, which includes results from a number of different \textit{ab initio} methods on organic molecules comprised of C, N, O and H. The DFTBML models can reproduce the predictions of CCSD(T)*/CBS to about 3 kcal/mol, which is comparable to the accuracy of DFT (see Figure \ref{fig:quant_chem_comparison}). We also show that 20000 molecular configurations are sufficient to train the model. This saturation of performance with increasing data suggests that the accuracy is limited by the SKF-DFTB model form itself, not by the amount of training data. The data requirements of DFTBML are considerably below the $\sim$1M data points typically used to train deep learning models, which is significant given that the generation of \textit{ab initio} training data is a primary computational bottleneck in model development. This opens the possibility of using trained SEQC models as replacements for DFT, substantially reducing computational cost without, as in traditional SEQC models, sacrificing accuracy or, as in many ML models, sacrificing interpretability.

\section{Results and discussion}\label{sec:all_results}

\subsection{Experimental design} \label{subsec:training_details}

To explore the performance of DFTBML, we train the model under various conditions. To aid comparisons, it is useful to introduce a standard notation for the resulting parameter sets. To evaluate the generalization of the DFTBML models, we consider both near- and far-transfer, with the difference being the degree to which the model is being transferred to larger systems. For near-transfer, where the training and testing data contain systems with 1 - 8 heavy atoms, we use ``DFTBML'' followed by the energy target (DFT for wB97x/def2-TZVPP; CC for CCSD(T)*/CBS) and the number of configurations in the training set, e.g. ``DFTBML CC 20000''. For far-transfer, where the training data has molecules with 1~- 5 heavy atoms while the test data has molecules with 6 - 8 heavy atoms, we use ``Transfer'' as the prefix, e.g. ``Transfer CC 20000''. We also consider results obtained when only a short-range repulsive potential is trained to the data, with the electronic parameters being those of Auorg\cite{Fihey2015SCC-DFTBCompounds}. For these models, we use ``Repulsive'' as a prefix, e.g. ``Repulsive CC 20000''.

\subsection{Effects of regularization on model performance}\label{sec:unregularized_model_performance}

\begin{figure*}[ht!]
    \centering
    \includegraphics[width=1.0\textwidth]{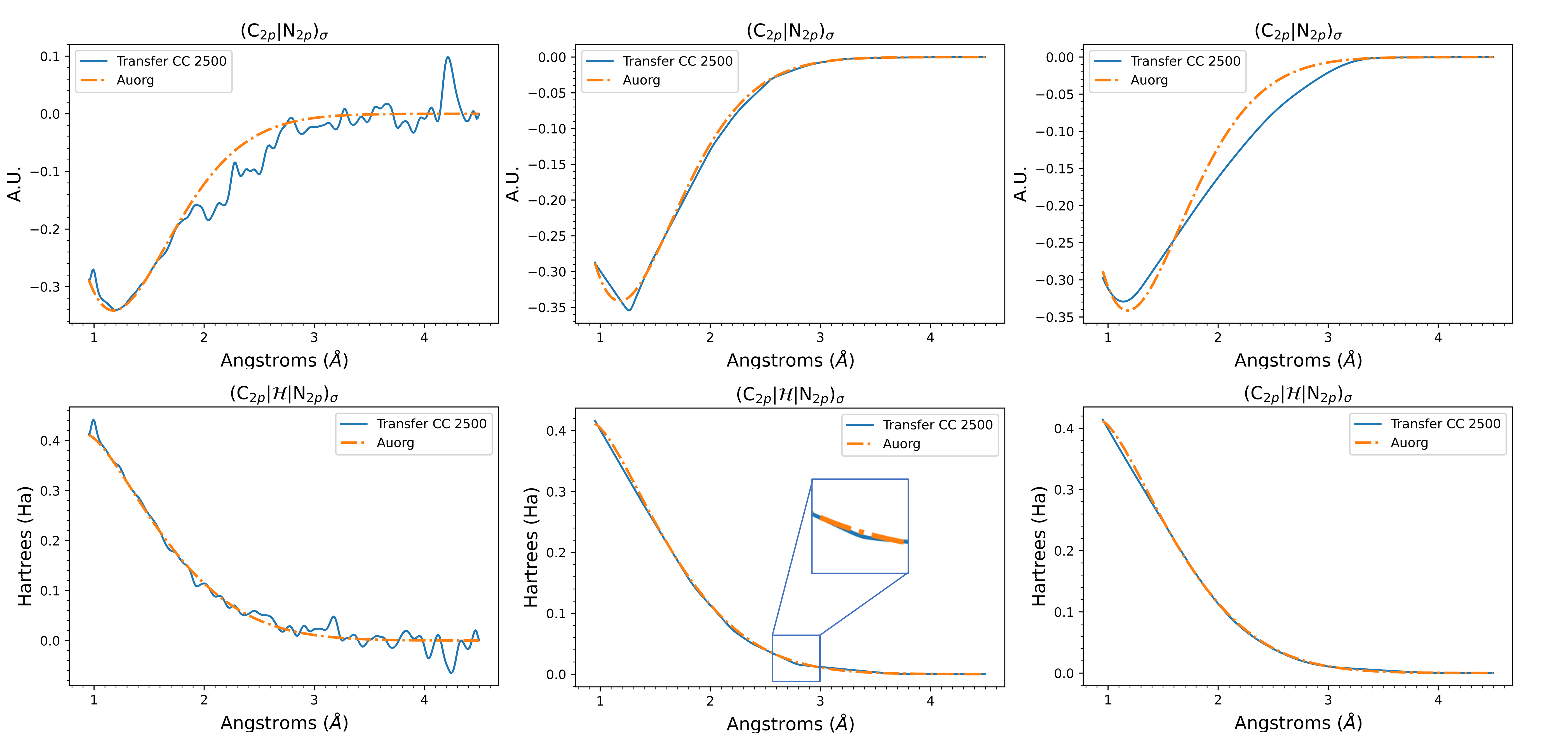}
    \caption{Effects of regularization on the $(C_{2p}|N_{2p})_\sigma$ overlaps (\textbf{S}, top row) and Hamiltonian elements ($\mathbf{H_1}$, bottom row): no regularization (left column), convex penalty that constrains sign of second derivative (middle column), and convex plus smoothing that penalizes the magnitude of the third derivative (right column). The Auorg reference functions (orange, dashed lines) are included for comparison to the trained functions (blue).}
    \label{fig:reg_change_comparison}
\end{figure*}

A challenge with developing the DFTBML model was creating an effective regularization scheme that would prevent overfitting without degrading model performance by being too restrictive. Without regularization, the resulting functions show highly oscillatory behavior (left column of Figure~\ref{fig:reg_change_comparison}). Previous work\cite{Li2018,Zhou2022DeepMechanics} penalized deviations from a set of physically-derived reference parameters, e.g. deviation from the Auorg parameter set of DFTB. This approach to regularization is problematic because it may overly bias the training towards the reference parameters and does not prevent non-physical behaviors such as oscillation of a trained function around the smooth form of the reference function\cite{Li2018}. A commonly used approach for smoothing splines applies a penalty to the magnitude of the second derivative\cite{Rice1983SmoothingDeconvolution,Eilers1996FlexiblePenalties}. However, for DFTBML, such a smoothing penalty substantially degrades performance of the models because there is no reason to expect the second derivative to have a limited magnitude. 

We instead adapt an approach from Akshay et al.\cite{Akshay2021} which is motivated by the shape of the functions in reference parameter sets, such as those of Auorg in Figure~\ref{fig:reg_change_comparison}. For the Hamiltonian ($\mathbf{H_1}$) matrix elements, the functions decay smoothly to zero and have an upward curvature. To enforce this behavior, we apply a ``convex'' penalty that enforces the second derivative of the trained potentials, evaluated on a dense grid of 500 points, to have a physically motivated sign. For overlaps ($\mathbf{S}$), there can also be an inflection point associated with nodes in the atomic orbitals (upper panels of Figure \ref{fig:reg_change_comparison}). We therefore extend the convex penalty to allow a single inflection point, whose location is optimized during training. The results indicate that, although inclusion of an inflection point improves model performance, the results are not sensitive to its precise location (see Section S12.3 of the Supporting Information). The magnitude of the weighting factor for these convex penalties does not require fine tuning beyond being large enough to prevent violations of the constraints without being so large that it leads to numerical instabilities in gradient descent optimization. The convex penalty successfully removes oscillatory behavior (middle column of Figure~\ref{fig:reg_change_comparison}). However, the resulting functions exhibit non-physical, piecewise-linear behavior, which is more pronounced in the overlap integrals but also present in the Hamiltonian matrix elements (see inset in Figure~\ref{fig:reg_change_comparison}).

To remove this piecewise-linear behavior, we apply a ``smoothing'' penalty to the third derivative, based on the sum of squares of the third derivative evaluated on a grid of 500 points. Our use of a fifth-order spline for $\mathbf{H_1}$ and $\mathbf{S}$ is motivated by the high order needed for the spline to have a continuous third derivative. The magnitude of the penalty is adjusted to remove the piecewise-linear behavior while minimizing degradation of the model performance (see Section S12.4 of the Supporting Information). The short-range repulsion ($\mathbf{R}$) does not exhibit piecewise-linear behavior, so a smoothing penalty is not applied and we use a third-order spline for $\mathbf{R}$.

It is somewhat surprising, given the highly non-physical behavior observed without regularization, that the effects of regularization on model performance are not more dramatic (Table~\ref{Tab:change_with_regularization}). For near-transfer, the performance of the unregularized model (4.97 kcal/mol) is a factor of two better than the Auorg reference model (10.55 kcal/mol). This is despite the highly oscillatory behavior of the functions and the fact that the test data and training data have molecules with disjoint empirical formulas. This suggests coupling between potentials, with oscillations in one potential cancelling out the effects of oscillations in another potential. The effects of regularization are more pronounced for far-transfer, but even here, the performance of the unregularized model (10.08 kcal/mol) is comparable to that of the Auorg reference model (11.81 kcal/mol).

It is also noteworthy that, although addition of the convex penalty leads to substantial improvements in model performance on test data, addition of the smoothing penalty has a much smaller effect and even slightly degrades performance for the far-transfer experiments. So based on the typical approach of ML, where regularization is adjusted to optimize performance on test data, the smoothing penalty would not be viewed as necessary. However, the resulting functions of Figure \ref{fig:reg_change_comparison} suggest that a smoothing penalty is needed to obtain physically reasonable functional forms. This illustrates a general finding of this work, that the level of regularization needed to achieve a physically reasonable model goes beyond that needed to achieve good transfer\cite{Matlock2021DeepChemistry,Hutchison2021EvaluationOptimization,Haghighatlari2019ThinkingChemistry,Welborn2018TransferabilityBasis} from train to test data. 

\begin{table}[!ht]
    \centering
    \caption{Effects of regularization on near-transfer (DFTBML CC 2500), i.e. training and testing on molecules with up to eight heavy, and far-transfer (Transfer CC 2500), i.e. training on molecules with 1 - 5 heavy atoms and testing on molecules with 6 - 8 heavy atoms.}
    \label{Tab:change_with_regularization}
    \sisetup{round-mode=places}
    \large
    \resizebox{\columnwidth}{!}{\begin{tabular}{lHHS[round-precision=2]S[round-precision=3]S[round-precision=3]}
        \toprule
        \multicolumn{6}{c}{\textbf{Near-transfer: DFTBML CC 2500}}\\
        \textbf{Parameterization} & \textbf{Outliers} & \textbf{Nonconverged} & \textbf{MAE energy (kcal/mol)} & \textbf{MAE dipole (e\AA)} & \textbf{MAE charge (e)} \\ \midrule
        Auorg & 0 & 0 & 10.548555105 & 0.07888197 & 0.084601841\\ 
        MIO & 0 & 0 & 10.693563047 & 0.078904702 & 0.084754859 \\ 
        No regularization & 1 & 1 & 4.9697219981992005 & 0.03706377152432995 & 0.056110038790930836 \\ 
        Convex only & 0 & 0 & 3.173976603334923 & 0.04113430390634895 & 0.05998812731519756 \\ 
        Convex with smoothing & 0 & 1 & 2.947089603 & 0.035830429 & 0.053621256 \\ \midrule
        \multicolumn{6}{c}{\textbf{Far-transfer: Transfer CC 2500}}\\
        \textbf{Parameterization} & \textbf{Outliers} & \textbf{Nonconverged} & \textbf{MAE energy (kcal/mol)} & \textbf{MAE dipole (e\AA)} & \textbf{MAE charge (e)} \\ \midrule
        Auorg & 0 & 0 & 11.81432903 & 0.088503081 & 0.087564107 \\ 
        MIO & 0 & 0 & 11.864115851 & 0.088528894 & 0.087712701 \\ 
        No regularization & 0 & 0 & 10.084486430 & 0.048864148 & 0.060749176 \\ 
        Convex only & 0 & 0 & 4.701075424 & 0.051328511 & 0.064316908 \\ 
        Convex with smoothing & 0 & 1 & 4.830947015 & 0.051354795 & 0.065185255 \\ \bottomrule
    \end{tabular}}
\end{table}

\subsection{Model performance} \label{sec:diff_dset_size} 

\begin{figure}[ht!]
    \centering
    \includegraphics[width=1.0\columnwidth]{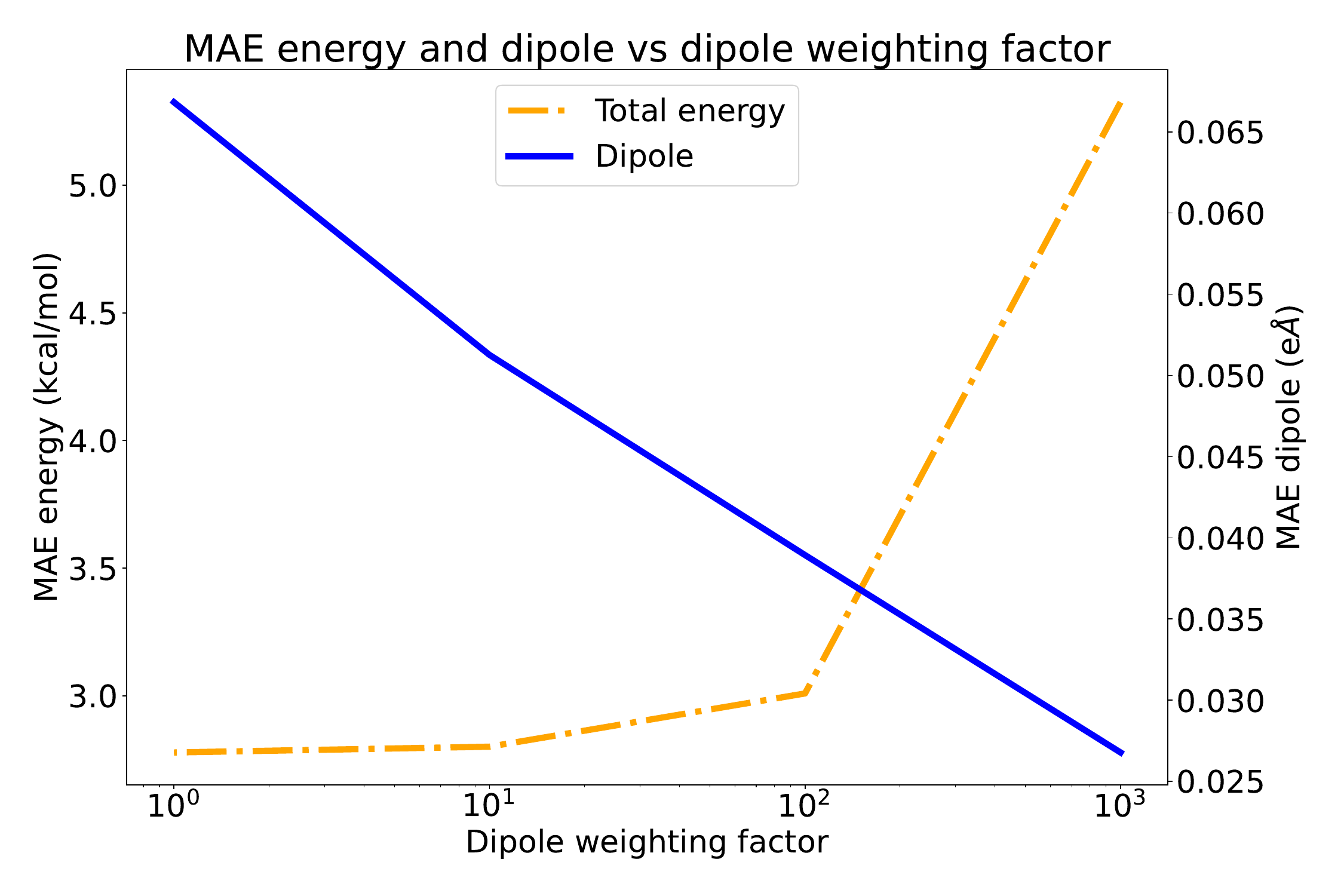}
    \caption{Tradeoff between MAE in total energy and dipoles as a function of the dipole weighting factor for DFTBML CC 2500. A weighting factor of 100 was chosen to improve performance on dipoles while only marginally impacting performance on total energy. More details on hyperparameter sensitivities can be found in Section S12 of the Supporting Information.}
    \label{fig:target_tradeoff_overlay}
\end{figure}

By changing the weights applied to the energy and dipole components of the loss function, we can also explore the tradeoff between fitting these properties. As the weight applied to the dipole component increases, we initially see large improvements in dipole with little impact on the energy. However, beyond a weight of $10^2$, the error for energy increases rapidly (see Figure \ref{fig:target_tradeoff_overlay}). For all reported results, we use a dipole weighting factor of $10^2$. 

\begin{figure*}[h!]
    \centering
    \includegraphics[width=1.0\textwidth]{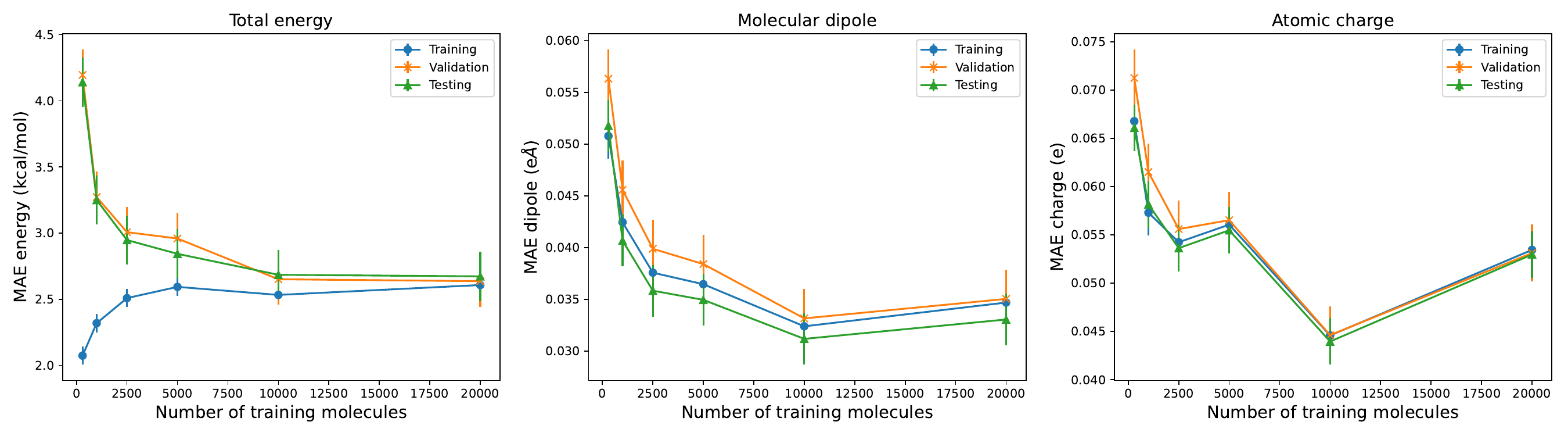}
    \caption{Final training, validation, and testing losses for each of the physical targets as a function of the size of the dataset used for training. Results are for training to the CC energy target. Error bars are shown as $\pm\frac{\sigma}{3}$ where $\sigma$ is the standard deviation of the errors calculated separately for the training, validation, and testing values.}
    \label{fig:target_convergence_v_size}
\end{figure*}

To examine how the performance of DFTBML varies with amount of training data, models were trained on datasets with between 300 and 20000 molecular configurations (Figure \ref{fig:target_convergence_v_size} and Table \ref{Tab:cc_exp_quant}). Each model was assessed against a standard set of 10000 test molecules. More complete results are provided in the Supporting Information, Section S13, including results from training to both CC and DFT targets and learning curves for each experiment. 

For the energy, the training and test errors converge at about 20000 configurations, indicating that the training is saturated and additional data is unlikely to improve performance. For dipoles and charges, the training, validation, and testing losses track each other closely. This likely reflects the high weighting of energy in the loss function such that specialization of the model to the training data occurs only for the energy. It is a bit unusual that, for dipoles and charges, the test error is smaller than the train error. However, this behavior inverts if the train and test sets are switched, suggesting that the training set has somewhat more difficult configurations than the test set.

\begin{table}[!ht]
    \centering
    \caption{Performance of various models on the CC energies of the 10000 molecule test set.}
    \label{Tab:cc_exp_quant}
    \sisetup{round-mode=places}
    \large
    \resizebox{\columnwidth}{!}{\begin{tabular}{lHHS[round-precision=2]S[round-precision=3]S[round-precision=3]}
    \toprule
        \textbf{Parameterization} & \textbf{Outliers} & \textbf{Nonconverged} & \textbf{MAE energy (kcal/mol)} & \textbf{MAE dipole (e\AA)} & \textbf{MAE charge (e)} \\ \midrule
        Auorg & 0 & 0 & 10.548555105 & 0.07888197 & 0.084601841\\ 
        MIO & 0 & 0 & 10.693563047 & 0.078904702 & 0.084754859 \\ 
        GFN1-xTB & 0 & 0 & 10.659580098 & 0.135582684 & 0.102679051 \\ 
        GFN2-xTB & 0 & 0 & 13.032294013 & 0.152755742 & 0.088788801 \\ 
        Repulsive CC 20000 & 3 & 0 & 5.407935998 & 0.078899922 & 0.084598111 \\
        DFTBML CC 20000 & 0 & 0 & 2.672659848 & 0.033047873 & 0.052958396 \\ 
        DFTBML CC 2500 & 0 & 1 & 2.947089603 & 0.035830429 & 0.053621256 \\ 
        DFTBML CC 300 & -4 & 0 & 4.140990797 & 0.05176159 & 0.066075539 \\ \bottomrule
    \end{tabular}}
\end{table}

DFTBML substantially improves upon the standard DFTB parameterizations, Auorg\cite{Fihey2015SCC-DFTBCompounds} and MIO\cite{Elstner1998Self-consistent-chargeProperties}, as well as both GFN1-xTB and GFN2-xTB\cite{Grimme2017A1-86,Bannwarth2019GFN2-xTBContributions,Koopman2019CalculationMethods} (Table~\ref{Tab:cc_exp_quant}). Auorg is a more direct comparison than MIO as both Auorg and DFTBML are shell-resolved, where Coulombic interactions differ between atomic shells (e.g. 2s versus 2p). Compared to Auorg, DFTBML CC 20000 gives a percent improvement of approximately 75\% for total energy, 58\% for dipole, and 38\% for atomic charges. The improvement is largest for total energy, which is consistent with the greater emphasis being placed on total energy in the loss function. Comparison with ``Repulsive 20000" in Table~\ref{Tab:cc_exp_quant} indicates that only half of the improvement arises from the short-range repulsive potential, emphasizing the benefits of training both the electronic and repulsive components. Similar results are observed when fitting to the DFT total energy target (see Supporting Information Section S13), suggesting that the performance of DFTBML is not strongly dependent on the level of \textit{ab initio} theory used to generate the target quantities. 

The results of Table~\ref{Tab:cc_exp_quant} are on the standard test set of 10000 molecules. For the full ANI-1CCX dataset, the performance of 2.90 kcal/mol for ``DFTBML CC 20000'' is comparable to that of 3.19 kcal/mol for DFT wB97x/def2-TZVPP in Figure \ref{fig:quant_chem_comparison}. Examination of the orbital energies and interaction functions confirms that the parameters are physically reasonable (see Supporting Information Section S9). This suggests that SKF-DFTB is a sufficiently flexible model form that, when trained to \textit{ab initio} data, the resulting model has accuracy comparable to that of commonly used high-cost methods such as DFT. 

We next consider two experiments that help reveal the extent to which DFTBML is learning the physics of the interactions present in these systems. The first is the far-transfer experiments discussed above, where the model is trained on 2500 configurations with up to five heavy atoms and tested on molecules with 6 - 8 heavy atoms (Table \ref{Tab:transfer}). Because the functions being learned by DFTBML go to zero beyond 4.5 \AA{}, and such distances are present in molecules with up to five heavy atoms, we may expect the performance in far-transfer experiments to be close to that of near-transfer. For far-transfer, DFTBML improves on Auorg by 59\% for energy, 43\% for dipole, and 26\% for charges. For near-transfer with 2500 training configurations, the analogous improvements are 72\% for energy, 54\% for dipole, and 36\% for charges. These results suggest that DFTBML can learn from molecules with only up to five heavy atoms, as expected based on the range of the interactions being learned. The somewhat better performance seen in near-transfer may reflect the greater chemical diversity present in molecules with up to eight heavy atoms.  

\begin{figure}[ht!]
    \centering
    \includegraphics[width=1.0\columnwidth]{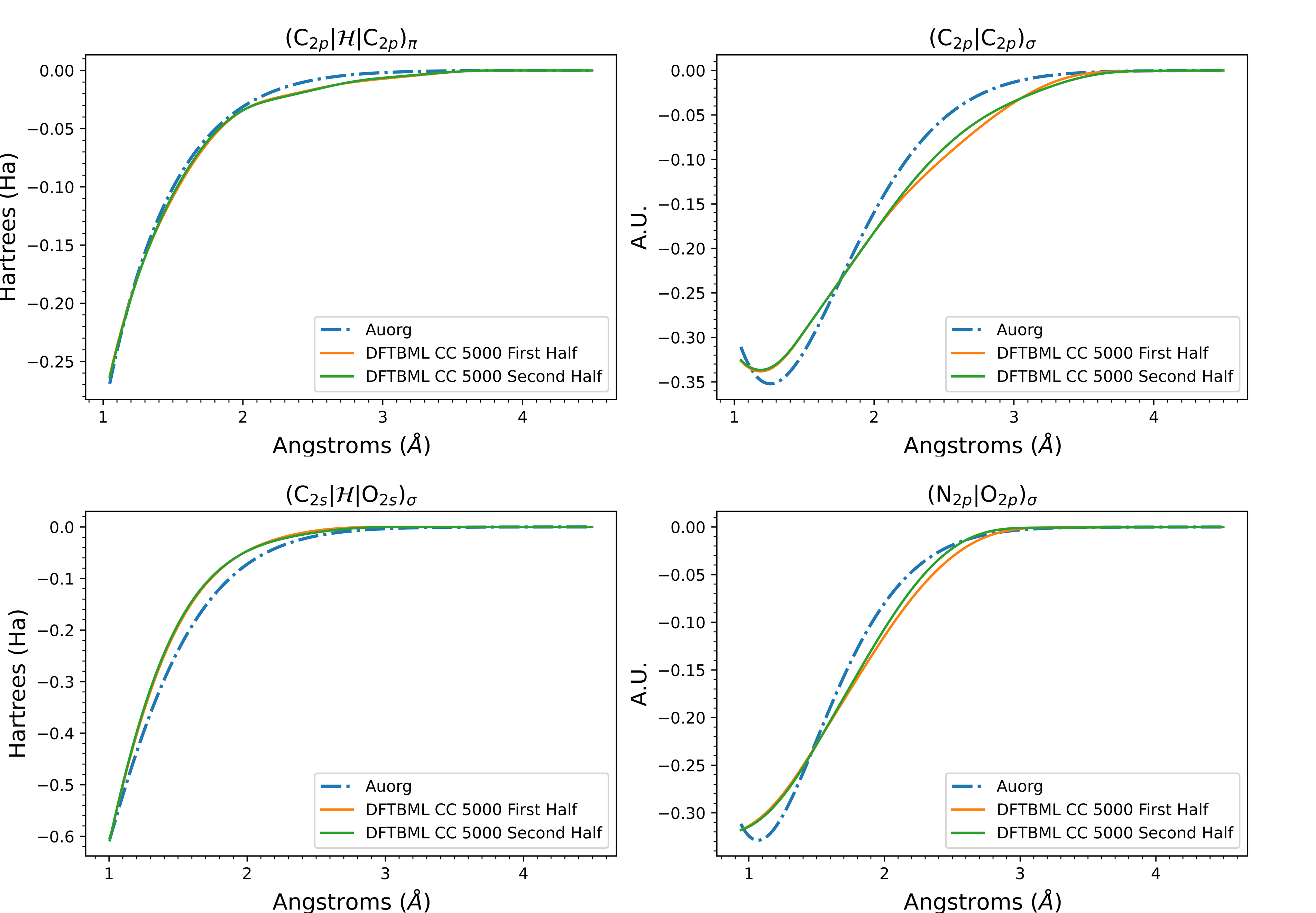}
    \caption{Example splines for Hamiltonian elements ($\mathbf{H_1}$, left) and overlap elements (\textbf{S}, right) generated from DFTBML training on disjoint data sets: 5000 First Half and 5000 Second Half. The Auorg potentials are included for reference.}
    \label{fig:reprod_example_splines}
\end{figure}

A second experiment, that explores the extent to which DFTBML is learning the underlying physics, examines the sensitivity of the model parameters to training data. For this, we train to two non-overlapping sets of training molecules, obtained by splitting the dataset ``DFTBML CC 10000'' into two halves. The performance of the resulting models are in close agreement on all targets (Table~\ref{Tab:reproducibility}), as are the resulting model forms (see Figure \ref{fig:reprod_example_splines}). That the resulting models are not sensitive to the specific data used to train the model suggests that DFTBML is learning the underlying physical interactions. 

\begin{table}[!ht]
    \centering
    \caption{Performance of various models on the test data used for far-transfer experiments, where DFTBML is trained on molecules with up to five heavy atoms and tested on molecules with 6 - 8 heavy atoms.}
    \label{Tab:transfer}
    \sisetup{round-mode=places}
    \large
    \resizebox{\columnwidth}{!}{\begin{tabular}{lHHS[round-precision=2]S[round-precision=3]S[round-precision=3]}
    \toprule
        \textbf{Parameterization} & \textbf{Outliers} & \textbf{Nonconverged} & \textbf{MAE energy (kcal/mol)} & \textbf{MAE dipole (e\AA)} & \textbf{MAE charge (e)} \\ \midrule
        Auorg CC & 0 & 0 & 11.81432903 & 0.088503081 & 0.087564107 \\ 
        MIO CC & 0 & 0 & 11.864115851 & 0.088528894 & 0.087712701 \\ 
        Auorg DFT & 0 & 0 & 13.2546505 & 0.088503081 & 0.087564107 \\ 
        MIO DFT & 0 & 0 & 13.047158534 & 0.088528894 & 0.087712701 \\ 
        GFN1-xTB CC & 0 & 0 & 10.854720214 & 0.146972956 & 0.103969430 \\ 
        GFN2-xTB CC & 0 & 0 & 13.508626798 & 0.165065903 & 0.090688288 \\ 
        GFN1-xTB DFT & 0 & 0 & 10.142095148 & 0.146972956 & 0.103969430 \\ 
        GFN2-xTB DFT & 0 & 0 & 12.256613247 & 0.165065903 & 0.090688288 \\ 
        Transfer CC 2500 & 0 & 1 & 4.81771769 & 0.050938791 & 0.065010125 \\ 
        Transfer DFT 2500 & 0 & 0 & 4.785119285 & 0.049683176 & 0.062830391 \\ \bottomrule
    \end{tabular}}
\end{table}

\begin{table}[!ht] 
    \centering
    \caption{Performance of DFTBML trained on two disjoint training sets.}
    \label{Tab:reproducibility}
    \sisetup{round-mode=places}
    \large
    \resizebox{\columnwidth}{!}{\begin{tabular}{lHHS[round-precision=2]S[round-precision=3]S[round-precision=3]}
    \toprule
        \textbf{Parameterization} & \textbf{Outliers} & \textbf{Nonconverged} & \textbf{MAE energy (kcal/mol)} & \textbf{MAE dipole (e\AA)} & \textbf{MAE charge (e)} \\ \midrule
        Auorg & 0 & 0 & 10.548555105 & 0.07888197 & 0.084601841\\ 
        MIO & 0 & 0 & 10.693563047 & 0.078904702 & 0.084754859 \\ 
        GFN1-xTB & 0 & 0 & 10.659580098 & 0.135582684 & 0.102679051 \\ 
        GFN2-xTB & 0 & 0 & 13.032294013 & 0.152755742 & 0.088788801 \\ 
        DFTBML CC 5000 First Half & 0 & 0 & 2.825952121 & 0.037237587 & 0.058284592 \\ 
        DFTBML CC 5000 Second Half & 0 & 0 & 2.90015058 & 0.037661643 & 0.051032672 \\ \bottomrule
    \end{tabular}}
\end{table}

\subsection{COMP6 benchmark performance}\label{sec:COMP6_benchmark}
To explore transfer of the model to molecules well outside the above training and testing data, we report the performance of DFTBML on the COMP6 benchmark suite developed by Isayev and colleagues\cite{Smith2018LessLearning}. The benchmark suite contains a series of chemically diverse molecules, including an expansion on the S66x8 benchmark, frames obtained from molecular dynamics using the ANI-1x potential, subsets of molecules with various numbers of heavy atoms, and pharmacologically relevant structures (Table~\ref{Tab:COMP6_experiment_parameters}). 

Results are presented for Auorg and MIO, GFN1-xTB and GFN2-xTB, along with several of the DFTBML parameter sets discussed above.  To compare atomization energies, a linear reference energy term is re-fit for each comparison (see Equation~\ref{eqn:ref_energy_main} in Section \ref{sec:experimental_details}). Tables \ref{Tab:COMP6_tot_ener}, \ref{Tab:COMP6_dipole}, and \ref{Tab:COMP6_charge} show the performance for energy, dipoles, and charges, respectively. Energies are reported per atom to aid comparisons across test sets that contain molecules with vastly different sizes, and to allow comparison to published results on HIPNN+SEQM\cite{Zhou2022DeepMechanics}, an alternative approach to semiempirical machine learning that uses neural networks. 

\begin{table}[!ht]
    \centering
    \caption{Details on the DFTBML parameterizations (first three rows) and the COMP6 benchmarks (remaining rows) on which these are tested. }
    \label{Tab:COMP6_experiment_parameters}
    \small
    \resizebox{\columnwidth}{!}{\begin{tabular}{ll}
    \toprule
        \textbf{Parameterization/COMP6 set} & \textbf{Description}\\ \midrule 
        DFTBML CC/DFT & Trained on 20000 molecules, 1 - 8 heavy atoms \\ 
        Repulsive CC/DFT & Trained on 20000 molecules, 1 - 8 heavy atoms \\ 
        Transfer CC/DFT & Trained on 2500 molecules, 1 - 5 heavy atoms \\ 
        ANI MD & 1791 molecules with 11 - 158 heavy atoms \\ 
        Drugbank & 13379 molecules with 3 - 65 heavy atoms \\ 
        GDB 7 - 13 & 83670 molecules total; each GDB $n$ contains molecules with $n$ heavy atoms \\ 
        S66x8 & 528 molecules with 2 - 16 heavy atoms \\ 
        Tripeptide & 1979 molecules with 17 - 37 heavy atoms \\ \bottomrule
    \end{tabular}}
\end{table}

\begin{table}[!ht]
    \centering
    \caption{MAE of total energy in eV/atom for DFTBML and various models on the COMP6 benchmark suite. The lowest MAEs for each column are in bold. Numbers for HIPNN+SEQM are from Zhou et al.\cite{Zhou2022DeepMechanics}.}
    \label{Tab:COMP6_tot_ener}
    \resizebox{\columnwidth}{!}{\begin{tabular}{llllll}
    \toprule
        \textbf{Parameterization} & \textbf{Ani MD} & \textbf{GDB} & \textbf{Drugbank} & \textbf{S66x8} & \textbf{Tripeptide} \\ \midrule
        Auorg & 0.0056 & 0.0258 & 0.0151 & 0.0121 & 0.0078 \\ 
        MIO & 0.0050 & 0.0252 & 0.0143 & 0.0100 & 0.0077 \\ 
        GFN1-xTB & 0.0051 & 0.0216 & 0.0124 & 0.0116 & 0.0052 \\ 
        GFN2-xTB & 0.0092 & 0.0214 & 0.0122 & 0.0118 & 0.0062 \\ 
        DFTBML CC & 0.0048 & 0.0112 & 0.0086 & \textbf{0.0071} & 0.0043 \\ 
        Repulsive CC & 0.0064 & 0.0170 & 0.0120 & 0.0128 & 0.0066 \\ 
        DFTBML DFT & 0.0040 & 0.0082 & \textbf{0.0070} & 0.0072 & \textbf{0.0034} \\ 
        Repulsive DFT & 0.0053 & 0.0151 & 0.0107 & 0.0124 & 0.0052 \\ 
        Transfer CC & 0.0042 & 0.0126 & 0.0094 & 0.0101 & 0.0052 \\ 
        Transfer DFT & \textbf{0.0032} & 0.0089 & 0.0072 & 0.0077 & 0.0041 \\ 
        HIPNN+SEQM & 0.0110 & \textbf{0.0070} & 0.0090 & 0.0140 & 0.0070 \\ \bottomrule
    \end{tabular}}
\end{table}

\begin{table}[!ht]
    \centering
    \caption{MAE of dipole in e\AA{} for DFTBML and various models on the COMP6 benchmark suite. The lowest MAEs for each column are in bold.}
    \label{Tab:COMP6_dipole}
    \resizebox{\columnwidth}{!}{\begin{tabular}{llllll}
    \toprule
        \textbf{Parameterization} & \textbf{Ani MD} & \textbf{GDB} & \textbf{Drugbank} & \textbf{S66x8} & \textbf{Tripeptide} \\ \midrule
        Auorg & 0.167 & 0.105 & 0.113 & 0.062 & 0.128 \\ 
        MIO & 0.168 & 0.105 & 0.113 & 0.062 & 0.128 \\ 
        GFN1-xTB & 0.170 & 0.161 & 0.208 & 0.129 & 0.311 \\ 
        GFN2-xTB & 0.205 & 0.182 & 0.243 & 0.146 & 0.371 \\ 
        DFTBML CC & \textbf{0.104} & \textbf{0.040} & \textbf{0.053} & \textbf{0.031} & 0.073 \\ 
        Repulsive CC & 0.167 & 0.105 & 0.113 & 0.062 & 0.128 \\ 
        DFTBML DFT & 0.106 & 0.042 & 0.055 & 0.033 & \textbf{0.070} \\ 
        Repulsive DFT & 0.167 & 0.105 & 0.113 & 0.062 & 0.128 \\ 
        Transfer CC & 0.119 & 0.057 & 0.066 & 0.043 & 0.079 \\ 
        Transfer DFT & 0.115 & 0.054 & 0.063 & 0.040 & 0.074 \\ \bottomrule
    \end{tabular}}
\end{table}

\begin{table}[!ht]
    \centering
    \caption{MAE for charge, in e, for DFTBML and various models on the COMP6 benchmark suite. The lowest MAEs for each column are in bold.}
    \label{Tab:COMP6_charge}
    \resizebox{\columnwidth}{!}{\begin{tabular}{llllll}
    \toprule
        \textbf{Parameterization} & \textbf{Ani MD} & \textbf{GDB} & \textbf{Drugbank} & \textbf{S66x8} & \textbf{Tripeptide} \\ \midrule
        Auorg & 0.071 & 0.071 & 0.065 & 0.065 & 0.085 \\ 
        MIO & 0.071 & 0.071 & 0.065 & 0.065 & 0.085 \\ 
        GFN1-xTB & 0.090 & 0.096 & 0.090 & 0.094 & 0.100 \\ 
        GFN2-xTB & 0.076 & 0.084 & 0.074 & 0.084 & 0.087 \\ 
        DFTBML CC & \textbf{0.053} & \textbf{0.047} & \textbf{0.048} & \textbf{0.045} & \textbf{0.056} \\ 
        Repulsive CC & 0.071 & 0.071 & 0.065 & 0.065 & 0.085 \\ 
        DFTBML DFT & 0.057 & 0.051 & 0.052 & 0.048 & 0.060 \\ 
        Repulsive DFT & 0.071 & 0.071 & 0.065 & 0.065 & 0.085 \\ 
        Transfer CC & 0.058 & 0.054 & 0.052 & 0.049 & 0.064 \\ 
        Transfer DFT & 0.057 & 0.053 & 0.051 & 0.047 & 0.062 \\ \bottomrule
    \end{tabular}}
\end{table}

Molecules that are outliers or for which Self Consistent Field (SCF) iterations failed to converge are excluded from the comparisons. Such situations were rare, with DFTBML having less than five such instances for each test set, and xTB methods having a few hundred for the Drugbank test set. Because the COMP6 benchmarks include atomic charges but not molecular dipoles, the comparisons are to the dipole computed from charges. From the results shown in Tables~\ref{Tab:COMP6_tot_ener}~- \ref{Tab:COMP6_charge}, DFTBML models performed the best in every case except for the GDB test set, where HIPNN+SEQM performs slightly better. For total molecular energy, DFTBML performed better when trained against DFT data than CC data, which is not surprising given that the energies in the COMP6 datasets are from DFT with the wB97x functional and the 6-31G(d) basis set. For dipoles and charges, the DFTBML model trained to CC energies performs somewhat better than that trained to DFT energies (Tables~\ref{Tab:COMP6_dipole} and~\ref{Tab:COMP6_charge}). This is somewhat surprising given that, in the CC training data, the dipoles and charges are from DFT. 

In Table \ref{Tab:COMP6_tot_ener}, the DFTBML parameters for ``Transfer DFT/CC'' were trained on 2500 molecules with up to five heavy atoms while the parameters for ``DFTBML DFT/CC'' were trained on 20000 molecules with up to eight heavy atoms. Comparison of the results shows that training to a larger set of data does tend to improve performance on the COMP6 tests, but the improvements are modest. This further illustrates that reasonable DFTBML models can be obtained with relatively small amounts of training data. 

\section{Conclusion}

Here, we develop and evaluate a semiempirical quantum chemical model that can learn from large data sets while maintaining a physics-based and interpretable form. The resulting DFTBML model reduces the prediction error on the ANI-1CCX dataset, relative to standard DFTB parameterizations, by up to 75\% for energy, 58\% for dipoles, and 38\% for atomic charges. The model also transfers well to the COMP6 benchmark suite, with DFTBML improving substantially on standard DFTB parameterizations and outperforming both GFN1-xTB and GFN2-xTB. The performance of DFTBML is also somewhat better than HIPNN+SEQM\cite{Zhou2022DeepMechanics} on all but one of the COMP6 benchmarks. HIPNN+SEQM is similar to DFTBML in that the approach uses data to improve the parameters of a semiempirical Hamiltonian. HIPNN+SEQM uses a neural network to make a subset of the parameters in the PM3 Hamiltonian functions of the environment of the atom. The neural networks provide the flexibility needed for the model to learn from training data; however, the neural networks function as black boxes that are difficult to interpret. Here, the ability of the semiempirical model to learn from data is imparted by the use of a flexible form for the one-dimensional functions that describe the dependence of the interactions on interatomic distance. Regularizations are applied to ensure these functions have reasonable physical forms, such that the model is physics-based and interpretable. As a result, DFTBML is able to learn from the data and achieve a performance equivalent to HIPNN+SEQM while maintaining an interpretable form. 

The interpretability of the DFTBML model has two related aspects. The first is emphasized in this work, that the Hamiltonian and model parameters can be examined to understand the physics that is included, and excluded, from the model predictions. The other aspect relates to the intermediate quantities, such as orbital energies and populations, that come from a physics-based model. Chemists often use this additional information to make sense of the results they obtain from quantum chemical calculations\cite{Lu2012Multiwfn:Analyzer} and gain insights that go beyond numerical predictions for energy, dipole, and other specific targets.

An alternative approach to integrating ML into the DFTB model form has been explored by Fan et al.\cite{Fan2022ObtainingLearning,McSloy2023TBMaLTLearning}. DFTB obtains the electronic parameters from DFT solutions for isolated atoms that are placed in a confinement potential to include effects from surrounding atoms. The approach explored by Fan et al. uses ML to make the confinement potential a function of the atomic environment. This helps ensure the energies, interactions, and overlaps of the DFTB Hamiltonian are consistent, in that they can be traced back to atomic orbitals. This approach has been shown to improve the accuracy of DFTB for charges, dipoles and charge population analysis on molecules with up to five heavy atoms. The improvement in accuracy for dipole moments is comparable to that of the DFTBML models reported here. 

In addition to providing an interpretable model, the data requirements for DFTBML are substantially smaller than the millions of molecules needed for deep learning. Reasonable results are obtained when DFTBML is trained to as few as 300 molecules and the training saturates at about 20000 molecules. Given that generation of the \textit{ab initio} training data is a main computational bottleneck in model development, this reduction in required training data is a substantial practical advantage over deep learning. The saturation of DFTBML observed with greater amounts of training data also suggests that the performances reported here reflect the limit of a DFTB-SKF model, and that further enhancements in accuracy may require improvements to the model Hamiltonian itself, or the use of context-sensitive parameters such as in HIPNN+SEQM \cite{Zhou2022DeepMechanics}. Such extensions to the model may also help with training to multiple targets, relaxing the current tradeoff between the accuracy of energy and dipole targets (see Figure \ref{fig:target_tradeoff_overlay}).

The work here is restricted to SCF solutions for distorted structures of organic molecules consisting of C, H, N, and O. Future work includes extensions to additional elements such as transition metals, additional properties\cite{Gaus2011DFTB3:SCC-DFTB} such as excitation energies\cite{Stratmann1998AnMolecules,Yang2017ChargeSystems} and reaction barriers\cite{Kroonblawd2018GeneratingModel,Priyadarsini2021Comparative2D-surface}, and inclusion of additional interactions such as dispersion\cite{Stohr2016Communication:Materials,Mortazavi2018StructureBinding} and solvent interactions\cite{Stohr2019QuantumInteractions,Gregory2022AIons}. 

\section{Experimental details}\label{sec:experimental_details}

\begin{figure}[ht!]
    \centering
    \includegraphics[width=1.0\columnwidth]{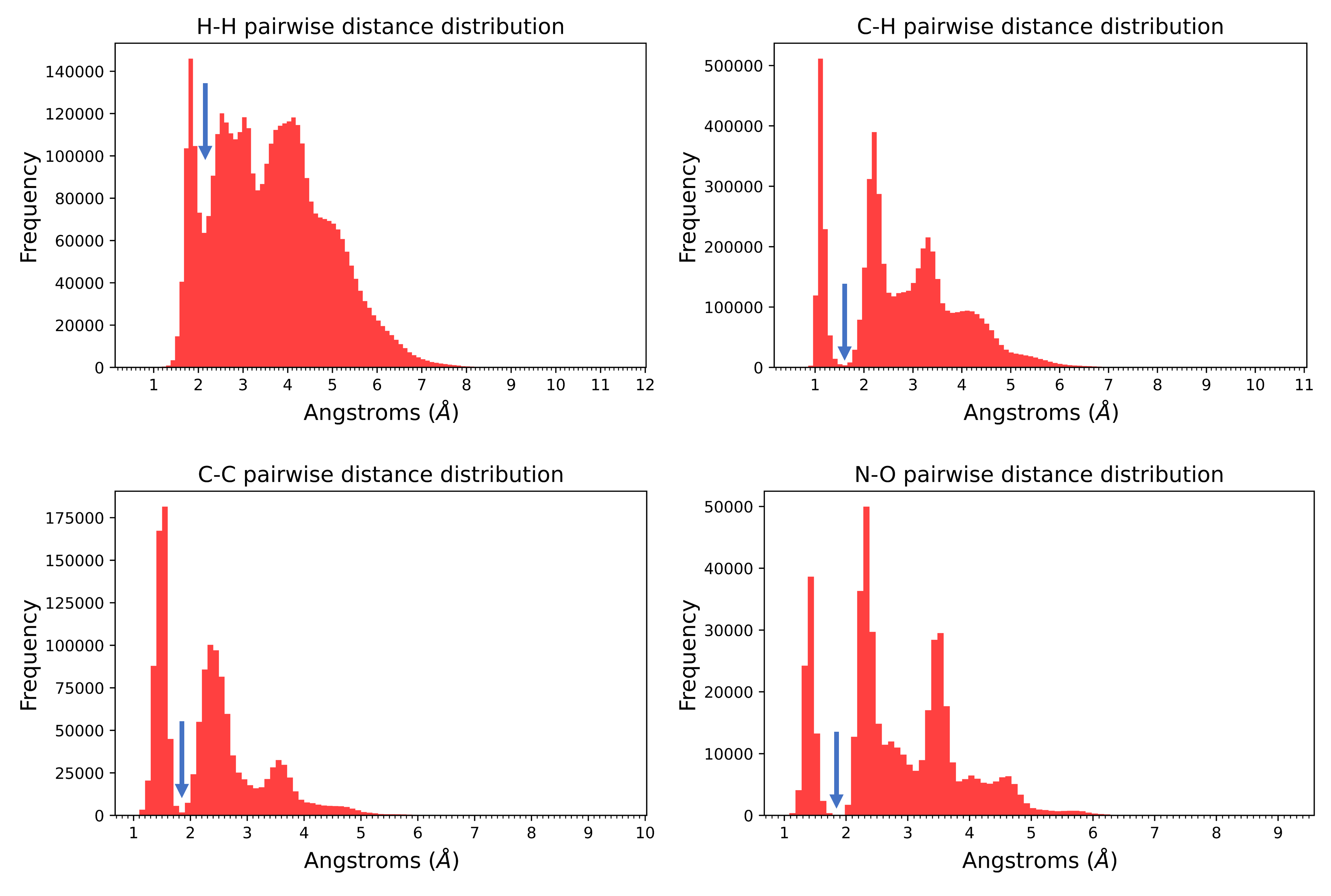}
    \caption{Distributions of internuclear distances between H-H, C-C, C-H, and N-O in the cleaned ANI-1CCX dataset for molecules with up to eight heavy atoms. Repulsive interactions are truncated beyond nearest-neighbor interactions (blue arrows) with a lower bound of 0 \AA{}. Electronic interactions go to longer range (4.5 \AA{}) with a lower bound slightly lower than the shortest distance in a given distribution. Precise cutoffs for electronic and repulsive splines can be found in Tables S2 and S3 of the Supporting Information, respectively.}
    \label{fig:ANI-1ccx_dist_hists}
\end{figure}

The DFTB method uses a physics based procedure to derive Hamiltonian matrix elements for a valence-only minimal atomic basis set. \textit{Ab initio} data on molecules is used only to determine an empirical pairwise-additive repulsive potential that accounts for interactions between core electrons not included in the electronic Hamiltonian. Here, we instead fit all aspects of the model to \textit{ab initio} data while retaining the following restrictions imposed by the SKF file format: the atomic orbital energies are trained constants; the one-electron Hamiltonian matrix elements ($\mathbf{H_1}$), overlap integrals ($\mathbf{S}$), and repulsive potentials ($\mathbf{R}$) are functions of only interatomic distance; and Coulombic interactions ($\mathbf{G}$) use a model form that depends only on Hubbard parameters associated with the atomic shells\cite{Elstner1998Self-consistent-chargeProperties}. We use fifth-order splines for the electronic (Hamiltonian and overlap) functions and third-order splines for the repulsive potentials, with distance ranges specified by analyzing distributions of pairwise distances (see Figure \ref{fig:ANI-1ccx_dist_hists}). Boundary conditions are applied only at the upper limit, where we force both the function and its derivative to go to zero at large interatomic separations. No boundary conditions are imposed at the lower limit. In addition to retaining a physics-based and interpretable model form, SKF-DFTB has the advantage that trained models can be easily distributed through SKF files that are supported by many computational chemistry packages\cite{Hourahine2020DFTB+Simulations,Frisch2016GaussianC.01,amber22}. All test results quoted here were obtained from DFTB+ using SKF files produced by our training code. This approach gives a stronger guarantee about the validity of the quoted model performances.

\begin{figure}[!ht]
    \centering
    \includegraphics[width=1.0\columnwidth]{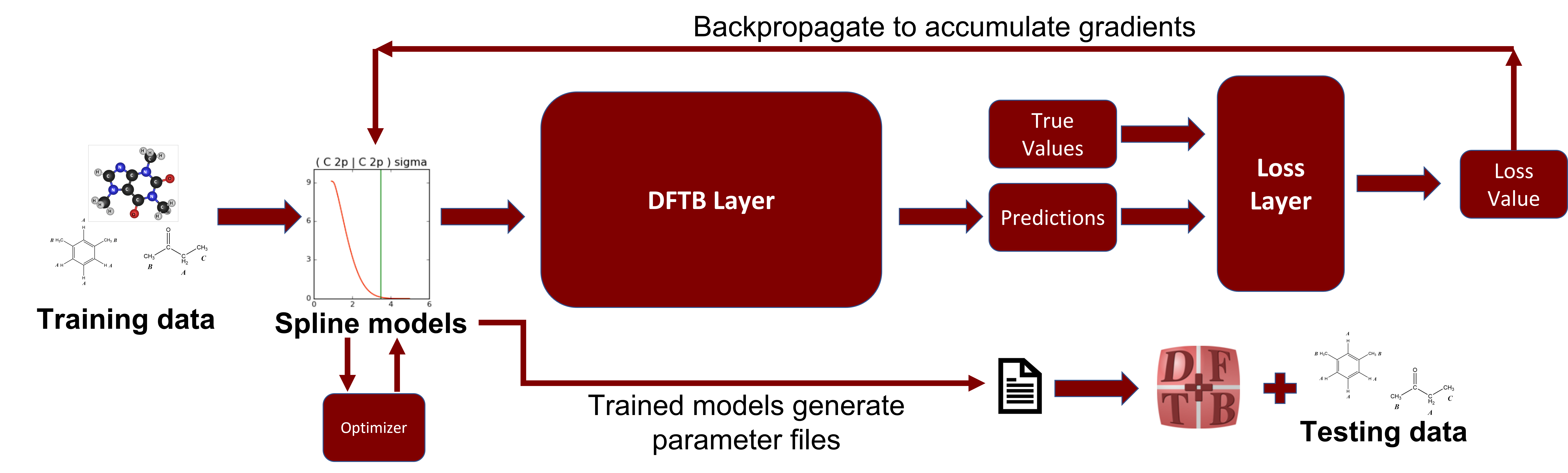}
    \caption{High-level overview of the DFTBML model workflow. Note that model testing uses DFTB+ and is external to model training (lower right).}
    \label{fig:model_workflow}
\end{figure}

Figure \ref{fig:model_workflow} gives an overview of the DFTBML model structure implemented using PyTorch\cite{Paszke2017AutomaticPyTorch}. The DFTB layer\cite{Li2018} is central to the DFTBML model as it solves the quantum chemical system for the desired properties on each forward pass. The training and validation data are randomly divided into batches which each contain 10 configurations. During each epoch of the training process, the batches are randomly shuffled and fed through the model. Since a training experiment can consist of thousands of epochs, a precomputation is used to calculate and save quantities that do not depend on trained model parameters. This significantly decreases the training time but does come at the cost of increased memory usage and fixed batch compositions. The PyTorch implementation does, however, remove restrictions in the previous TensorFlow implementation, which required all batches to have the same sequence of empirical formulas\cite{Li2018}. 

\begin{figure}[ht]
    \centering
    \includegraphics[width=1.0\columnwidth]{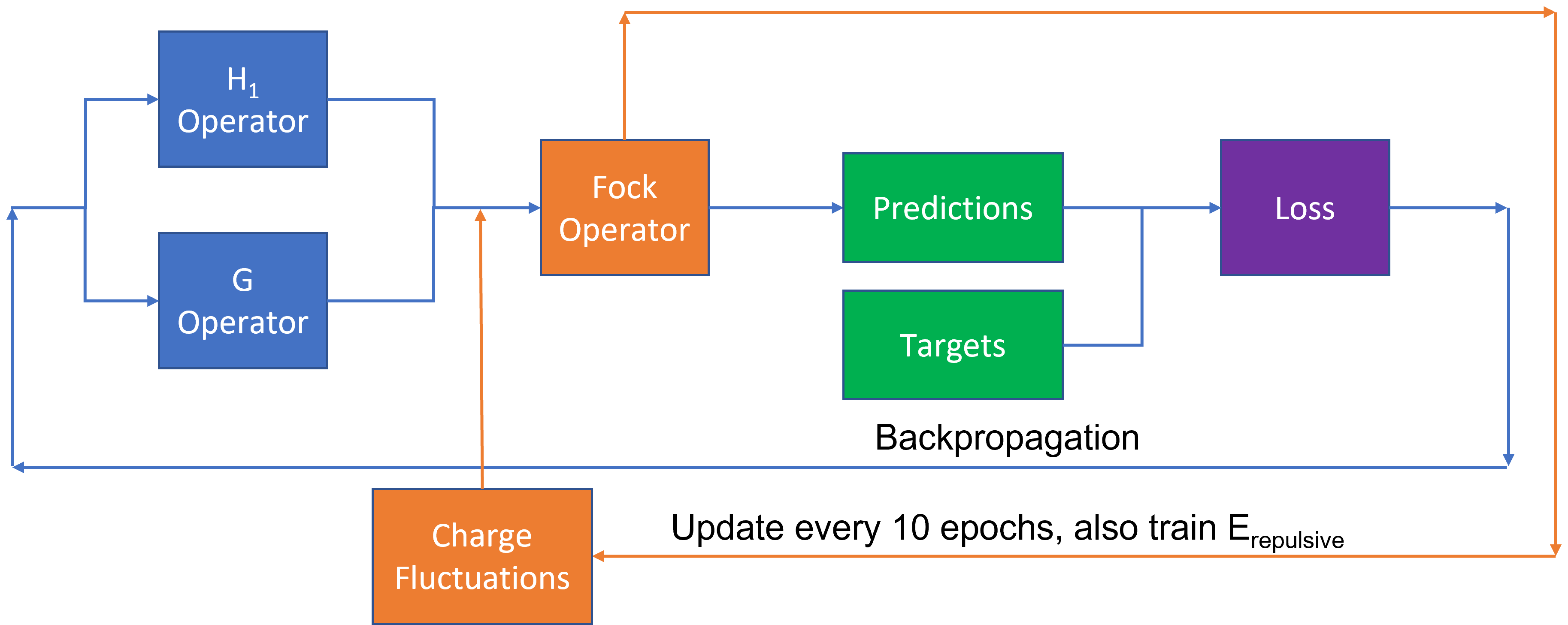}
    \caption{Schematic illustration of inverting the SCF (orange arrows) and training (blue arrows) loops of the DFTBML workflow. In the outer loop, the charge fluctuations needed for the Fock operator are updated based on the current model parameters. The repulsive model is updated on the same schedule as the charge fluctuations.}
    \label{fig:scf_inversion_loop}
\end{figure}

To enable efficient backpropagation through SCF calculations during training, the SCF and training loops are inverted as shown in Figure \ref{fig:scf_inversion_loop}. This loop inversion scheme avoids backpropagation through multiple SCF cycles and instead moves the update of the charge fluctuations, required for the construction of the Fock operator, outside of the gradient descent steps used to improve the model parameters\cite{Li2018}. SCF calculations are performed every 10 epochs throughout training. Updates to the repulsive model are also done every 10 epochs. Because the repulsive model and associated regularizations are linear, convex optimization can be used to find the global optimum for the entire training set. Implementation details of the repulsive model can be found in Section S4 of the Supporting Information. 

\begin{figure}[ht]
    \centering
    \includegraphics[width=1.0\columnwidth]{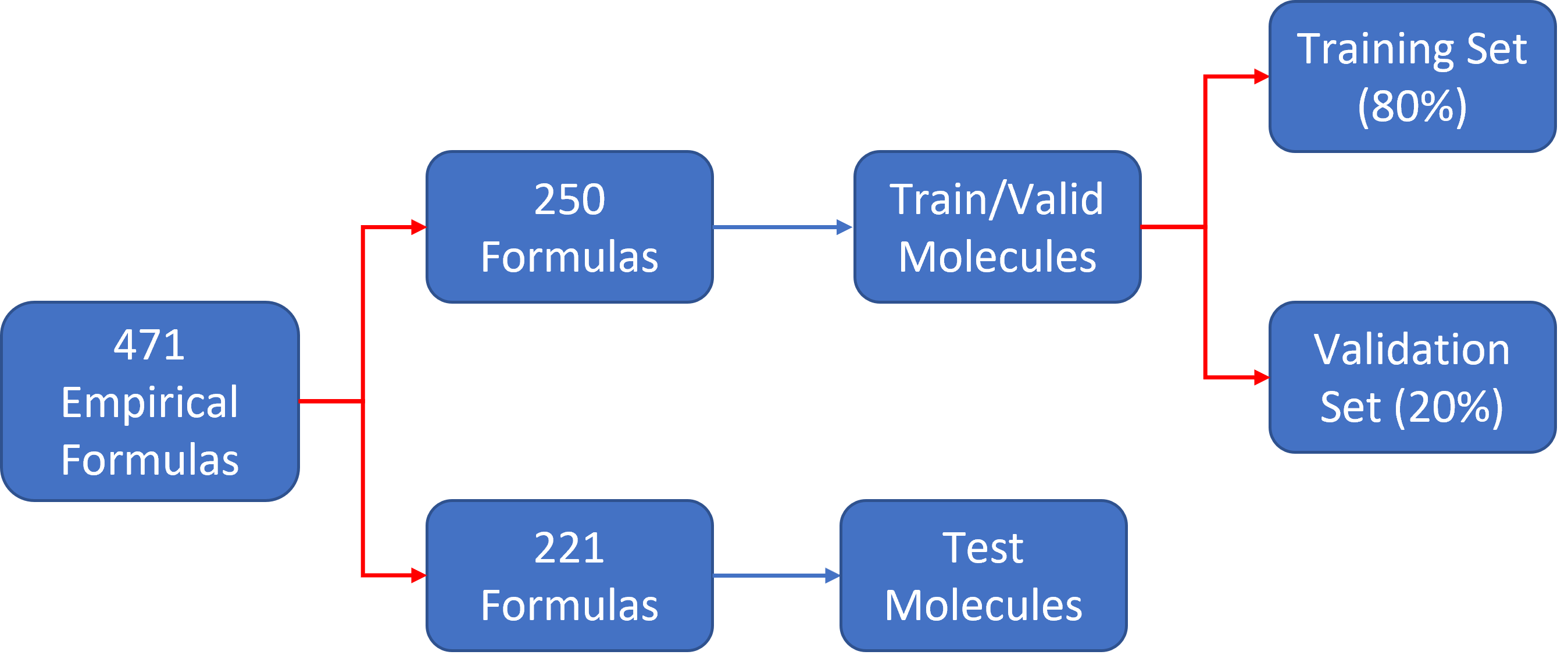}
    \caption{High-level overview of the method used to generate datasets. Red arrows indicate random sampling. Molecules are divided based on their empirical formulas, ensuring no mixing between training and testing data.}
    \label{fig:DsetGen}
\end{figure}

For developing DFTBML, we use the ANI-1CCX dataset\cite{Smith2020TheMolecules} which contains organic molecules with only C, H, N, and O. We focus here only on molecules with up to eight heavy (non-hydrogen) atoms and we retain only configurations which have complete entries for all fields, resulting in 471 unique empirical formulas with a total of 232310 molecular configurations. The division of the data into training, validation, and testing data is shown schematically in Figure \ref{fig:DsetGen}. We divide molecules by empirical formulas to ensure that there is no overlap between training and testing data. A more detailed explanation can be found in Section S10 of the Supporting Information.

Comparisons between different quantum chemical methods are done on atomization energies. This is implemented through a linear reference energy correction which has the following form,
\begin{align}
    \label{eqn:ref_energy_main}
    E_{ref} = \sum_{Z} N_Z C_Z + C_0
\end{align}
where the sum is over elements, $N_Z$ is the number of times element $Z$ appears in the molecule, $C_Z$ is a coefficient for element $Z$, and $C_0$ is a constant term. The coefficients are obtained through a least squares fit of Equation~\ref{eqn:ref_energy_main} to the energy differences between the two quantum chemical methods being compared. The reported MAEs refer to the residuals from this least-squares fit. While training DFTBML, the reference energy is incorporated into the repulsive potential (see Section S4 of the Supporting Information).

All experiments presented use the ADAM optimizer with a learning rate of 1E-05 and the default values for all other parameters\cite{Kingma2014Adam:Optimization}. All models were trained for 2500 epochs, with a learning rate scheduler that reduces the learning rate by a factor of 0.9 when a plateau is detected in performance improvements. The loss function combines the root-mean-square error for multiple targets, with weights of: 6270 Ha$^{-1}$ for total energy (Ha), 100 (e\AA{})$^{-1}$ for dipoles (e\AA{}), and 1 e$^{-1}$ for charges (e). The sensitivity of the results to the number of knots in the splines and weights for the regularization penalties are provided in Section S12 of the Supporting Information. Results for xTB use the standard implementations of GFN1-xTB and GFN2-xTB\cite{Grimme2017A1-86,Bannwarth2019GFN2-xTBContributions,Koopman2019CalculationMethods} distributed via Anaconda. In reporting model performance, outliers are removed based on a threshold of 20 standard deviations above the mean error for total energy. The highest percentage of outliers for DFTBML was less than 0.05\%. Per atom and per heavy atom results are provided in the Supporting Information to facilitate comparison to other studies including HIPNN+SEQM\cite{Zhou2022DeepMechanics}.

\clearpage
\bibliography{references.bib,misc.bib}

\clearpage
For Table of Contents only.

\includegraphics[width=1.0\columnwidth]{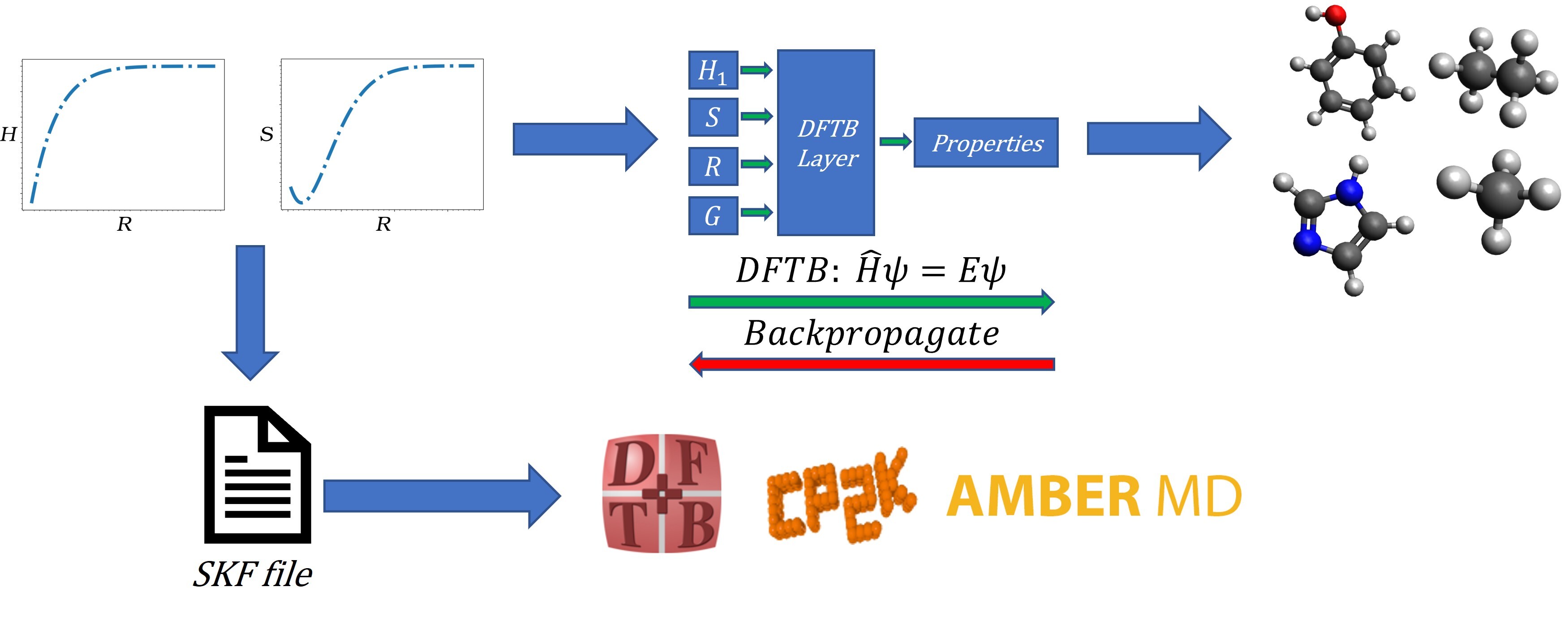}

\end{document}


\maketitle
\tableofcontents

\section{DFTBML model timing}\label{sec:timing}

For reporting the DFTBML timings, the total process time in hours is used which is the sum of the total system and user CPU time. A distinction is made between SCF and non-SCF epoch timings, since the SCF cycle, which is performed every 10 epochs, is done outside of the PyTorch code used to update parameters. Table \ref{Tab:experiment_timings} gives the average epoch time for both SCF and non-SCF epochs ($\mu_{SCF}$ and $\mu_{non-SCF}$) and the standard deviation of the epoch times for SCF and non-SCF epochs ($\sigma_{SCF}$ and $\sigma_{non-SCF}$). The percentage of the total time spent performing SCF  calculations (\%$_{SCF}$) is also reported and is approximated as:
\begin{align}
    \label{eqn:DFTBML_timing_estimation}
    \%_{SCF} \approx \left(\frac{\left(\mu_{SCF} - \mu_{non-SCF}\right)*250}{T_{tot}}\right)*100
\end{align}

Where $T_{tot}$ is the total amount of time taken for the experiment. The factor of 250 is used because each experiment is run for 2500 epochs and charge updates happen every 10 epochs, meaning a total of 250 epochs where there is an additional SCF time added on top of the normal epoch runtime. This quantity is not the exact breakdown but serves to give an idea of how much time is spent in the SCF cycle relative to the rest of the training process. The repulsive update calculation that happens every 10 epochs is not included in the measured timings since solving the convex optimization problem takes negligible time relative to charge updates even when training on 20000 molecules. All experiments were run using 16 CPU cores.

\begin{table}[!ht]
    \centering
    \caption{Total CPU process time used for different experiments in hours}
    \label{Tab:experiment_timings}
    \sisetup{round-mode=places}
    \resizebox{\textwidth}{!}{\begin{tabular}{lS[round-precision=2]S[round-precision=4]S[round-precision=2]S[round-precision=4]S[round-precision=2]S[round-precision=2]}
    \toprule
        \textbf{Parameterization} & \boldmath{$\mu_{non-SCF}$} & \boldmath{$\sigma_{non-SCF}$} & \boldmath{$\mu_{SCF}$} & \boldmath{$\sigma_{SCF}$} & \boldmath{$T_{tot}$} & \boldmath{$\%_{SCF}$} \\ \midrule
        DFTBML 20000 CC & 0.812488818 & 0.10341768 & 1.805206314 & 0.238482331 & 2279.401418 & 10.88791873 \\ 
        DFTBML 10000 CC & 0.488510654 & 0.016861131 & 1.125306706 & 0.050511048 & 1380.475648 & 11.53218555 \\ 
        DFTBML 5000 CC & 0.241269118 & 0.008908348 & 0.534541732 & 0.021258215 & 676.490949 & 10.83800953 \\ 
        DFTBML 2500 CC & 0.120208755 & 0.00398104 & 0.27044037 & 0.008454145 & 338.0797908 & 11.1091833 \\ 
        DFTBML 1000 CC & 0.052579062 & 0.00303422 & 0.127786527 & 0.007490315 & 150.2495217 & 12.51376117 \\ 
        DFTBML 300 CC & 0.022825617 & 0.000719352 & 0.06752258 & 0.001960978 & 68.23828239 & 16.37532557 \\ \midrule
        DFTBML 20000 DFT & 0.876105942 & 0.1216941 & 1.93629984 & 0.274065561 & 2455.313329 & 10.79489413 \\ 
        DFTBML 10000 DFT & 0.37299232 & 0.006635578 & 0.829500403 & 0.016599501 & 1046.60782 & 10.90446858 \\ 
        DFTBML 5000 DFT & 0.242396153 & 0.008551628 & 0.538917181 & 0.016933434 & 680.1206403 & 10.89957465 \\ 
        DFTBML 2500 DFT & 0.119952137 & 0.003259832 & 0.273049673 & 0.007788659 & 338.1547255 & 11.31860099 \\ 
        DFTBML 1000 DFT & 0.05254871 & 0.002058086 & 0.12737422 & 0.005502375 & 150.0781536 & 12.46442401 \\ 
        DFTBML 300 DFT & 0.023261844 & 0.000557023 & 0.06745256 & 0.001439361 & 69.20228946 & 15.96432565 \\ \bottomrule
    \end{tabular}}
\end{table}

From Table \ref{Tab:experiment_timings}, it is evident that the amount of time taken scales roughly linearly with the amount of data used, where doubling the data approximately doubles $T_{tot}$. There is also no systematic difference between the timing performance when training to the CC total energy target versus the DFT total energy target. Furthermore, the percentage of time spent performing the required SCF calculations takes up less than 20\% for all experiments, and $\%_{SCF}$ increases as the total amount of training time decreases, which is expected.  
\clearpage

\section{Spline Implementation}\label{sec:spline_detailed_deriv}
For a spline in a B-spline basis of order $k$, a prediction $y$ for a given input $x$ is generated through the linear combination of the set of $N$ B-spline basis functions $\{b_{j}^k\}$ using the set of coefficients $\{\beta_j\}$ as follows:
\begin{align}
    y = \sum_{i = 1}^N \beta_i b_{i}^k(x) + \beta_0
\end{align}

For DFTBML, the input $x$ corresponds to an interatomic distance. Instead of predicting a single value at a time, we now wish to predict a vector of values $\mathbf{y}$, essentially performing the following transformation:
\begin{align}
    y_j &= \sum_{i = 0}^N \beta_i b_{i}^k(x_j), \forall j
\end{align}
Where the constant term $\beta_0$ is subsumed into the summation and $b_{0}^k = 1$. Now, we can transform this into matrix form by introducing a second index $j$ as follows:
\begin{align}
    \label{eqn:spline_matrix_multiplication_first_part}
    y_j &= \sum_{i = 0}^N \beta_i b_{i,j}^k, \forall j\\
    \label{eqn:spline_basis_matrix_rearrangement}
    \mathbf{y} &= (b_{j,i}^k) \boldsymbol\beta
\end{align}

The quantity $(b_{j,i}^k)$ is the spline basis matrix, and $\boldsymbol\beta$ is the coefficient vector. Setting $\mathbf{A} = (b_{j,i}^k)$ and $\mathbf{x} = \boldsymbol\beta$, we get the matrix multiplication 
\begin{align}
    \label{eqn:SI_basic_spline_form}
    \mathbf{y} = \mathbf{Ax} + \mathbf{c}
\end{align}

Where $\mathbf{c}$ is used to account for boundary conditions specifying fixed values. 

Predictions for higher derivatives can be obtained for the spline models with the same coefficient vector $\mathbf{x}$ for each derivative. From Equation \ref{eqn:spline_matrix_multiplication_first_part}, predicting the value of a higher derivative can be done as follows:
\begin{align}
    \label{eqn:spline_higher_derivative_derivation}
    y_j^{(n)} &= \frac{d^n}{dx^n} \sum_{i = 0}^N \beta_i b_{i,j}^k = \sum_{i=0}^N \beta_i b_{i,j}^{(n),k}, \forall j\\
    \mathbf{y}^{(n)} &= (b_{j,i}^{(n),k})\boldsymbol\beta
\end{align}
Where $n$ is the order of the derivative to calculate. Setting the matrix $\mathbf{A}^{(n)} = (b_{j,i}^{(n),k})$ gives a similar result to Equation \ref{eqn:SI_basic_spline_form}, where:
\begin{align}
    \label{eqn:spline_matmul_higher_deriv}
    \mathbf{y}^{(n)} = \mathbf{A}^{(n)}\mathbf{x} + \mathbf{c}^{(n)}
\end{align}

\clearpage

\section{The DFTB method}

A derivation of the DFTB method can be found in work by Elstner et al.\cite{Elstner1998Self-consistent-chargeProperties,Elstner2014DensityBinding} and in previous work by Li et al.\cite{Li2018}. For DFTBML, we include a classical pairwise repulsive term as well as a reference energy correction which takes the following form:
\begin{align}
    \label{eqn:ref_ener}
    E_{ref} = \sum_{z\in \{m\}} N_z C_z + C_0
\end{align}
Where $\{m\}$ is the set of all atomic numbers needed to describe a given molecule, $N_z$ is the number of times atom $z$ appears in the molecule, $C_z$ is the coefficient for atom $z$, and $C_0$ is a constant term. The coefficients are obtained through a least squares fit. In comparing two different quantum chemical methods, disagreements refer to to the residuals from this least-squares fit. 

The parameters of DFTB are the Hubbard parameters that model the coulombic interactions within electron shells, on-site energies for each type of atomic orbital (e.g. 2s on C, 2p on N), neutral orbital occupations for each element, and the Hamiltonian and overlap integrals which form the Hamiltonian and overlap operator matrices.

Formally, the Hamiltonian integrals can be written as:
\begin{align}
    \label{eqn:Hamiltonian_potential}
    \Big<\phi_{\mu}(\mathbf{r})\Big|-\frac{1}{2}\nabla^2 + \nu_{\text{eff}}[n^{\alpha}(\mathbf{r})] + \nu_{\text{eff}}[n^{\beta}(\mathbf{r} - \mathbf{r_0})]
    \Big|\phi_{\nu}(\mathbf{r} - \mathbf{r_0})\Big>, \mu \in \alpha, \nu \in \beta\\
    \label{eqn:Hamiltonian_superposition}
    \Big<\phi_{\mu}(\mathbf{r})\Big|-\frac{1}{2}\nabla^2 + \nu_{\text{eff}}[n^{\alpha}(\mathbf{r}) + n^{\beta}(\mathbf{r} - \mathbf{r_0})]
    \Big|\phi_{\nu}(\mathbf{r} - \mathbf{r_0})\Big>, \mu \in \alpha, \nu \in \beta
\end{align}
Where the first case is the potential case and the second case is the superposition case. $n^\alpha$ and $n^\beta$ are the atomic densities and $\nu_{\text{eff}}$ is the effective potential. $\phi_\mu$ and $\phi_\nu$ are the atomic valence basis functions for atom $\alpha$ and $\beta$, respectively, and $\mathbf{r}$ is the internuclear distance vector. 

The overlap integrals can be written as:
\begin{align}
    \label{eqn:overlap_integral}
    \Big<\phi_{\mu}(\mathbf{r})\Big|\phi_{\nu}(\mathbf{r - r_0})\Big>,\mu \in \alpha, \nu \in \beta
\end{align}
For both the Hamiltonian and overlap integrals, the integrals are evaluated in orientations corresponding to $\sigma$, $\pi$, and, for $d$ orbitals, $\delta$. 

In traditional DFTB, approximate atomic orbitals are used to explicitly evaluate the integrals of Eqs.~\ref{eqn:Hamiltonian_potential} through~\ref{eqn:overlap_integral}. In DFTBML, these integrals are instead derived from fits to the training data. 

\clearpage

\section{Repulsive model formulation}\label{sec:reference+repulsive}

The DFTB layer\cite{Li2018} handles calculations of the electronic energy. However, the total molecular energy is comprised of the electronic ($E_{elec}$), repulsive ($E_{rep}$), and reference ($E_{ref}$) energies, i.e. $E_{tot} = E_{elec} + E_{rep} + E_{ref}$. Here, we provide an overview of the repulsive model which accounts for the $E_{rep}$ and $E_{ref}$ contributions. 

The repulsive energy is modeled using pairwise additive models. Consider a set of molecules where each molecule contains $D$ atoms with $M$ many atom types. Then, the repulsive energy of a single molecule is:
\begin{align}
    E_{rep_{mol}} = \sum_{i<j}^D f_{Z_i,Z_j}(|r_i-r_j|), Z_i, Z_j \in M
\end{align}
Where $f_{Z_i,Z_j}$ denotes the pairwise function describing a repulsive interaction between atoms $Z_i$ and $Z_j$ that only depends on their internuclear distance $|r_i - r_j|$. In the DFTBML implementation, the set of functions $\{f_{Z_i,Z_j}\}$ is modeled using cubic splines represented in a B-spline basis. Rather than training the reference energy separately, it is incorporated into the repulsive energy so that the full formulation is as follows:
\begin{align}
    \label{eqn:new_rep}
    E_{newrep_{mol}} = \sum_{i<j}^D f_{Z_i,Z_j}(|r_i-r_j|) +  \sum_{z}^M N_z C_z + C_0
\end{align}
The first term in Equation \ref{eqn:new_rep} can be expressed as a matrix multiplication between a matrix that depends on the distances present in the molecule, and a vector that contains the trainable parameters associated with the spline, $\mathbf{x}$ and $\mathbf{c}$ of Equation \ref{eqn:SI_basic_spline_form}.  The relation between the energy and the trainable parameters in Eq.~\ref{eqn:new_rep} is linear and so may, for each molecule, be written as,
\begin{align}
    \label{eqn:total_rep_equation}
    E^{mol}_{rep+ref} = \boldsymbol\gamma_{mol}\mathbf{x}
\end{align}
where $\mathbf{x}$ is a vector holding all training parameters and $\boldsymbol\gamma$ describes the linear relation between the energy of the molecule and these parameters. 

Unlike the computation of the electronic energy through the DFTB layer, the repulsive energy is linear in model parameters and optimization does not require a gradient descent procedure. We instead uses a quadratic programming approach\cite{Vandenberghe2010TheSolvers} to solve for the globally optimal solution, using the CVXOPT package in Python. Quadratic programming aims to solve the following program for the coefficient vector $\mathbf{x}$:

\begin{align}
    \label{eqn:repulsive_quadratic_program}
    &\underset{\mathbf{x}\in \{\mathbf{R}^n\}}{\mathrm{argmin}} \left(\frac{1}{2}\right) \mathbf{x}^T\mathbf{P}\mathbf{x} + \mathbf{q}^T\mathbf{x}\\
    \label{eqn:repulsive_inequality_constraint}
    &\mathbf{Gx}\preceq\mathbf{h}\\
    \label{eqn:repulsive_equality_constraint}
    &\mathbf{Ax}=\mathbf{b}
\end{align}
Where Equation \ref{eqn:repulsive_quadratic_program} is linear least squares optimization, Equation \ref{eqn:repulsive_inequality_constraint} specifies an inequality constraint on the coefficient vector $\mathbf{x}$ and Equation \ref{eqn:repulsive_equality_constraint} specifies an equality constraint. 
For repulsive potentials, the spline model is regularized by forcing the function to be monotonically decaying. This constraint, that the first derivative be negative, can be written as the linear inequality of Equation \ref{eqn:repulsive_inequality_constraint}. For our application here, we do not apply the equality constraint of Equation \ref{eqn:repulsive_equality_constraint}. 

The matrices $\mathbf{P}$ and $\mathbf{q}$ are: 
\begin{align}
    \mathbf{P} &= \sum_{mol} \boldsymbol\gamma_{mol}^T\boldsymbol\gamma_{mol}\\
    \mathbf{q} &= -\sum_{mol} \left(\frac{1}{N_{mol}}\right)\mathbf{y}_{mol}^T\boldsymbol\gamma_{mol}
\end{align}
Where the sum is over all molecular configurations, and $\mathbf{y}_{mol}$ is the target that the repulsive energy is being trained to. The target for the repulsive potential is the difference between the total molecular energy and the electronic energy. Both the predicted and target molecular energies are divided by the number of heavy atoms, so that the optimization is performed on the energy per heavy atom. 

\clearpage

\section{Parameterization of the repulsive potential in SKF-DFTB}\label{sec:SKF_repulsive_format}
The repulsive interaction is modeled as a cubic spline with a fifth degree polynomial to describe the final interval. For distances below where the spline begins, the repulsive energy is assumed to have the following form near the repulsive wall:
\begin{align}
    e^{-a_1r+a_2} + a_3
\end{align}
Where $a_1$, $a_2$, and $a_3$ are constants specified in the file. In DFTBML, we do not train these constants and instead use the spline repulsive to describe all relevant repulsive interactions, thus spanning a range which encompasses all physically relevant distances.

The remaining form of the spline repulsive is a series of coefficients $c_0$, $c_1$, $c_2$, and $c_3$ which specify the following cubic function spanning the distance of the associated interval:
\begin{align}
    \label{eqn:repulsive_skf_form}
    c_0 + c_1(r-r_0) + c_2(r-r_0)^2 + c_3(r-r_0)^3
\end{align}
The final series of coefficients specifies a fifth order polynomial with two additional coefficients $c_4$ and $c_5$:
\begin{align}
    \label{eqn:repulsive_skf_final_polynomial}
    c_0 + c_1(r-r_0) + c_2(r-r_0)^2 + c_3(r-r_0)^3 + c_4(r-r_0)^4 + c_5(r-r_0)^5
\end{align}
However, in DFTBML, this fifth order polynomial is reduced to a cubic one by setting $c_4 = c_5 = 0$ since the entirety of the repulsive potential is modeled using a cubic spline in a B-spline basis. 

\clearpage

\section{Spline regularization}\label{sec:SI_loss_reg}
Formally, the loss for each batch takes on a Root-Mean-Square (RMS) definition as follows:
\begin{align}
    \label{eqn:master_loss_fxn}
    Loss = \sum_{prop} w_{prop} \sqrt{\frac{1}{N_{prop}}\sum_{i}^{N_{prop}}|Pred_i - Target_i|^2} + L_{form}
\end{align}
Where the summation goes over all properties of interest and $Pred$ and $Target$ are the predicted and target values for each property, respectively. $N_{prop}$ is the number of predicted values for each property, and $w_{prop}$ is a weighting factor used to target certain attributes. The $w_{prop}$ values are treated as hyperparameters throughout training. Multiple properties are considered in the loss function because DFTBML is a multitask learning model, where multiple targets are simultaneously optimized. Of interest here is total molecular energy, molecular dipole, and atomic charge. 

The final term in Equation \ref{eqn:master_loss_fxn}, $L_{form}$, is a regularization term for the electronic splines used to model elements of the Hamiltonian and overlap operator elements. $L_{form}$ contains two terms, with one governing the curvature of the spline and the other penalizing the magnitude of the third derivative as follows:
\begin{align}
    \label{eqn:SI_l_form_simplified_form}
    L_{form} &= w_{convex} * L_{convex} + w_{TD} * L_{TD}\\
    \label{eqn:SI_second_deriv}
    L_{convex} &= \sqrt{\frac{1}{K} \sum_{mod}^K \frac{|ReLU(dsgn(mod)(\mathbf{y}^{(2)}_{mod}\circ\mathbf{p}_{mod}))|^2}{N_{mod}}}\\
    \label{eqn:SI_third_deriv}
    L_{TD} &= \sqrt{\frac{1}{K} \sum_{mod}^K \sqrt{\frac{|\mathbf{y}^{(3)}_{mod}|^2}{N_{mod}}}}
\end{align}

Where $K$ is the number of spline models contained in the batch, $\mathbf{y}^{(2)}_{mod}$ is the vector of predicted second derivative values for the current indexed model, $\mathbf{y}^{(3)}_{mod}$ is the vector of predicted third derivative values for the current indexed model, and $N_{mod}$ is the number of values predicted for the given model for the second or third derivatives. $ReLU$ is the rectified linear unit function and $dsgn(\cdot)$ evalutes to 1 or $-1$ for the given model depending on the sign of the model's curvature. Because the spline models are univariate curvatures as a function of distance, each model is either mostly concave up or concave down. The $dsgn(\cdot)$ function returns 1 for a negative integral and -1 for a positive integral so that applying $ReLU$ selects the correct values to penalize.

\begin{figure}[ht!]
    \centering
    \includegraphics[width=1.0\columnwidth]{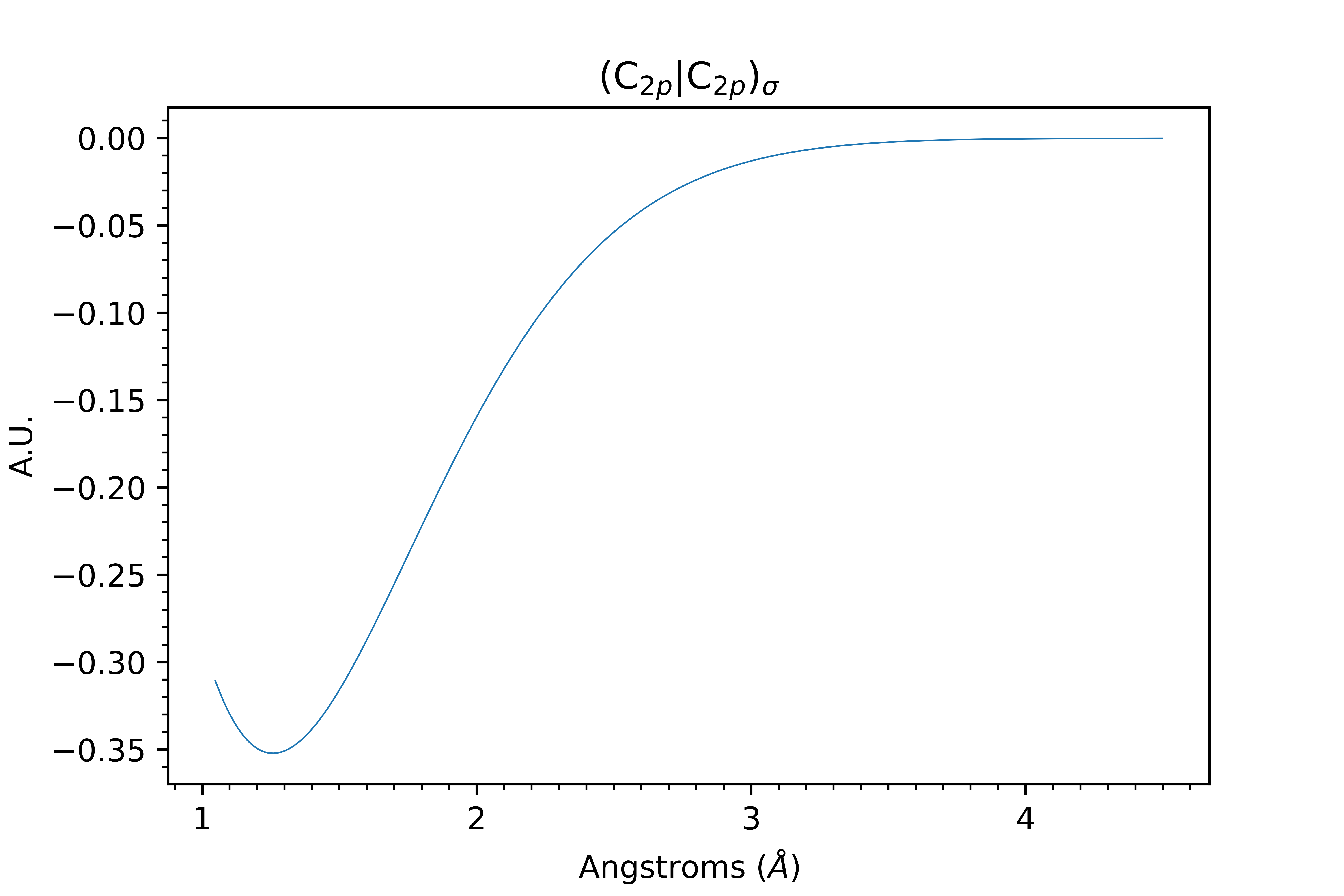}
    \caption{The spline used to model the overlap interaction between two carbon 2p orbitals in the $\sigma$ orientation. The y-axis is represented in arbitrary units for the overlap integral.}
    \label{fig:Auorg_CC_overlap}
\end{figure}

For the overlap operator, there are cases where an inflection point can exist at short range. This inflection arises from the nodal structure of the atomic orbitals. For those cases where an inflection point is allowed, the penalty vector $\mathbf{p}_{mod}$ is multiplied element-wise into the predictions of the second derivative before the functions $dsgn(\cdot)$ and $ReLU$ are applied. The penalty vector $\mathbf{p}_{mod}$ is used to account for functional forms which have a change to upward curvature at short range, a phenomenon seen in the overlap integrals of the Auorg and MIO parameterizations. Figure \ref{fig:Auorg_CC_overlap} shows an example of this with the $\big(C_{2p}|C_{2p})_\sigma$ overlap matrix element.

For the derivation of $\mathbf{p}_{mod}$, the first step is to define the functional form of the inflection point. In DFTBML, the inflection point is defined as follows:
\begin{align}
    \label{eqn:inflect_point_master_form}
    r_{inflect} = r_{l} + \left(\frac{r_{h} - r_{l}}{2}\right)\left(\frac{2 \arctan{x}}{\pi} + 1\right)
\end{align}
Where $r_l$ and $r_h$ are the boundary values for a given model and $x$ is the variable which is optimized during the gradient descent procedure. In this way, the inflection point is tied to a single variable which simplifies the training process. During training, the current value of the inflection point is calculated using Equation \ref{eqn:inflect_point_master_form} and that value of $r_{inflect}$ is used to calculate $\mathbf{p}_{mod}$ as follows:
\begin{align}
    \label{eqn:penalty_vector_formula}
    p_i = \arctan\left({10 (r_{i} - r_{inflect})}\right), \forall i
\end{align}
Where the $r_i$ are the distances corresponding to the predictions of the second derivative. This method means that for all distances in $\mathbf{p}_{mod}$ that are greater than the inflection point, you have a positive value whereas for all distances smaller than the inflection point, you have some negative value. Multiplying this into the second derivative vector allows for the sign of the second derivative to change once across the inflection point in a smooth and differentiable manner. 

\clearpage

\section{Backpropagation and degeneracy}
One of the key steps in the DFTB layer is the formation and diagonalization of the Fock matrix. In evaluating the gradients needed to backpropagate through the eigensystem, singularities arise for degenerate orbitals. These singularities are not related to the occupation of the orbitals, and arise even if the degenerate orbitals are fully occupied or are completely unoccupied. In the original work with the DFTB layer\cite{Li2018}, the effects of these singularities were reduced by removing symmetric molecular configurations from the training data, with the symmetry being measured by the separation between orbital energies computed using the MIO DFTB parameters\cite{Elstner1998Self-consistent-chargeProperties}. 

Here, we use a more general approach based on the eigenvalue broadening of Seeger et al.\cite{Seeger2017}. In the forward pass through the model, the symmetric eigendecomposition is as follows:
\begin{align}
    &(\mathbf{U},\boldsymbol\lambda) = syevd(\mathbf{A})\\
    &\mathbf{A} = \mathbf{U}^T(diag(\boldsymbol\lambda))\mathbf{U}\\
    \label{eqn:unitary_U}
    &\mathbf{U}^T\mathbf{U} = \mathbf{I}
\end{align}
Where $syevd(\cdot)$ represents the symmetric eigenvalue decomposition function, $\mathbf{A}$ is the matrix we are decomposing, $\mathbf{U}$ is the matrix of eigenvectors, $\boldsymbol\lambda$ is the matrix of eigenvalues, and $\mathbf{I}$ is the identity matrix. Since $\mathbf{A}$ is specified as a symmetric matrix, the eigenvectors $\mathbf{U}$ are unitary, as seen in Equation \ref{eqn:unitary_U}. 

The backward pass is performed as follows:
\begin{align}
    &\mathbf{\bar{A}} = \mathbf{U}^T(sym(\mathbf{\bar{U}}\mathbf{U}^T\circ\mathbf{F}) + \bar{\boldsymbol\Lambda})\mathbf{U}\\
    \label{eqn:F_mat_eq}
    &F_{ij} = \frac{I_{\{i\neq{}j\}}}{h(\lambda_i - \lambda_j)}\\
    \label{eqn:broadening_eq}
    &h(t) = max(|t|, \epsilon)sgn(t)
\end{align}
Where $\mathbf{U}$ and $\boldsymbol\lambda$ are the outputs from the forward pass and $\mathbf{\bar{U}}$ are the gradients for the matrix $\mathbf{U}$ and $\mathbf{\bar{\Lambda}}$ is the diagonal matrix formed from the gradients for the eigenvalues, ${\mathbf{\bar{\boldsymbol\lambda}}}$. $sgn(\cdot)$ is the sign function which returns $\pm 1$ depending on the sign of $t$. Equations \ref{eqn:F_mat_eq} and \ref{eqn:broadening_eq} specify a conditional eigenvalue broadening behavior, whereby if an eigengap between two eigenvalues $\lambda_i$ and $\lambda_j$, $i\neq{j}$, is smaller than a fixed constant $\epsilon$, the value of $\epsilon$ is substituted instead. This represents a tradeoff between accuracy and stability, since the backward pass is now able to handle vanishing eigengaps but the results of the backward pass can only be treated as approximate rather than exact. For this study, we use a value of 1E-12 for $\epsilon$, a small scalar which provides the numerical stability to process all systems of interest without overly compromising on accuracy.  

\clearpage

\section{Tabulated spline cutoffs}
This section presents the tabulated cutoffs $r_h$ and $r_l$ for both the electronic and repulsive splines. Values for these cutoffs were determined through a distribution analysis as explained in Section 4 of the main paper. It is important to note that when using the electronic potentials generated from DFTBML, the data that the potentials are applied to cannot contain configurations with internuclear distances lower then the $r_l$ value of a given atom pair. This is equivalent to extrapolating beyond the trained region of the spline and can lead to unreasonable or incorrect results. Distances greater than $r_h$ are allowed because the potentials converge to 0 past this distance. 

\begin{table}[!ht] 
    \centering
    \caption{Low- and high-end cutoffs for the electronic splines of DFTBML}
    \label{Tab:SI_spline_cutoffs_electronic}
    \large
    \begin{tabular}{ccc}
    \hline
    \textbf{Element pair} & $\mathbf{r_l}$ \textbf{(\AA)} & $\mathbf{r_h}$ \textbf{(\AA)}\\ \toprule
    H-H & 0.5 & 4.5 \\  
    C-C & 1.04 & 4.5 \\  
    H-C & 0.602 & 4.5 \\  
    N-N & 0.986 & 4.5 \\  
    C-N & 0.948 & 4.5 \\  
    H-N & 0.573 & 4.5 \\  
    H-O & 0.599 & 4.5 \\  
    C-O & 1.005 & 4.5 \\  
    N-O & 0.933 & 4.5 \\  
    O-O & 1.062 & 4.5 \\ \bottomrule  
    \end{tabular}
\end{table}

\begin{table}[t] 
    \centering
    \caption{Low- and high-end cutoffs for the repulsive splines of DFTBML}
    \label{Tab:SI_spline_cutoffs_repulsive}
    \large
    \begin{tabular}{ccc}
    \hline
    \textbf{Element pair} & $\mathbf{r_l}$ \textbf{(\AA)} & $\mathbf{r_h}$ \textbf{(\AA)}\\ \toprule
    H-H & 0 & 2.10 \\  
    C-C & 0 & 1.80 \\  
    H-C & 0 & 1.60 \\  
    N-N & 0 & 1.80 \\  
    C-N & 0 & 1.80 \\  
    H-N & 0 & 1.60 \\  
    H-O & 0 & 1.50 \\  
    C-O & 0 & 1.80 \\  
    N-O & 0 & 1.80 \\  
    O-O & 0 & 1.80 \\ \bottomrule  
    \end{tabular}
\end{table}

\clearpage

\section{Atomic energies and Hubbard parameters}\label{sec:SI_ener_hub_vals}
Presented here are some tables for the atomic energies and Hubbard parameters, as well as the extend to which they changed, relative to those of Auorg, during training. Notation wise, $E_s$ and $E_p$ are the energies for the s and p orbitals respectively, and $U_s$ and $U_p$ are the Hubbard parameters for the s and p orbitals. The energies of the d orbitals (e.g. $E_d$ and $U_d$) are excluded as they are zero for first- and second-row elements.  

\begin{table}[!ht]
    \centering
    \caption{Energies and Hubbard parameters for DFTBML CC 20000}
   \sisetup{round-mode=places}
    \begin{tabular}{lHS[round-precision=3]S[round-precision=3]HS[round-precision=3]S[round-precision=3]}
    \toprule
        \textbf{Element} & $\mathbf{E_d}$ & $\mathbf{E_p}$ & $\mathbf{E_s}$ & $\mathbf{U_d}$ & $\mathbf{U_p}$ & $\mathbf{U_s}$ \\ \midrule
        H & 0 & 0 & -0.214672137 & 0 & 0 & 0.368121594 \\ 
        C & 0 & -0.209390764 & -0.497412307 & 0 & 0.398727827 & 0.420445594 \\ 
        N & 0 & -0.275809967 & -0.669946992 & 0 & 0.442894004 & 0.67298264 \\ 
        O & 0 & -0.335928172 & -0.901812882 & 0 & 0.531970652 & 0.696193231 \\ \bottomrule
    \end{tabular}
\end{table}

\begin{table}[!ht]
    \centering
    \caption{Energies and Hubbard parameters for DFTBML DFT 20000}
    \sisetup{round-mode=places}
    \begin{tabular}{lHS[round-precision=3]S[round-precision=3]HS[round-precision=3]S[round-precision=3]}
    \toprule
        \textbf{Element} & $\mathbf{E_d}$ & $\mathbf{E_p}$ & $\mathbf{E_s}$ & $\mathbf{U_d}$ & $\mathbf{U_p}$ & $\mathbf{U_s}$ \\ \midrule
        H & 0 & 0 & -0.209430004 & 0 & 0 & 0.361908777 \\ 
        C & 0 & -0.206710745 & -0.495113477 & 0 & 0.378310292 & 0.391076089 \\ 
        N & 0 & -0.275542306 & -0.662954597 & 0 & 0.439827829 & 0.633720473 \\ 
        O & 0 & -0.332974397 & -0.901331333 & 0 & 0.53320308 & 0.732950458 \\ \bottomrule
    \end{tabular}
\end{table}

\begin{table}[!ht]
    \centering
    \caption{Energies and Hubbard parameters for DFTBML CC 2500}
    \sisetup{round-mode=places}
    \begin{tabular}{lHS[round-precision=3]S[round-precision=3]HS[round-precision=3]S[round-precision=3]}
    \toprule
        \textbf{Element} & $\mathbf{E_d}$ & $\mathbf{E_p}$ & $\mathbf{E_s}$ & $\mathbf{U_d}$ & $\mathbf{U_p}$ & $\mathbf{U_s}$ \\ \midrule
        H & 0 & 0 & -0.211129487 & 0 & 0 & 0.364882368 \\ 
        C & 0 & -0.204045465 & -0.497780051 & 0 & 0.397263514 & 0.408893711 \\ 
        N & 0 & -0.274296488 & -0.661481563 & 0 & 0.444051427 & 0.627680442 \\ 
        O & 0 & -0.334310946 & -0.892667388 & 0 & 0.512268947 & 0.571068497 \\ \bottomrule
    \end{tabular}
\end{table}

\begin{table}[!ht]
    \centering
    \caption{Energies and Hubbard parameters for DFTBML DFT 2500}
    \sisetup{round-mode=places}
    \begin{tabular}{lHS[round-precision=3]S[round-precision=3]HS[round-precision=3]S[round-precision=3]}
    \toprule
        \textbf{Element} & $\mathbf{E_d}$ & $\mathbf{E_p}$ & $\mathbf{E_s}$ & $\mathbf{U_d}$ & $\mathbf{U_p}$ & $\mathbf{U_s}$ \\ \midrule
        H & 0 & 0 & -0.20726535 & 0 & 0 & 0.368166423 \\ 
        C & 0 & -0.201138943 & -0.495451087 & 0 & 0.434869176 & 0.445000393 \\ 
        N & 0 & -0.275373996 & -0.661583763 & 0 & 0.469923892 & 0.666604195 \\ 
        O & 0 & -0.333805834 & -0.887586942 & 0 & 0.543215381 & 0.661425235 \\ \bottomrule
    \end{tabular}
\end{table}

\begin{table}[!ht]
    \centering
    \caption{Energies and Hubbard parameters for DFTBML CC 300}
    \sisetup{round-mode=places}
    \begin{tabular}{lHS[round-precision=3]S[round-precision=3]HS[round-precision=3]S[round-precision=3]}
    \toprule
        \textbf{Element} & $\mathbf{E_d}$ & $\mathbf{E_p}$ & $\mathbf{E_s}$ & $\mathbf{U_d}$ & $\mathbf{U_p}$ & $\mathbf{U_s}$ \\ \midrule
        H & 0 & 0 & -0.221935875 & 0 & 0 & 0.36314316 \\ 
        C & 0 & -0.192726794 & -0.508495156 & 0 & 0.364982754 & 0.364544928 \\ 
        N & 0 & -0.271740632 & -0.646341529 & 0 & 0.422676451 & 0.43709351 \\ 
        O & 0 & -0.332938651 & -0.880486223 & 0 & 0.511086345 & 0.538297925 \\ \bottomrule
    \end{tabular}
\end{table}

\begin{table}[!ht]
    \centering
    \caption{Energies and Hubbard parameters for DFTBML DFT 300}
    \sisetup{round-mode=places}
    \begin{tabular}{lHS[round-precision=3]S[round-precision=3]HS[round-precision=3]S[round-precision=3]}
    \toprule
        \textbf{Element} & $\mathbf{E_d}$ & $\mathbf{E_p}$ & $\mathbf{E_s}$ & $\mathbf{U_d}$ & $\mathbf{U_p}$ & $\mathbf{U_s}$ \\ \midrule
        H & 0 & 0 & -0.226977679 & 0 & 0 & 0.374091683 \\ 
        C & 0 & -0.196650114 & -0.507915153 & 0 & 0.380049454 & 0.375665919 \\ 
        N & 0 & -0.264261028 & -0.64416762 & 0 & 0.387454924 & 0.500065038 \\ 
        O & 0 & -0.332452115 & -0.878957824 & 0 & 0.483966274 & 0.526585996 \\ \bottomrule
    \end{tabular}
\end{table}


\begin{table}[!ht]
    \centering
    \caption{Changes in energies and Hubbard parameters for DFTBML CC 20000}
    \sisetup{round-mode=places}
    \begin{tabular}{lHS[round-precision=3]S[round-precision=3]HS[round-precision=3]S[round-precision=3]}
    \toprule
        \textbf{Element} & $\mathbf{E_d}$ & $\mathbf{E_p}$ & $\mathbf{E_s}$ & $\mathbf{U_d}$ & $\mathbf{U_p}$ & $\mathbf{U_s}$ \\ \midrule
        H & 0 & 0 & 0.023928407 & 0 & 0 & -0.051495833 \\ 
        C & 0 & -0.015035584 & 0.007479458 & 0 & 0.03406133 & 0.055779096 \\ 
        N & 0 & -0.015081884 & -0.029946992 & 0 & 0.012006046 & 0.242094682 \\ 
        O & 0 & -0.003796399 & -0.022980298 & 0 & 0.036566482 & 0.200789061 \\ \bottomrule
    \end{tabular}
\end{table}

\begin{table}[!ht]
    \centering
    \caption{Changes in energies and Hubbard parameters for DFTBML DFT 20000}
    \sisetup{round-mode=places}
    \begin{tabular}{lHS[round-precision=3]S[round-precision=3]HS[round-precision=3]S[round-precision=3]}
    \toprule
        \textbf{Element} & $\mathbf{E_d}$ & $\mathbf{E_p}$ & $\mathbf{E_s}$ & $\mathbf{U_d}$ & $\mathbf{U_p}$ & $\mathbf{U_s}$ \\ \midrule
        H & 0 & 0 & 0.02917054 & 0 & 0 & -0.057708649 \\ 
        C & 0 & -0.012355565 & 0.009778289 & 0 & 0.013643795 & 0.026409592 \\ 
        N & 0 & -0.014814223 & -0.022954597 & 0 & 0.008939871 & 0.202832516 \\ 
        O & 0 & -0.000842623 & -0.022498749 & 0 & 0.03779891 & 0.237546288 \\ \bottomrule
    \end{tabular}
\end{table}

\begin{table}[!ht]
    \centering
    \caption{Changes in energies and Hubbard parameters for DFTBML CC 2500}
    \sisetup{round-mode=places}
    \begin{tabular}{lHS[round-precision=3]S[round-precision=3]HS[round-precision=3]S[round-precision=3]}
    \toprule
        \textbf{Element} & $\mathbf{E_d}$ & $\mathbf{E_p}$ & $\mathbf{E_s}$ & $\mathbf{U_d}$ & $\mathbf{U_p}$ & $\mathbf{U_s}$ \\ \midrule
        H & 0 & 0 & 0.027471057 & 0 & 0 & -0.054735058 \\ 
        C & 0 & -0.009690285 & 0.007111715 & 0 & 0.032597017 & 0.044227213 \\ 
        N & 0 & -0.013568404 & -0.021481563 & 0 & 0.013163469 & 0.196792484 \\ 
        O & 0 & -0.002179172 & -0.013834804 & 0 & 0.016864777 & 0.075664327 \\ \bottomrule
    \end{tabular}
\end{table}

\begin{table}[!ht]
    \centering
    \caption{Changes in energies and Hubbard parameters for DFTBML DFT 2500}
    \sisetup{round-mode=places}
    \begin{tabular}{lHS[round-precision=3]S[round-precision=3]HS[round-precision=3]S[round-precision=3]}
    \toprule
        \textbf{Element} & $\mathbf{E_d}$ & $\mathbf{E_p}$ & $\mathbf{E_s}$ & $\mathbf{U_d}$ & $\mathbf{U_p}$ & $\mathbf{U_s}$ \\ \midrule
        H & 0 & 0 & 0.031335194 & 0 & 0 & -0.051451003 \\ 
        C & 0 & -0.006783764 & 0.009440679 & 0 & 0.070202679 & 0.080333896 \\ 
        N & 0 & -0.014645912 & -0.021583763 & 0 & 0.039035934 & 0.235716237 \\ 
        O & 0 & -0.00167406 & -0.008754358 & 0 & 0.047811211 & 0.166021064 \\ \bottomrule
    \end{tabular}
\end{table}

\begin{table}[!ht]
    \centering
    \caption{Changes in energies and Hubbard parameters for DFTBML CC 300}
    \sisetup{round-mode=places}
    \begin{tabular}{lHS[round-precision=3]S[round-precision=3]HS[round-precision=3]S[round-precision=3]}
    \toprule
        \textbf{Element} & $\mathbf{E_d}$ & $\mathbf{E_p}$ & $\mathbf{E_s}$ & $\mathbf{U_d}$ & $\mathbf{U_p}$ & $\mathbf{U_s}$ \\ \midrule
        H & 0 & 0 & 0.016664669 & 0 & 0 & -0.056474266 \\ 
        C & 0 & 0.001628386 & -0.00360339 & 0 & 0.000316256 & -0.000121569 \\ 
        N & 0 & -0.011012548 & -0.006341529 & 0 & -0.008211507 & 0.006205552 \\ 
        O & 0 & -0.000806877 & -0.001653639 & 0 & 0.015682175 & 0.042893755 \\ \bottomrule
    \end{tabular}
\end{table}

\begin{table}[!ht]
    \centering
    \caption{Changes in energies and Hubbard parameters for DFTBML DFT 300}
    \sisetup{round-mode=places}
    \begin{tabular}{lHS[round-precision=3]S[round-precision=3]HS[round-precision=3]S[round-precision=3]}
    \toprule
        \textbf{Element} & $\mathbf{E_d}$ & $\mathbf{E_p}$ & $\mathbf{E_s}$ & $\mathbf{U_d}$ & $\mathbf{U_p}$ & $\mathbf{U_s}$ \\ \midrule
        H & 0 & 0 & 0.011622865 & 0 & 0 & -0.045525743 \\ 
        C & 0 & -0.002294934 & -0.003023387 & 0 & 0.015382956 & 0.010999421 \\ 
        N & 0 & -0.003532945 & -0.00416762 & 0 & -0.043433033 & 0.069177081 \\ 
        O & 0 & -0.000320342 & -0.00012524 & 0 & -0.011437896 & 0.031181826 \\ \bottomrule
    \end{tabular}
\end{table}

\clearpage

\section{Detailed dataset generation scheme}\label{sec:SI_dset_generation_detailed}

\begin{figure}[h]
    \centering
    \includegraphics[width=0.95\columnwidth]{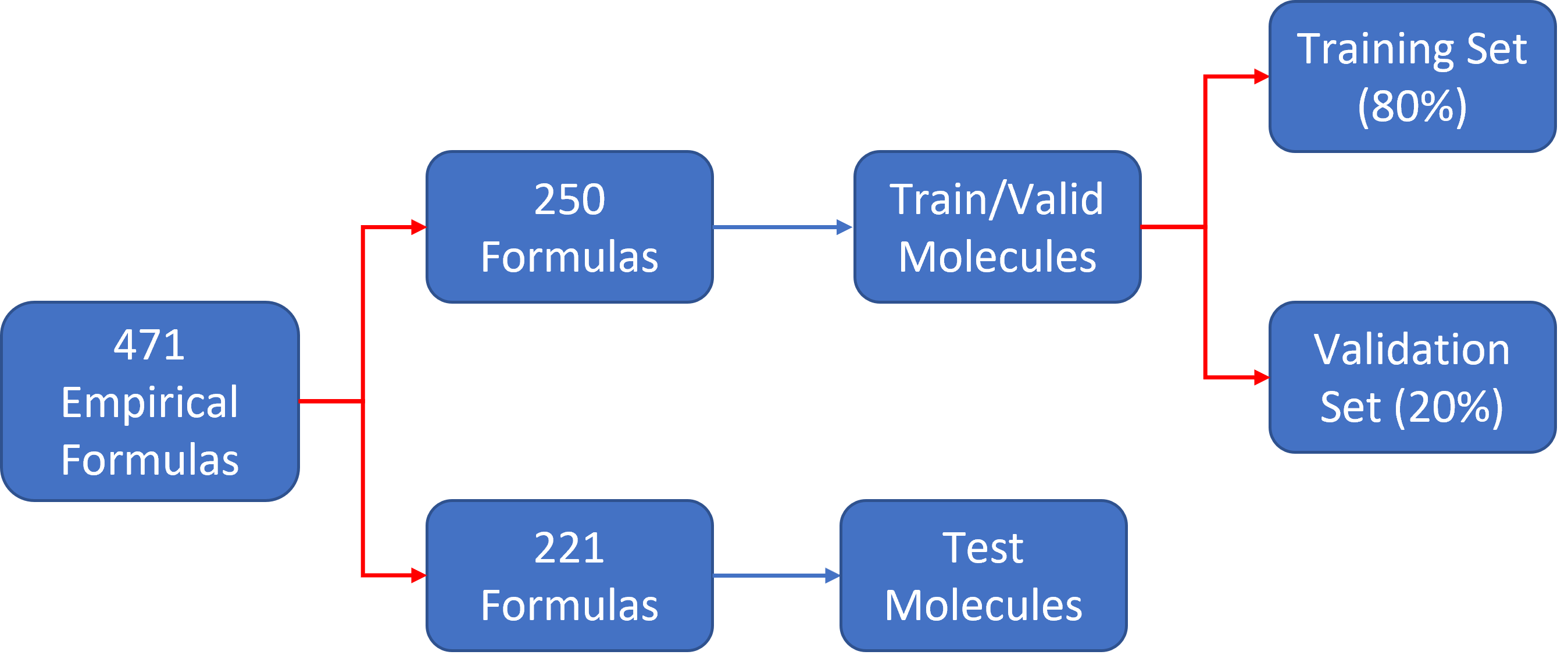}
    \caption{High-level overview of the method used to generate the initial dataset. Red arrows indicate random sampling.}
    \label{fig:supplemental_DsetGen}
\end{figure}

For a more detailed explanation of the dataset generation scheme for the base dataset, the workflow consists of the following steps:

\begin{enumerate}
    \item A subset of the 471 empirical formulas is randomly chosen to be used for the training and validation sets. The remaining empirical formulas are used exclusively for the test set, and this ensures that no molecules used during the training process have the same empirical formula as those used during testing. By separating the test set from the training and validation sets based on empirical formula, we have a stricter assessment of the model's performance when evaluating the model's predictions on the test set. 
    \item All the molecules that can potentially be used for the training and validation sets are gathered and organized by their empirical formulas. A specified variable indicates the maximum number of configurations to include for each empirical formula such that each empirical formula is represented as uniformly as possible. This is important because the ANI-1CCX dataset includes different numbers of configurations for different empirical formulas, and this number varies from one through several thousand. Standardizing the number of configurations for each empirical formula ensures that the random sample is unbiased in favor of certain empirical formulas over others. Furthermore, the maximum number of configurations for each empirical formula is set such that the product between the number of empirical formulas and the maximum number of configurations for each formula is as close to the total number of training and validation molecules as possible.
    \item The set of all possible molecules for the training and validation sets determined from step 2 is randomly shuffled. A random sampling without replacement is performed to obtain the set of training molecules followed by a second random sampling of the remaining molecules to obtain the set of validation molecules. The number of training molecules and the number of validation molecules is set such that the training molecules comprise 80\% of the total and the validation molecules comprise 20\%. 
    \item The test set is constructed systematically from the remaining empirical formulas such that every empirical formula has as equal as possible a number of configurations in the final set. This ensures that the test set gives a comprehensive representation of all the formulas included. 
    \item The test, train, and validation sets are saved to a specified directory and prepared for precomputation.
\end{enumerate}

Once the base dataset is obtained, other datasets are generated as variations of this parent set. To generate larger datasets, the above workflow is repeated using the same empirical formulas as those found in the training set of the base dataset rather than randomly sampling from the original 471 formulas but with more configurations per formula and more molecules sampled in total. Both the training and validation sets are expanded when moving to datasets of larger size, but the test set is kept the same. To generate smaller datasets, a smaller training set is created by randomly sampling from the base dataset training set. For smaller datasets, both the validation and test sets remain unchanged. 

For creating datasets with different partitioning schemes, the two we have focused on here are separating based on the number of heavy atoms and separating a larger dataset into two smaller datasets. When separating the data based on the number of heavy atoms, the above workflow is repeated except that the empirical formulas used for the training and validation sets are those containing fewer than or equal to some limit of heavy atoms and those used for the test set are those containing strictly greater than the limit of heavy atoms. This way, the training and validation sets are concerned solely with lighter molecules while the test set contains only the heavier molecules. When separating a larger dataset into two smaller datasets, the training set is split in half while the validation and test sets are copied over. 

In terms of dealing with different physical targets, we are only concerned with varying the target used for the total molecular energy. In addition to assessing the performance of the model against CC level energies, we also wish to assess the performance of the model against DFT level energies. For this reason, a CC and DFT version of each dataset is generated for most of the experiments presented here, and the DFT version is usually copied over from the CC version with the only difference being the total energy value and the level of theory used. 

\clearpage

\section{Outlier exclusion method}
For removing outliers, the following process is applied on the predicted and target values for total molecular energy:
\begin{enumerate}
    \item The absolute differences between the predicted total molecular energies and target total molecular energies, $D_{ener}$, are calculated.
    \item The mean $\mu$ and standard deviation $\sigma$ are calculated for the differences.
    \item Because all values are positive, the number of standard deviations between the maximum value and the mean is calculated. If $\frac{max(D_{ener}) - \mu}{\sigma} \geq 20$, the maximum value is removed from the set of differences.
    \item Steps 2 and 3 are repeated until the set of differences is consistent and does not contain any values above 20 standard deviations from the mean. 
\end{enumerate}

\clearpage

\section{Hyperparameter investigation}\label{sec:SI_hyperparam_testing}
Presented here is a detailed discussion of the hyperparameter sensitivity analyses. The following hyperparameters are of particular interest:
\begin{itemize}
    \item \textbf{The weighting factors for the charges and dipoles:} The focus of DFTBML is to predict quantum chemical properties at a level of accuracy approaching that of \textit{ab initio} CC theory. A great emphasis has been placed on predicting the total molecular energy, but also important is the performance of the model in predicting molecular dipoles and atomic charge. Different values of the weighting factors for these two physical targets are tested to determine optimal values. Because charge and dipole are coupled together, emphasis is placed on reproducing the observable quantities of total energy and molecular dipole.
    \item \textbf{The number of knots (control points) of the spline:} For splines, the knots or control points define the sequence of intervals where every interval is spanned by one polynomial function. Continuity conditions are enforced at these control points to ensure an overall continuous model. 
    \item \textbf{The position of the inflection point:} The initial position of the inflection point for splines modeling overlap matrix elements. Inflection points were implemented as a way to account for the change in curvature of overlap integrals, a phenomenon most commonly observed when dealing with the overlap of two p orbitals.
    \item \textbf{The weighting factor \boldmath{$w_{TD}$}:} One of the most important parameters for regularizing the functional form of these splines is the weighting factor for the penalty on the magnitude of the third derivative (see Section \ref{sec:SI_loss_reg}), as it is instrumental in ensuring the splines come out as smooth after training. 
\end{itemize}
The hyperparameters chosen for each of these categories is as follows:
\begin{itemize}
    \item \textbf{Energy, charge, and dipole weighting factors:} The following weighting factors ($w_{prop}$ of Equation \ref{eqn:master_loss_fxn}) are used: 6270 Ha$^{-1}$ for total energy (Ha), 100 (e\AA{})$^{-1}$ for dipoles (e\AA{}), and 1 e$^{-1}$ for charges (e). Units are excluded in the subsequent discussion.
    \item \textbf{The number of knots for each spline:} The number of knots is set at 100.
    \item \textbf{The position of the inflection point:} The initial position of the inflection point is chosen to be 1/10 the total range of the spline.
    \item \textbf{The weighting factor for the third derivative penalty:} The factor is chosen to be $w_{TD} = 10$. 
\end{itemize}
The following sections show, in detail, the effect that altering these hyperparameter values has on the performance of the model both quantitatively and qualitatively as it relates to model interpretability. 

\subsection{The effect of the charge and dipole weighting factors}\label{subsec:charge_dipole_weighting_factors} 
In addition to producing accurate predictions of the total molecular energy, DFTBML also aims to learn other quantities of interest. In total, predictions for three quantities are simultaneously optimized: total molecular energy, molecular dipole, and atomic charge. Internally, the dipole is calculated from the atomic charge by the following matrix multiplication:
\begin{align}
    \label{eqn:ESP_dipole_calculation}
    \boldsymbol\mu = \mathbf{R}^T\mathbf{q}
\end{align}
Where $\boldsymbol\mu$ is the dipole, $\mathbf{R}$ is a matrix of cartesian coordinates, and $\mathbf{q}$ is a vector of the atomic charges. In that sense, the atomic charge and dipole are coupled together. In early experiments with the DFTBML model, all three targets were trained at once with an independent weighting factor for each. This created problems as while the model was able to optimize all three targets, performance suffered as the three targets competed each other. While it is recognized that different weighting regimes can be used to tune the accuracy for different targets, we have opted to focus on optimizing predictions of total molecular energy. Thus, a different approach is adopted where rather than introducing three hyperparameter weights with one per physical target, a greater emphasis is placed on the total energy and dipoles since these are observable quantities. Since charges and dipoles are linked, training the dipole prediction ability of DFTBML indirectly trains the ability to predict atomic charge.

\begin{figure}[ht!]
    \centering
    \includegraphics[width=\columnwidth]{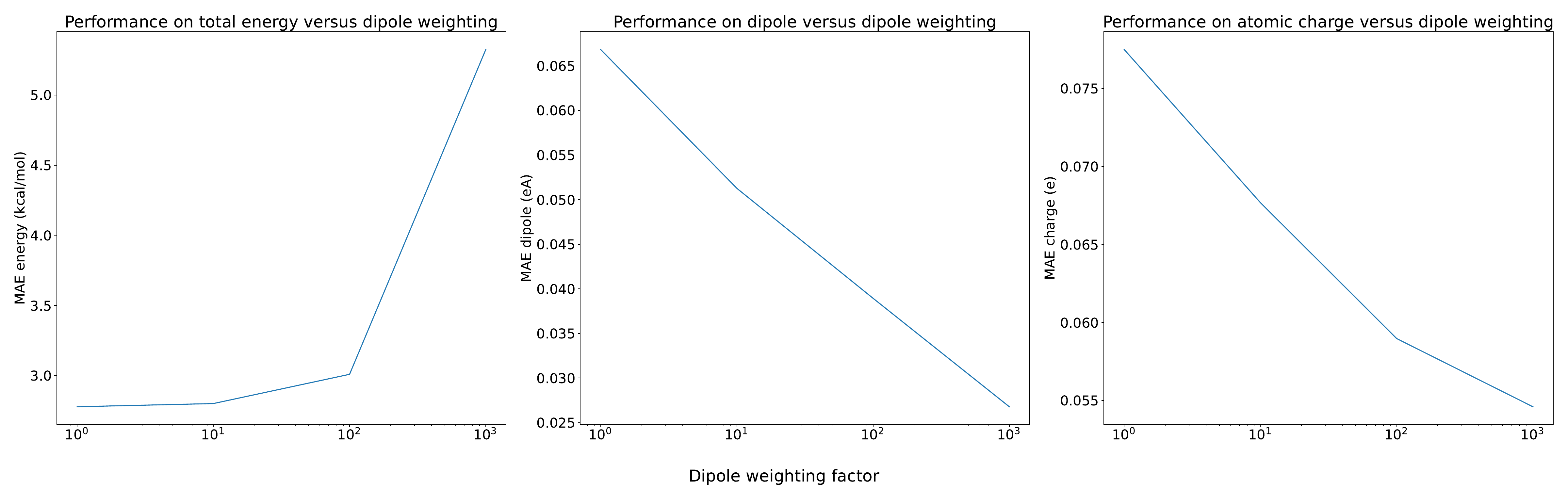}
    \caption{Performance on different physical targets as a function of the dipole weighting factor. The energy and charge weighting factors are fixed at 6270 and 1, respectively. The DFTBML CC 2500 dataset was used for the results presented here, and all the experiments were conducted for 1000 epochs with all other hyperparameters identical. These are the final numbers after any outliers have been excluded.}
    \label{fig:dipole_weight_comparison_logarithmic}
\end{figure}

To search for an optimal combination of weighting factors for the total energy and molecular dipole, the total energy was fixed with a weighting of 6270 and the charge was fixed with a weight of 1. The dipole weighting factor was then systematically varied over the value of 1, 10, 100, and 1000. The result of the experiments are shown in Figure \ref{fig:dipole_weight_comparison_logarithmic}.

\begin{figure}[ht!]
    \centering
    \includegraphics[width=\columnwidth]{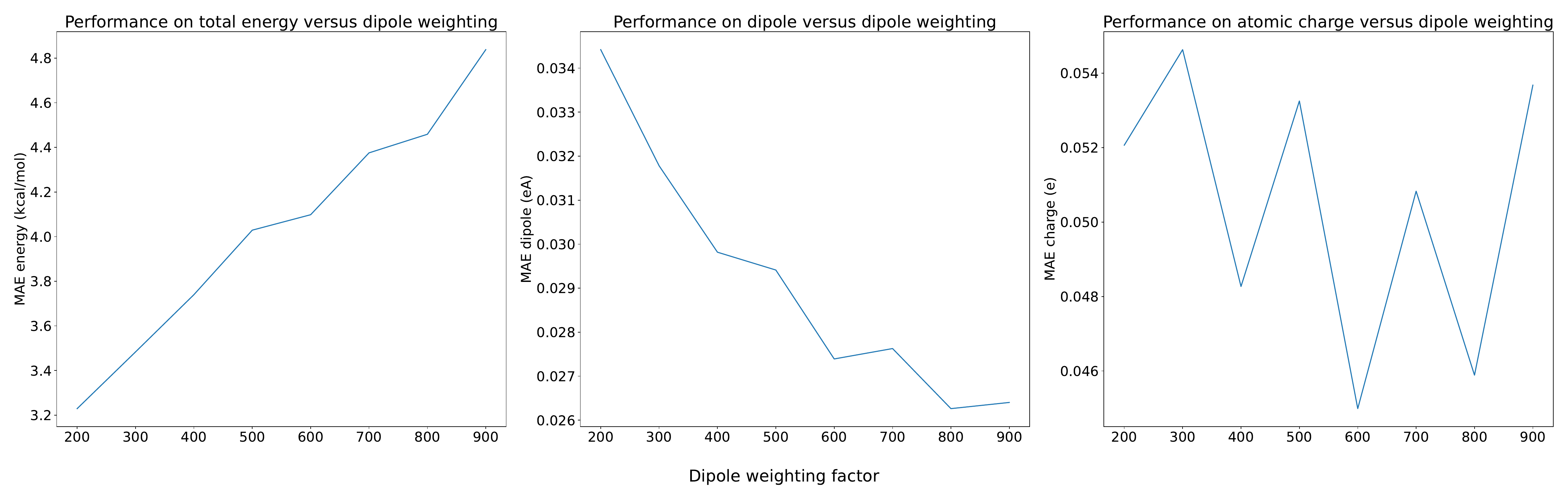}
    \caption{Performance on different physical targets as a function of the dipole weighting factor. The energy and charge weighting factors are fixed at 6270 and 1, respectively. The DFTBML CC 2500 dataset was used for the results presented here, and all the experiments were conducted for 1000 epochs with all other hyperparameters identical. These are the final numbers after any outliers have been excluded. The dipole weighting factor was scanned over the range of 200 to 900.}
    \label{fig:dipole_weight_comparison_100_1000}
\end{figure}

\begin{figure}[ht!]
    \centering
    \includegraphics[width=\columnwidth]{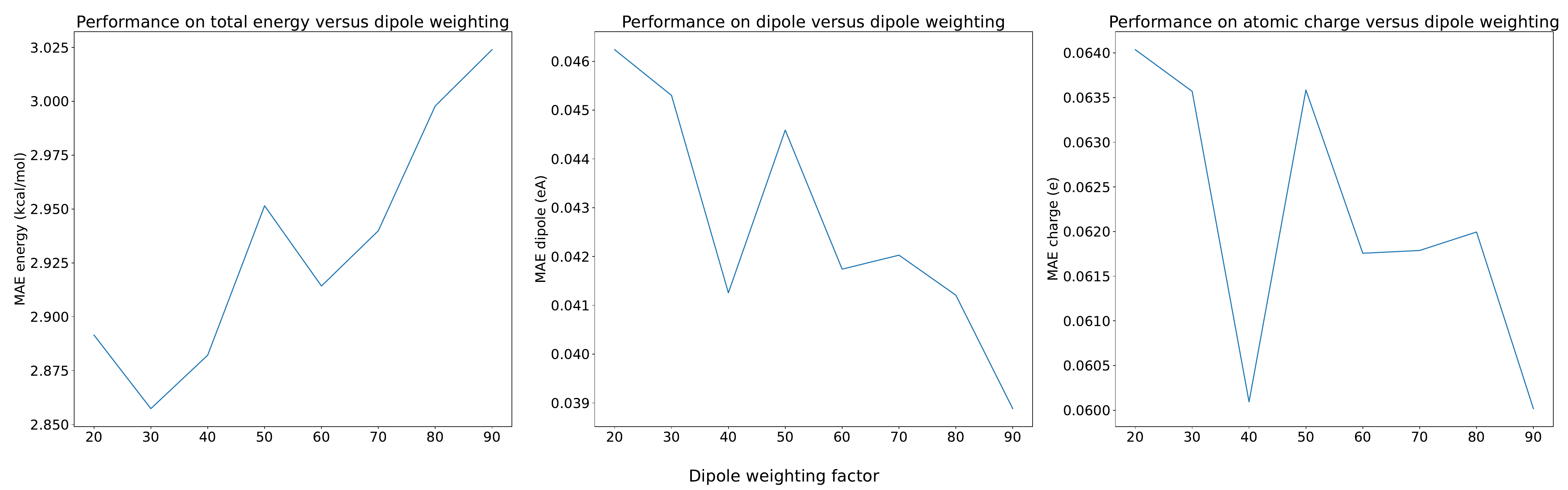}
    \caption{Performance on different physical targets as a function of the dipole weighting factor. The energy and charge weighting factors are fixed at 6270 and 1, respectively. The DFTBML CC 2500 dataset was used for the results presented here, and all the experiments were conducted for 1000 epochs with all other hyperparameters identical. These are the final numbers after any outliers have been excluded. The dipole weighting factor was scanned over the range of 20 to 90.}
    \label{fig:dipole_weight_comparison_10_100}
\end{figure}

It is clear that increasing the dipole weighting factor leads to an improvement in the model's performance on dipoles and charges while a corresponding degradation in the model's performance on total energy is observed. Because the plots in Figure \ref{fig:dipole_weight_comparison_logarithmic} are shown on a logarithmic scale, further analysis was done in the region from 10 to 100 and 100 to 1000 for the dipole weighting factor to confirm that the observed behavior was consistent. Figure \ref{fig:dipole_weight_comparison_100_1000} shows the results of testing the dipole weighting factor over the range of 200 to 900 with a step of 100 and Figure \ref{fig:dipole_weight_comparison_10_100} shows the results of testing the dipole weighting factor over the range of 20 to 90 with a step of 10.

It is apparent that in the cases presented in Figures \ref{fig:dipole_weight_comparison_100_1000} and \ref{fig:dipole_weight_comparison_10_100}, the general trend shown in Figure \ref{fig:dipole_weight_comparison_logarithmic} holds where reductions in the MAE for dipole and charges correspond with increases in the MAE for the total molecular energy. The erratic behavior of the charges in Figures \ref{fig:dipole_weight_comparison_100_1000} and \ref{fig:dipole_weight_comparison_10_100} is not surprising because of how little weight is placed on charges during the training process. No in-depth search was conducted over the interval spanning from 1 to 10 since increasing the dipole weighting factor by increments of 1 would not have a significant effect on the model's performance. 

Based on this hyperparameter search, we use a weight of 6270 for total energy, 100 for dipole, and 1 for charges. A weighting factor of 100 is chosen for dipoles for two reasons. First, it was observed that increasing the weighting factor beyond 100 towards 1000 led to an increase in the number of outliers and the number of molecules which failed to converge on the SCF cycle of the DFTB+ program. While in each case the number of molecules which had to be removed because of non-convergence or exceeding the outlier threshold was less than 0.1\%, the fact that the number of such occurrences increased with increasing dipole weight indicates the possibility of further instabilities from using parameters generated from this training approach. Second, a weight of 100 seems an optimal balance of performance for all three physical targets. Using a dipole weight of 100, the dipoles and charges can still be optimized to an extent without seeing a significant degradation in the model's performance on total energy. Furthermore, since going from a dipole weight of 1 or 10 to 100 does not result in a significant decrease in performance on total energy, it is worthwhile to use a relatively higher weight and gain some more performance on dipoles and charges rather than pursuing a marginal improvement in total energy. This investigation also motivated changing the number of epochs from 1000 to 2500 since 1000 epochs gave nearly full convergence of the targets but 2500 epochs was shown to give full convergence (see Section 2.3).

\subsection{The effect of the spline knot sequence} 
The number of knots chosen for the spline models define the number of intervals spanned by the polynomial basis functions such that for N knots, we have N - 1 intervals. The knots are initialized uniformly on the interval $[r_l, r_h]$ which defines the region over which the spline spans. Table \ref{Tab:cc_knot_hyperparam} shows the results from using 25, 50, 75, 100, 125, and 150 knots for training to the CC total energy target, and Table \ref{Tab:dft_knot_hyperparam} shows the results from using the same numbers of knots but training to the DFT total energy target. All the experiments shown were conducted using 2500 epochs with a 2500 molecule dataset and the performance was evaluated on a near-transfer test set of 10000 molecules. No outliers were detected throughout.

\begin{table}[!ht]
    \centering
    \caption{Performance of DFTBML models using different numbers of knots, when trained to the CC total energy target}
    \label{Tab:cc_knot_hyperparam}
    \sisetup{round-mode=places}
    \large
    \resizebox{\textwidth}{!}{\begin{tabular}{lcS[round-precision=2]S[round-precision=3]S[round-precision=3]}
    \toprule
        \textbf{Parameterization} & \textbf{Nonconverged} & \textbf{MAE energy (kcal/mol)} & \textbf{MAE dipole (e\AA)} & \textbf{MAE charge (e)} \\ \midrule
        DFTBML CC 2500, 150 knots & 0 & 3.124651693 & 0.039364035 & 0.058867106 \\ 
        DFTBML CC 2500, 125 knots & 0 & 2.998407819 & 0.036738301 & 0.057214098 \\ 
        DFTBML CC 2500, 100 knots & 0 & 2.995873393 & 0.038848605 & 0.059106972 \\ 
        DFTBML CC 2500, 75 knots & 1 & 2.92476632 & 0.035761465 & 0.058303256 \\
        DFTBML CC 2500, 50 knots & 1 & 2.855552808 & 0.036207319 & 0.058661473 \\ 
        DFTBML CC 2500, 25 knots & 0 & 2.825396635 & 0.036340539 & 0.055800884 \\ \bottomrule
    \end{tabular}}
\end{table}

\begin{table}[!ht]
    \centering
    \caption{Performance of DFTBML models using different numbers of knots, when trained to the DFT total energy target}
    \label{Tab:dft_knot_hyperparam}
    \sisetup{round-mode=places}
    \large
    \resizebox{\textwidth}{!}{\begin{tabular}{lcS[round-precision=2]S[round-precision=3]S[round-precision=3]}
    \toprule
        \textbf{Parameterization} & \textbf{Nonconverged} & \textbf{MAE energy (kcal/mol)} & \textbf{MAE dipole (e\AA)} & \textbf{MAE charge (e)} \\ \midrule
        DFTBML DFT 2500, 150 knots & 0 & 3.111569045 & 0.038744997 & 0.057427824 \\ 
        DFTBML DFT 2500, 125 knots & 0 & 3.080884021 & 0.040000005 & 0.054878446 \\ 
        DFTBML DFT 2500, 100 knots & 1 & 3.029951395 & 0.037784222 & 0.055933999 \\ 
        DFTBML DFT 2500, 75 knots & 0 & 2.931172999 & 0.034394388 & 0.044830669 \\ 
        DFTBML DFT 2500, 50 knots & 1 & 2.979522924 & 0.037053722 & 0.052455136 \\ 
        DFTBML DFT 2500, 25 knots & 0 & 2.924653526 & 0.037012624 & 0.057623805 \\ \bottomrule
    \end{tabular}}
\end{table}

\begin{figure}[ht!]
    \centering
    \includegraphics[width=\columnwidth]{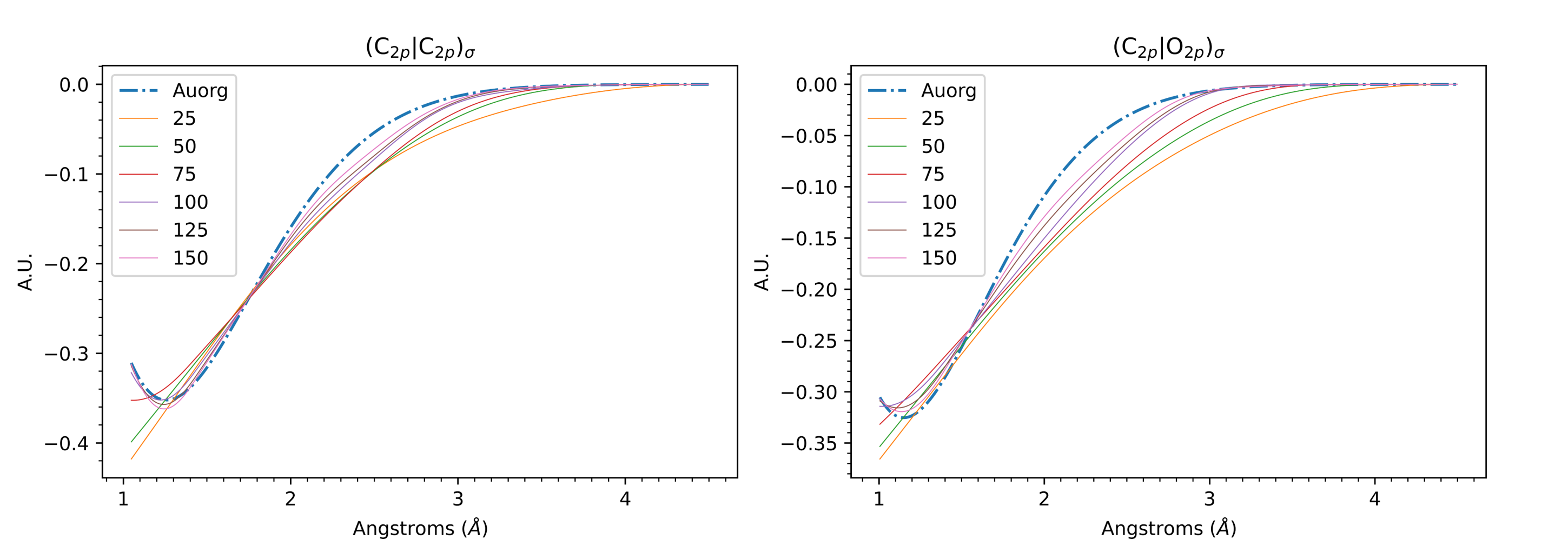}
    \caption{Overlay of spline models with different numbers of knots. The carbon-oxygen 2p overlap interaction is shown on the right and the carbon-carbon 2p overlap interaction is shown on the left. Both interactions are $\sigma$ in orientation and both plots are generated from the DFTBML CC results. Auorg curves are included for reference.}
    \label{fig:cc_knots_overlap}
\end{figure}

Numerically, the performance across the different numbers of knots is similar in terms of both total energy and dipole, although some deviations are observed for predicting the atomic charge. However, this is likely because the hyperparameters were set such that the optimization focus was on the total molecular energy. Figure \ref{fig:cc_knots_overlap} shows the overlay of a series of splines used to model two different overlap matrix elements. 

An interesting observation is that there is a clear difference in behavior between splines having 75 and fewer knots and those splines with 100 or more knots, whereby the splines with 75 or fewer knots have a tendency to resort to longer range interactions while also precluding the upward inflection at short range. In contrast to this, splines with 100 or more knots tend to keep interactions in the short range and are able to include the upward inflection in the functional forms. Furthermore, splines with 100+ knots all tend to give the same final form with close agreement, and their physical form more closely resembles that of the Auorg potentials, especially at short range. Based on the results here, it is clear that 100 knots is the optimal number since it is the minimal number of knots required to produce physically intuitive short range functional forms that incorporate an upward inflection and which incorporate features of the Auorg parameterization.

\subsection{The effect of the inflection point position} \label{subsec:SI_inflection_point_tuning}
For DFTBML, an inflection point follows the mathematical definition in that the curvature (i.e., sign of the second derivative) changes across the point. The implementation of the inflection point penalty is described in Section \ref{sec:SI_loss_reg}.

Because the inflection point variable is tied to the convex penalty and because the convex penalty converges quickly to 0 due to the nature of the inequality constraint, the inflection point shows little motility during training, typically moving less than 0.1 \AA{} from its starting point. Because of this, the initial value of the inflection point is considered a hyperparameter and is specified as some fraction of the range from $r_{l}$ to $r_{h}$ as follows:
\begin{align}
    r_{inflect} = r_{l} + \left(\frac{r_{h} - r_{l}}{x}\right)
\end{align}

\begin{figure}[ht!]
    \centering
    \includegraphics[width=\columnwidth]{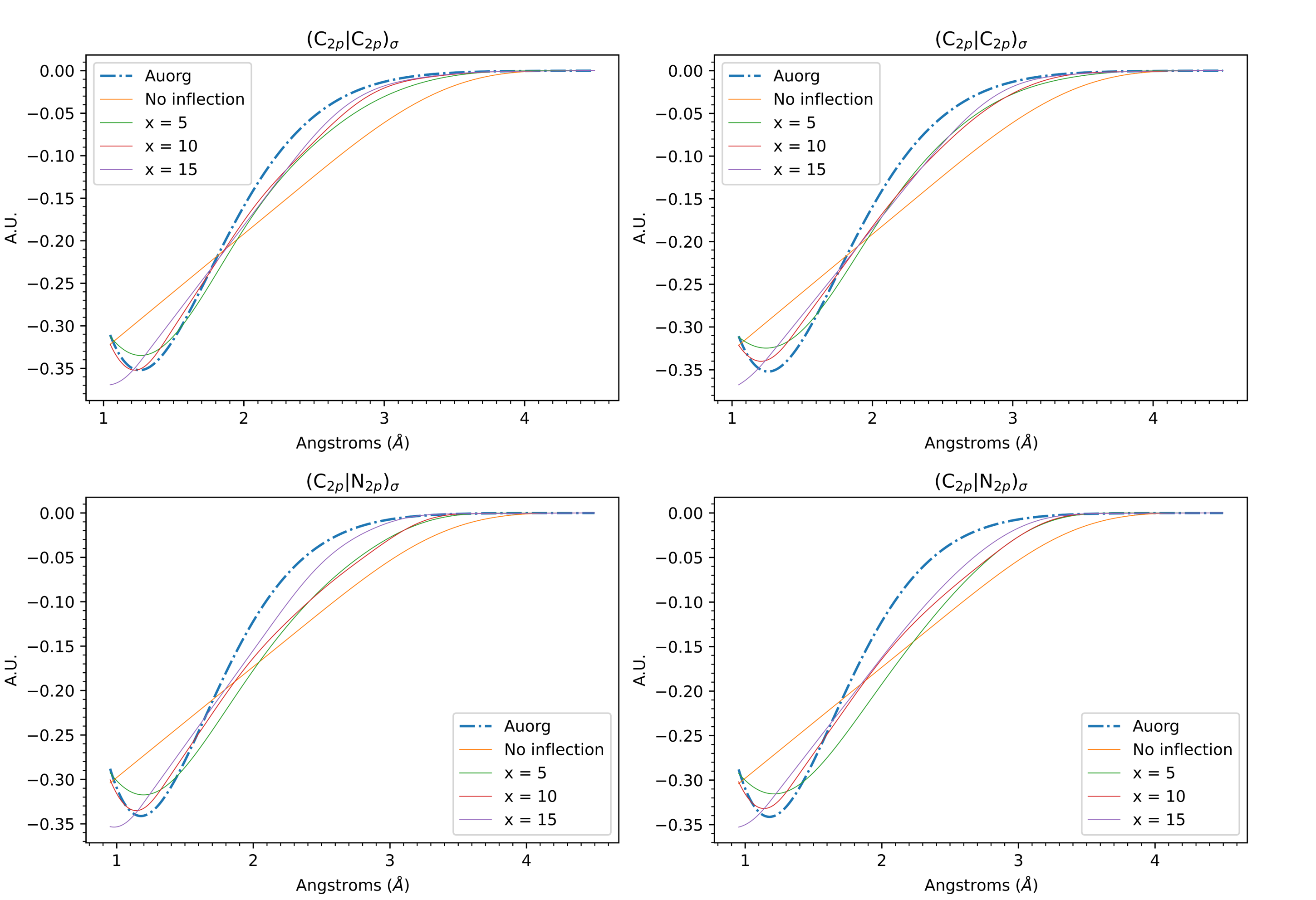}
    \caption{Spline models for the $(C_{2p}|C_{2p})_\sigma$ and $(C_{2p}|N_{2p})_\sigma$ integrals, trained using different initial values for the inflection point. The left column is from training to the CC total energy target and the right column is from training to the DFT total energy target.}
    \label{fig:SI_inflect_overlay}
\end{figure}

Where the denominator $x$ is varied, i.e. if the target is 1/10 the range, then $x$ is set to 10. To determine an optimal value of this hyperparameter, a series of near-transfer experiments were conducted including no inflection point, inflection point initialized at 1/15 the range, inflection point initialized at 1/10 the range, and inflection point initialized at 1/5 the range. The results of these experiments are shown in Table \ref{Tab:inflect_point_vals}, where each experiment was run for 2500 epochs and both CC and DFT energy targets were analyzed. Each experiment used a test set of 10000 molecules. Figure \ref{fig:SI_inflect_overlay} shows the overlaid results of these different inflection point runs for the $(C_{2p}|C_{2p})_\sigma$ and $(C_{2p}|N_{2p})_\sigma$ integrals where the upward curvature at short distances is evident. 

\begin{table}[!ht]
    \centering
    \caption{Performance of the model using different initial values for the inflection point}
    \label{Tab:inflect_point_vals}
    \sisetup{round-mode=places}
    \large
    \resizebox{\textwidth}{!}{\begin{tabular}{lcS[round-precision=2]S[round-precision=3]S[round-precision=3]}
    \toprule
        \textbf{Parameterization} & \textbf{Nonconverged} & \textbf{MAE energy (kcal/mol)} & \textbf{MAE dipole (e\AA)} & \textbf{MAE charge (e)} \\ \midrule
        DFTBML CC 2500, no inflection & 0 & 3.128334616 & 0.038611197 & 0.058853132 \\ 
        DFTBML CC 2500, x = 5 & 0 & 3.032331175 & 0.035305009 & 0.057814703 \\ 
        DFTBML CC 2500, x = 10 & 0 & 2.995873393 & 0.038848605 & 0.059106972 \\ 
        DFTBML CC 2500, x = 15 & 1 & 3.022220886 & 0.037667801 & 0.061014124 \\ \midrule
        DFTBML DFT 2500, no inflection & 0 & 3.210355452 & 0.03994526 & 0.054043658 \\ 
        DFTBML DFT 2500, x = 5 & 0 & 3.124477627 & 0.037896357 & 0.05699261 \\ 
        DFTBML DFT 2500, x = 10 & 1 & 3.029951395 & 0.037784222 & 0.055933999 \\ 
        DFTBML DFT 2500, x = 15 & 1 & 3.029104315 & 0.035701028 & 0.055101379 \\ \bottomrule
    \end{tabular}}
\end{table}

As observed in Table \ref{Tab:inflect_point_vals}, variations in the starting position of the inflection point has no major impact on the model's performance so long as it is initialized at relatively short range. This makes sense since in the short range region around 1.5 \AA, the curvature of the model is nearly zero, and so moving the inflection point along a region without changing curvature does not have an effect. In Figure \ref{fig:SI_inflect_overlay}, we can see that for the cases of $x$ = 5, 10, or 15 for the inflection point initial value, the functional forms are fairly similar. However, the case of no inflection point differs dramatically in that no upward curvature is allowed at shorter range. Since the positioning of the inflection point does not have a significant impact on the model's performance, an inflection point initialization value of 1/10 is set as the standard for all experiments.

\subsection{The effect of the third derivative penalty weight} \label{subsec:SI_third_derivative_weight_tuning}

\begin{figure}[ht!]
    \centering
    \includegraphics[width=\columnwidth]{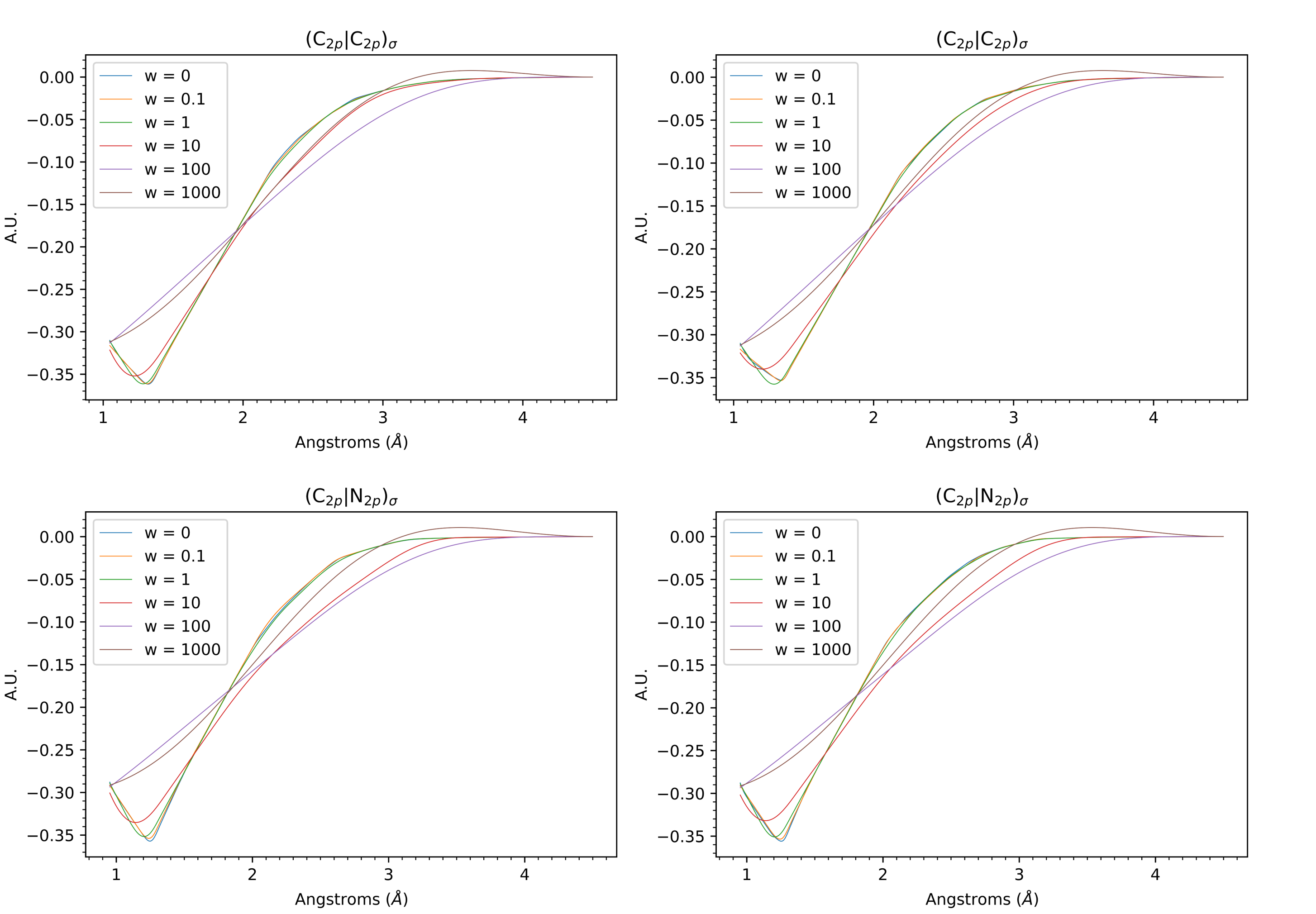}
    \caption{Spline models for the $(C_{2p}|C_{2p})_\sigma$ and $(C_{2p}|N_{2p})_\sigma$ integrals, trained using different weight values for the third derivative penalty. The left column is from training to the CC total energy target and the right column is from training to the DFT total energy target.}
    \label{fig:L2_spline_overlay}
\end{figure}

For the third derivative penalty, a weighting factor is applied to control how aggressively the magnitude of the third derivative is penalized. The mathematical form of the penalty is given in Equation \ref{eqn:SI_third_deriv}. The effect of the weighting factor $w_{TD}$ was investigated by systematically varying the value of $w_{TD}$ by multiples of 10. In total, six values were tested, with $w_{TD}$~= 0, 0.1, 1, 10, 100, and 1000. Tables \ref{Tab:L2_CC_results} and \ref{Tab:L2_DFT_results} below show the results for different experiments conducted with these different hyperparameter values and Figure \ref{fig:L2_spline_overlay} shows some resulting spline models. All the experiments conducted used 2500 epochs, 6270 for the energy weighting factor, 100 for the dipole weighting factor, 1 for the charge weighting factor, 100 knots, and an inflection point initialization value of 1/10 the total range. The test set consisted of 10000 molecules.

\begin{table}[!ht]
    \centering
    \caption{Performance of the model trained to CC total energy using different weight values for the third derivative penalty}
    \label{Tab:L2_CC_results}
    \sisetup{round-mode=places}
    \large
    \resizebox{\textwidth}{!}{\begin{tabular}{lccS[round-precision=2]S[round-precision=3]S[round-precision=3]}
    \toprule
        \textbf{Parameterization} & \textbf{Outliers} & \textbf{Nonconverged} & \textbf{MAE energy (kcal/mol)} & \textbf{MAE dipole (e\AA)} & \textbf{MAE charge (e)} \\ \midrule
        DFTBML CC 2500 $w_{TD}$ = 0 & 0 & 0 & 3.173976603 & 0.041134304 & 0.059988127 \\ 
        DFTBML CC 2500 $w_{TD}$ = 0.1 & 0 & 0 & 3.167574802 & 0.039067457 & 0.057265334 \\ 
        DFTBML CC 2500 $w_{TD}$ = 1 & 0 & 0 & 3.049332941 & 0.040128461 & 0.060253346 \\ 
        DFTBML CC 2500 $w_{TD}$ = 10 & 0 & 0 & 2.995873393 & 0.038848605 & 0.059106972 \\ 
        DFTBML CC 2500 $w_{TD}$ = 100 & 0 & 1 & 3.690281731 & 0.036573767 & 0.058445105 \\ 
        DFTBML CC 2500 $w_{TD}$ = 1000 & 0 & 1 & 5.598280984 & 0.042920525 & 0.06102969 \\ \bottomrule
    \end{tabular}}
\end{table}

\begin{table}[!ht]
    \centering
    \caption{Performance of the model trained to DFT total energy using different weight values for the third derivative penalty}
    \label{Tab:L2_DFT_results}
    \sisetup{round-mode=places}
    \large
    \resizebox{\textwidth}{!}{\begin{tabular}{lccS[round-precision=2]S[round-precision=3]S[round-precision=3]}
    \toprule
        \textbf{Parameterization} & \textbf{Outliers} & \textbf{Nonconverged} & \textbf{MAE energy (kcal/mol)} & \textbf{MAE dipole (e\AA)} & \textbf{MAE charge (e)} \\ \midrule
        DFTBML DFT 2500 $w_{TD}$ = 0 & 1 & 0 & 3.167360647 & 0.039389791 & 0.056059916 \\ 
        DFTBML DFT 2500 $w_{TD}$ = 0.1 & 1 & 0 & 3.03943632 & 0.035677693 & 0.044168233 \\ 
        DFTBML DFT 2500 $w_{TD}$ = 1 & 0 & 0 & 3.133640874 & 0.041218647 & 0.056985343 \\ 
        DFTBML DFT 2500 $w_{TD}$ = 10 & 0 & 1 & 3.029951395 & 0.037784222 & 0.055933999 \\ 
        DFTBML DFT 2500 $w_{TD}$ = 100 & 0 & 0 & 3.782811915 & 0.038961464 & 0.057575832 \\ 
        DFTBML DFT 2500 $w_{TD}$ = 1000 & 0 & 1 & 5.291445916 & 0.04161259 & 0.060088346 \\ \bottomrule
    \end{tabular}}
\end{table}

It is evident from Tables \ref{Tab:L2_CC_results} and \ref{Tab:L2_DFT_results} that variations of the $w_{TD}$ have no significant effect at lower values, but does significantly degrade model performance at the higher values of 100 and 1000. However, using values lower than 10 for the weighting factor does lead to the emergence of piecewise-linear behavior in the model which is undesirable, as seen in Figure \ref{fig:L2_spline_overlay}. Taking this into account, $w_{TD}$ = 10 is the optimal choice out of those tested, and it is the standard value used for experiments. 

\clearpage

\section{Tables and figures for Section 2} \label{sec:SI_additional_results}

\subsection{A note on learning curves}
In this section is presented the learning curves for all of the experiments performed for this study. The per-batch loss, as shown in Equation \ref{eqn:master_loss_fxn} in Section \ref{sec:SI_loss_reg}, is the quantity on which gradient descent steps are performed after every batch. However, the learning curves here report the average epoch loss, i.e. the loss averaged over the number of batches for the training and validation data for a given epoch, respectively, separated by target. Thus for a given target (i.e. total energy, dipoles, or charges), the values reported in the loss curves is as follows:
\begin{align}
    L_{prop_k} = \frac{1}{N_{batch}}\sum_j^{N_{batch}}\sqrt{\frac{1}{N_{prop_j}}\sum_{i}^{N_{prop_j}}|Pred_{i,j} - Target_{i,j}|^2}, \forall prop
\end{align}

Where $L_{prop_k}$ is the loss for the property $prop$ on epoch $k$. $N_{batch}$ is the number of batches in a given epoch, and the remaining terms have the same meaning as in Equation \ref{eqn:master_loss_fxn}. Due to the nature of its construction, this loss term gives back the appropriate units for each physical target, and the weighting factors $w_{prop}$ are divided out prior to reporting in the loss curves. The total energy loss is reported per heavy atom since internally, DFTBML trains on total energy per heavy atom as a way to account for the effect of molecule size. 

\subsection{Repulsive potential performance} \label{subsec:SI_solo_repulsive_full_results}

It is worthwhile to see how the repulsive potential performs on its own and to quantify the benefit of training the electronic portion of the model. For training the repulsive potential, the same datasets used for the experiments in Section 2.3 are repurposed, and the target that the repulsive model is being trained to is the difference between the true total molecular energy and the predicted electronic energy of the molecule obtained from a DFTB calculation using the Auorg parameters (see Section \ref{sec:reference+repulsive}). 

The naming conventions are established in Section 2.1 of the main paper. Full results can be seen in Tables \ref{Tab:SI_repulsive_solo_CC}, \ref{Tab:SI_repulsive_solo_DFT}, and \ref{Tab:SI_repulsive_solo_far_transfer}. All the experiments conducted used cubic splines with 50 knots and a vanishing boundary condition (zero, first, and second derivative all go to 0 at $r_h$). The results are also presented on a per heavy atom and per atom basis in Tables \ref{Tab:repulsive_solo_CC_per_atom}, \ref{Tab:repulsive_solo_DFT_per_atom}, and \ref{Tab:repulsive_solo_far_transfer_per_atom}.

\begin{table}[!ht] 
    \centering
    \caption{Performance of the repulsive potential trained to the CC total energy target}
    \label{Tab:SI_repulsive_solo_CC}
    \sisetup{round-mode=places}
    \large
    \resizebox{\textwidth}{!}{\begin{tabular}{lccS[round-precision=2]S[round-precision=3]S[round-precision=3]}
    \toprule
        \textbf{Parameterization} & \textbf{Outliers} & \textbf{Nonconverged} & \textbf{MAE energy (kcal/mol)} & \textbf{MAE dipole (e\AA)} & \textbf{MAE charge (e)} \\ \midrule
        Auorg & 0 & 0 & 10.548555105 & 0.07888197 & 0.084601841\\ 
        MIO & 0 & 0 & 10.693563047 & 0.078904702 & 0.084754859 \\ 
        GFN1-xTB & 0 & 0 & 10.659580098 & 0.135582684 & 0.102679051 \\ 
        GFN2-xTB & 0 & 0 & 13.032294013 & 0.152755742 & 0.088788801 \\ 
        Repulsive CC 20000 & 3 & 0 & 5.407935998 & 0.078899922 & 0.084598111 \\ 
        Repulsive CC 10000 & 4 & 0 & 5.408518381 & 0.078905702 & 0.08459702 \\ 
        Repulsive CC 5000 & 0 & 0 & 5.412641954 & 0.078881977 & 0.084601842 \\ 
        Repulsive CC 2500 & 0 & 0 & 5.422167647 & 0.078881977 & 0.084601842 \\ 
        Repulsive CC 1000 & 2 & 0 & 5.423583541 & 0.078885507 & 0.084609275 \\ 
        Repulsive CC 300 & 3 & 0 & 5.957531219 & 0.078880799 & 0.084611548 \\ \bottomrule
    \end{tabular}}
\end{table}

\begin{table}[!ht]
    \centering
    \caption{Performance of the repulsive potential trained to the DFT total energy target}
    \label{Tab:SI_repulsive_solo_DFT}
    \sisetup{round-mode=places}
    \large
    \resizebox{\textwidth}{!}{\begin{tabular}{lccS[round-precision=2]S[round-precision=3]S[round-precision=3]}
    \toprule
        \textbf{Parameterization} & \textbf{Outliers} & \textbf{Nonconverged} & \textbf{MAE energy (kcal/mol)} & \textbf{MAE dipole (e\AA)} & \textbf{MAE charge (e)} \\ \midrule
        Auorg & 0 & 0 & 11.95100350 & 0.07888197 & 0.08460184 \\ 
        MIO & 0 & 0 & 11.686877269 & 0.078904702 & 0.084754859 \\ 
        GFN1-xTB & 0 & 0 & 9.831820164 & 0.135582684 & 0.102679051 \\ 
        GFN2-xTB & 0 & 0 & 11.8196830253 & 0.152755742 & 0.088788801 \\ 
        Repulsive DFT 20000 & 0 & 0 & 5.714682975 & 0.078881977 & 0.084601842 \\ 
        Repulsive DFT 10000 & 0 & 0 & 5.75165196 & 0.078881977 & 0.084601842 \\ 
        Repulsive DFT 5000 & 0 & 0 & 5.740002075 & 0.078881977 & 0.084601842 \\ 
        Repulsive DFT 2500 & 0 & 0 & 5.735821536 & 0.078881977 & 0.084601842 \\ 
        Repulsive DFT 1000 & 3 & 0 & 5.739125959 & 0.078891511 & 0.084607684 \\ 
        Repulsive DFT 300 & 3 & 0 & 6.510409464 & 0.078880799 & 0.084611548 \\ \bottomrule
    \end{tabular}}
\end{table}

\begin{table}[!ht]
    \centering
    \caption{Performance of the repulsive potential in far-transfer}
    \label{Tab:SI_repulsive_solo_far_transfer}
    \sisetup{round-mode=places}
    \large
    \resizebox{\textwidth}{!}{\begin{tabular}{lccS[round-precision=2]S[round-precision=3]S[round-precision=3]}
    \toprule
        \textbf{Parameterization} & \textbf{Outliers} & \textbf{Nonconverged} & \textbf{MAE energy (kcal/mol)} & \textbf{MAE dipole (e\AA)} & \textbf{MAE charge (e)} \\ \midrule
        Auorg CC & 0 & 0 & 11.81432903 & 0.088503081 & 0.087564107 \\ 
        MIO CC & 0 & 0 & 11.864115851 & 0.088528894 & 0.087712701 \\ 
        Auorg DFT & 0 & 0 & 13.2546505 & 0.088503081 & 0.087564107 \\ 
        MIO DFT & 0 & 0 & 13.047158534 & 0.088528894 & 0.087712701 \\ 
        GFN1-xTB CC & 0 & 0 & 10.854720214 & 0.146972956 & 0.103969430 \\ 
        GFN2-xTB CC & 0 & 0 & 13.508626798 & 0.165065903 & 0.090688288 \\ 
        GFN1-xTB DFT & 0 & 0 & 10.142095148 & 0.146972956 & 0.103969430 \\ 
        GFN2-xTB DFT & 0 & 0 & 12.256613247 & 0.165065903 & 0.090688288 \\ 
        Repulsive Transfer CC 2500 & 0 & 0 & 7.172917109 & 0.088503081 & 0.087564107 \\ 
        Repulsive Transfer DFT 2500 & 0 & 0 & 7.394333283 & 0.088503081 & 0.087564107 \\ \bottomrule
    \end{tabular}}
\end{table}

\begin{table}[!ht]
    \centering
    \caption{Performance per atom and per heavy atom of the repulsive potential trained to the CC total energy target}
    \label{Tab:repulsive_solo_CC_per_atom}
    \sisetup{round-mode=places}
    \large
    \resizebox{\textwidth}{!}{\begin{tabular}{lS[round-precision=2]S[round-precision=2]}
    \toprule
        \textbf{Parameterization} & \textbf{MAE energy (kcal mol\boldmath{$^{-1} N_{heavy}^{-1}$})} & \textbf{MAE energy (kcal mol\boldmath{$^{-1} N_{atom}^{-1}$})} \\ \midrule
        Auorg & 1.708164528 & 0.971925778 \\ 
        MIO & 1.740547040 & 0.975882385 \\ 
        GFN1-xTB & 1.826831025 & 1.053434836 \\ 
        GFN2-xTB & 2.233148843 & 1.330365437 \\ 
        Repulsive CC 20000 & 0.94822248 & 0.544627432 \\ 
        Repulsive CC 10000 & 0.943338299 & 0.541630778 \\ 
        Repulsive CC 5000 & 0.94042816 & 0.538779452 \\ 
        Repulsive CC 2500 & 0.94318568 & 0.54232137 \\ 
        Repulsive CC 1000 & 0.960647504 & 0.550408607 \\ 
        Repulsive CC 300 & 1.021271244 & 0.582387107 \\ \bottomrule
    \end{tabular}}
\end{table}

\begin{table}[!ht]
    \centering
    \caption{Performance per atom and per heavy atom of the repulsive potential trained to the DFT total energy target}
    \label{Tab:repulsive_solo_DFT_per_atom}
    \sisetup{round-mode=places}
    \large
    \resizebox{\textwidth}{!}{\begin{tabular}{lS[round-precision=2]S[round-precision=2]}
    \toprule
        \textbf{Parameterization} & \textbf{MAE energy (kcal mol\boldmath{$^{-1} N_{heavy}^{-1}$})} & \textbf{MAE energy (kcal mol\boldmath{$^{-1} N_{atom}^{-1}$})} \\ \midrule
        Auorg & 1.923890536 & 1.098888961 \\ 
        MIO & 1.880101281 & 1.069081415 \\ 
        GFN1-xTB & 1.706850090 & 0.971032201 \\ 
        GFN2-xTB & 2.053740821 & 1.210437811 \\ 
        Repulsive DFT 20000 & 1.007904876 & 0.572299926 \\ 
        Repulsive DFT 10000 & 1.017514653 & 0.577009375 \\ 
        Repulsive DFT 5000 & 1.003558108 & 0.57004717 \\ 
        Repulsive DFT 2500 & 1.004037919 & 0.57179305 \\ 
        Repulsive DFT 1000 & 1.021685556 & 0.582878088 \\ 
        Repulsive DFT 300 & 1.122753048 & 0.642316283 \\ \bottomrule
    \end{tabular}}
\end{table}

\begin{table}[!ht]
    \centering
    \caption{Performance per atom and per heavy atom of the repulsive potential in far-transfer}
    \label{Tab:repulsive_solo_far_transfer_per_atom}
    \sisetup{round-mode=places}
    \large
    \resizebox{\textwidth}{!}{\begin{tabular}{lS[round-precision=2]S[round-precision=2]}
    \toprule
        \textbf{Parameterization} & \textbf{MAE energy (kcal mol\boldmath{$^{-1} N_{heavy}^{-1}$})} & \textbf{MAE energy (kcal mol\boldmath{$^{-1} N_{atom}^{-1}$})} \\ \midrule
        Auorg CC & 1.644885647 & 0.970457777 \\ 
        MIO CC & 1.651700437 & 0.966793710 \\ 
        Auorg DFT & 1.846739579 & 1.096933306 \\ 
        MIO DFT & 1.815826239 & 1.074810495 \\ 
        GFN1-xTB CC & 1.515169543 & 0.909799076 \\ 
        GFN2-xTB CC & 1.883735044 & 1.155594396  \\ 
        GFN1-xTB DFT & 1.412688788 & 0.831117175  \\ 
        GFN2-xTB DFT & 1.705860984 & 1.030303543  \\ 
        Repulsive Transfer CC 2500 & 0.988402117 & 0.598910645 \\ 
        Repulsive Transfer DFT 2500 & 1.020551595 & 0.614436344 \\ \bottomrule
    \end{tabular}}
\end{table}

Unlike when training the electronic model, the only target of interest here is the performance on total molecular energy since training the repulsive potential only introduces an energy correction term. As seen in Tables \ref{Tab:SI_repulsive_solo_CC} and \ref{Tab:SI_repulsive_solo_DFT}, the MAE on dipoles and atomic charges remains consistent despite the change in the size of the dataset. An additional observation is that the improvement in performance for total energy saturates quickly and there are no significant improvements obtained by consistently increasing the amount of training data beyond 1000 molecules. In training to either the CC or DFT total energy target, the performance from using 2500 through 20000 training molecules remains remarkably steady, leveling out around 5.4 kcal/mol in the CC case and 5.7 kcal/mol in the DFT case. A degradation of performance associated with a lack of training data is only observed when dropping to 300 training molecules in both cases. 

In terms of the repulsive potential's performance in the far-transfer experiments, the repulsive potential alone performs worse than the full DFTBML model (see Table 3 of the main paper), but it still performs better than the xTB methods and standard DFTB parameterizations. Collectively, the results presented show that training only the repulsive potential does provide an improvement to the performance of base DFTB, but it does not perform as well as the full DFTBML model. Furthermore, the repulsive potential is not as sensitive to the quantity of training data as the full model. This may be partly related to the use of a quadratic programming approach to find the global minimum, as opposed to the gradient descent approach used to train the full DFTBML model. 

\begin{figure}[ht!]
    \centering
    \includegraphics[width=1.0\columnwidth]{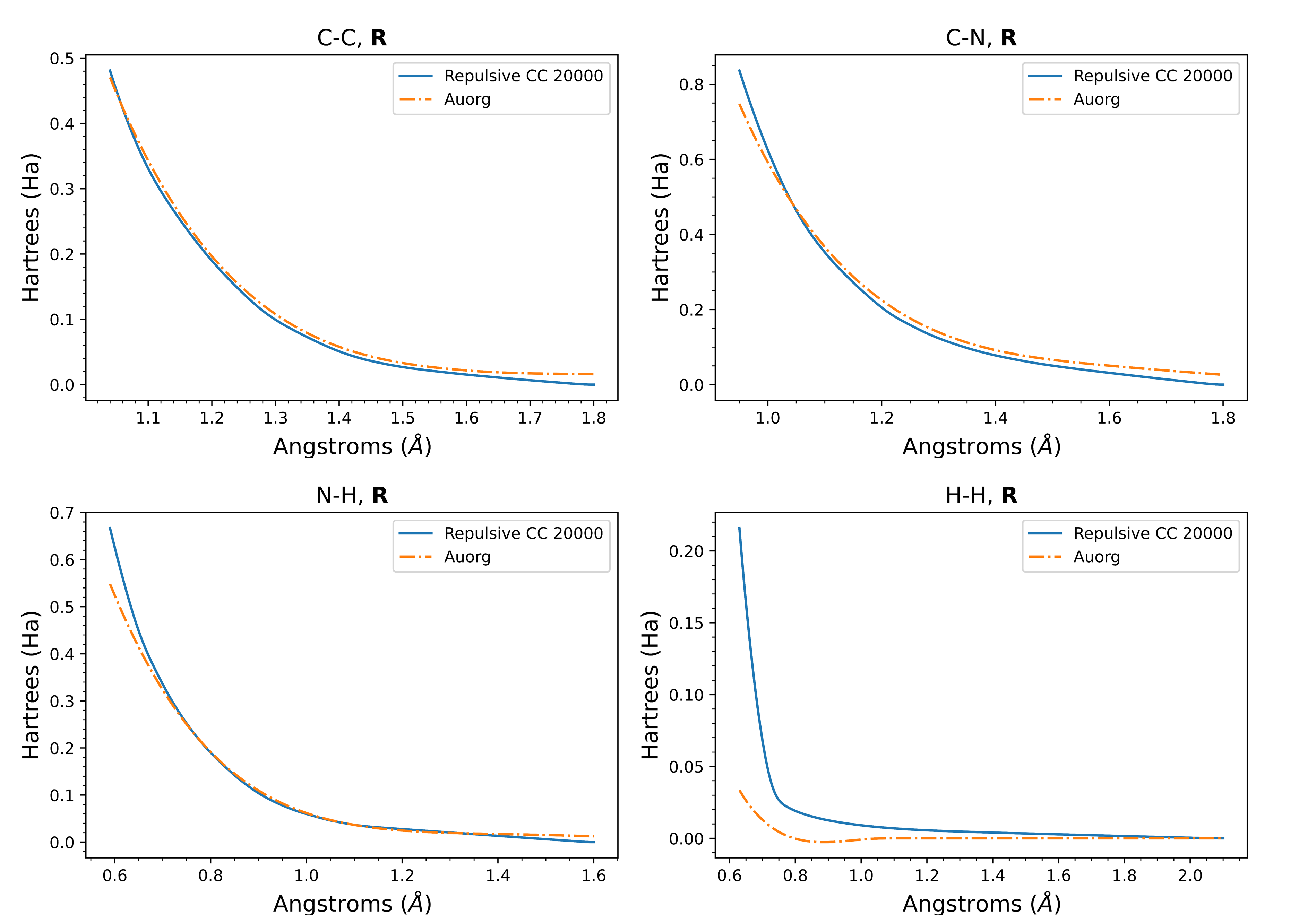}
    \caption{Four different repulsive potentials from the Repulsive CC 20000 experiment. The Auorg repulsive potentials are overlaid for reference.}
    \label{fig:repulsive_potential_only}
\end{figure}

The functional forms of the models used for the repulsive potential are simpler than the forms for the models used for the Hamiltonian or overlap integrals since the repulsive potential is a classical interaction intended to include interactions between the core electrons that are excluded from the DFTB Hamiltonian. As such, the potentials are constrained to have a positive second derivative at all points, giving a smooth and exponentially decaying form. A few examples of the repulsive potential are shown in Figure \ref{fig:repulsive_potential_only}. In total there are 10 such potentials needed to specify all pairwise interactions between only C, H, N, and O atoms. 

\clearpage

\subsection{Unregularized model performance}

\begin{figure}[ht!]
    \centering
    \includegraphics[width=\columnwidth]{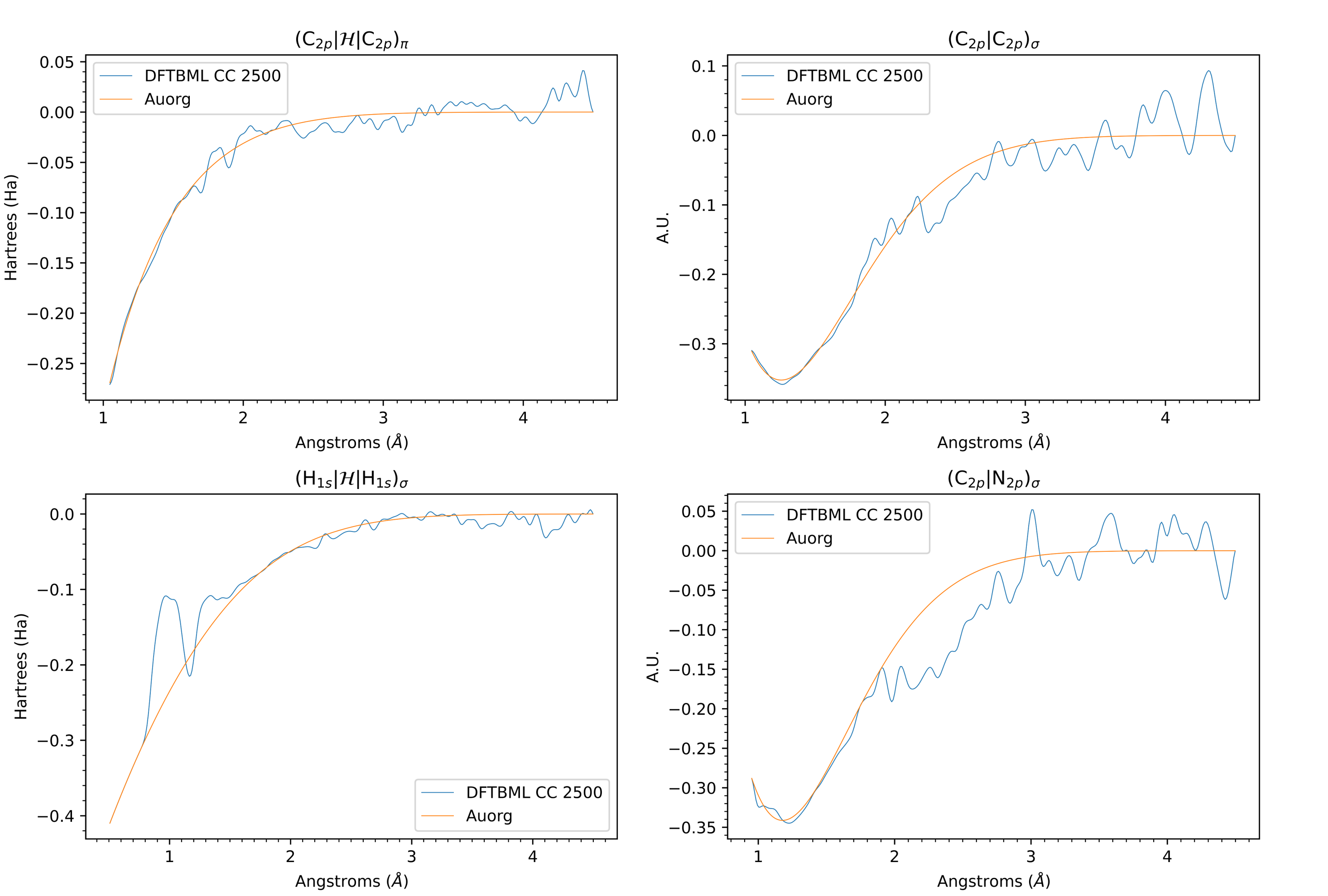}
    \caption{Examples of unregularized splines obtained from training using the DFTBML CC 2500 dataset. The Auorg potentials are included for reference.}
    \label{fig:unregularized_splines}
\end{figure}

\clearpage

\subsection{Model performance when training on different dataset sizes}\label{sec:SI_different_dset_tables_figs}

\begin{figure*}[h!]
    \centering
    \includegraphics[width=1.0\textwidth]{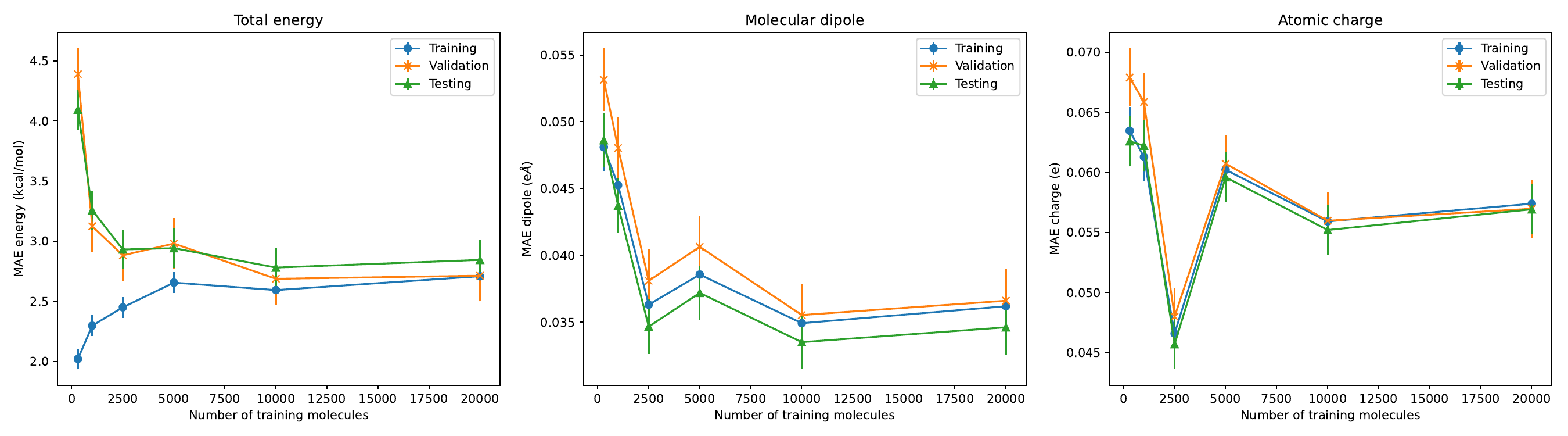}
    \caption{Final training, validation, and test loss for each of the physical targets as a function of the size of the dataset used for training. Results were obtained using the datasets containing the DFT total energy targets. Error bars are shown as $\pm\frac{\sigma}{3}$ where $\sigma$ is the standard deviation of the errors calculated separately for the training, validation, and testing values.}
    \label{fig:target_convergence_v_size_DFT}
\end{figure*}

\begin{figure}[ht!]
    \centering
    \includegraphics[width=\columnwidth]{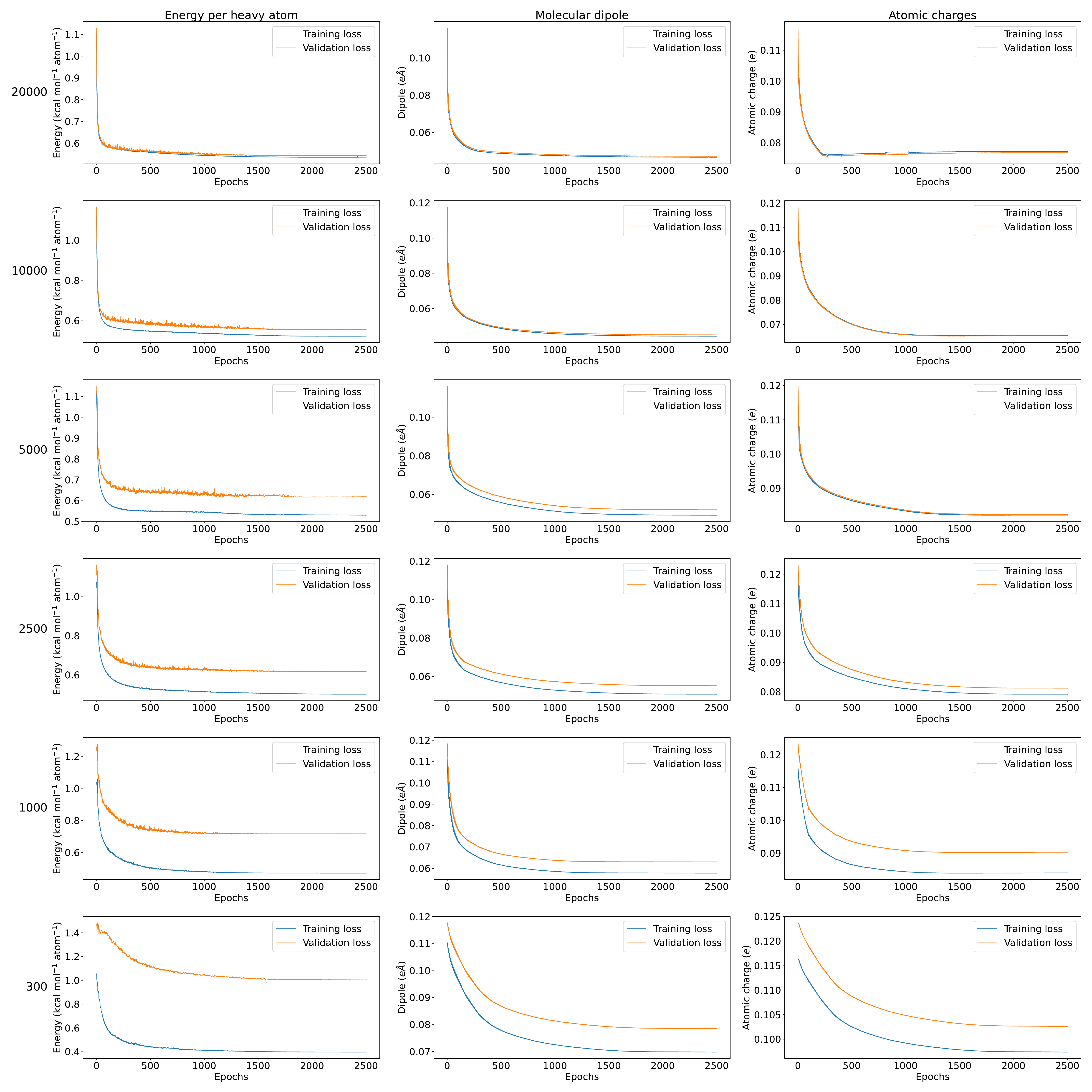}
    \caption{Learning curves for total energy per heavy atom, molecular dipole, and atomic charge when training to the CC energy target for total energy. The label for each row indicates the number of training molecules used for each experiment.}
    \label{fig:CC_learning_curves}
\end{figure}

\begin{figure}[ht!]
    \centering
    \includegraphics[width=\columnwidth]{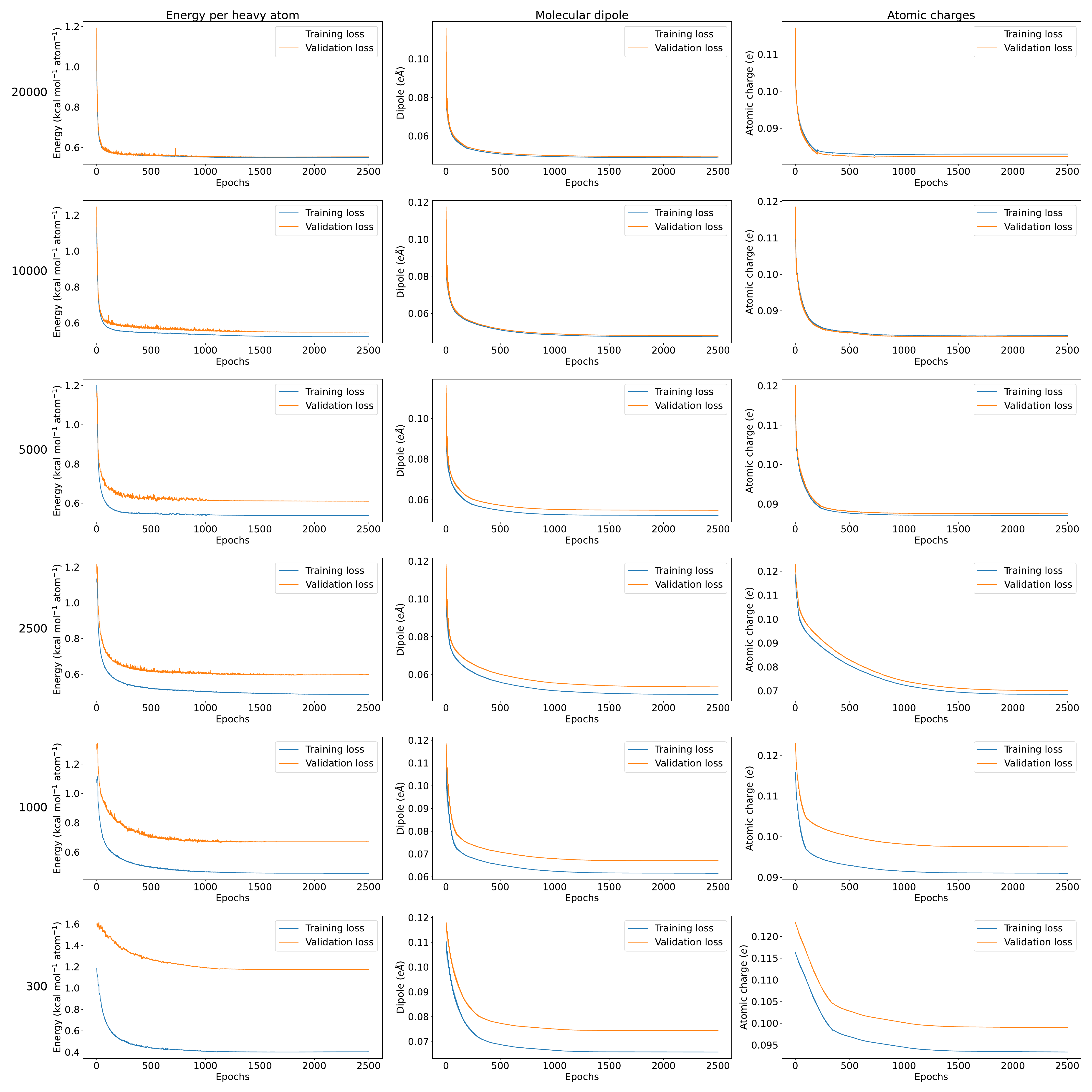}
    \caption{Learning curves for total energy per heavy atom, molecular dipole, and atomic charge when training to the DFT energy target for total energy. The label for each row indicates the number of training molecules used for each experiment.}
    \label{fig:WT_learning_curves}
\end{figure}

\begin{table}[!ht]
    \centering
    \caption{Performance of different parameterizations trained against CC energy target}
    \label{Tab:SI_cc_exp_quant}
    \sisetup{round-mode=places}
    \large
    \resizebox{\textwidth}{!}{\begin{tabular}{lccS[round-precision=2]S[round-precision=3]S[round-precision=3]}
    \toprule
        \textbf{Parameterization} & \textbf{Outliers} & \textbf{Nonconverged} & \textbf{MAE energy (kcal/mol)} & \textbf{MAE dipole (e\AA)} & \textbf{MAE charge (e)} \\ \midrule
        Auorg & 0 & 0 & 10.548555105 & 0.07888197 & 0.084601841\\ 
        MIO & 0 & 0 & 10.693563047 & 0.078904702 & 0.084754859 \\ 
        GFN1-xTB & 0 & 0 & 10.659580098 & 0.135582684 & 0.102679051 \\ 
        GFN2-xTB & 0 & 0 & 13.032294013 & 0.152755742 & 0.088788801 \\ 
        DFTBML CC 20000 & 0 & 0 & 2.672659848 & 0.033047873 & 0.052958396 \\ 
        DFTBML CC 10000 & 0 & 0 & 2.684503777 & 0.031177289 & 0.043952657 \\ 
        DFTBML CC 5000 & 0 & 1 & 2.843044101 & 0.034946179 & 0.055470864 \\ 
        DFTBML CC 2500 & 0 & 1 & 2.947089603 & 0.035830429 & 0.053621256 \\ 
        DFTBML CC 1000 & -2 & 0 & 3.249747978 & 0.040668527 & 0.058148915 \\ 
        DFTBML CC 300 & -4 & 0 & 4.140990797 & 0.05176159 & 0.066075539 \\ \bottomrule
    \end{tabular}}
\end{table}

\begin{table}[!ht]
    \centering
    \caption{Performance of different parameterizations trained against DFT energy target}
    \label{Tab:SI_dft_exp_quant}
    \sisetup{round-mode=places}
    \large
    \resizebox{\textwidth}{!}{\begin{tabular}{lccS[round-precision=2]S[round-precision=3]S[round-precision=3]}
    \toprule
        \textbf{Parameterization} & \textbf{Outliers} & \textbf{Nonconverged} & \textbf{MAE energy (kcal/mol)} & \textbf{MAE dipole (e\AA)} & \textbf{MAE charge (e)} \\ \midrule
        Auorg & 0 & 0 & 11.95100350 & 0.07888197 & 0.08460184 \\ 
        MIO & 0 & 0 & 11.686877269 & 0.078904702 & 0.084754859 \\ 
        GFN1-xTB & 0 & 0 & 9.831820164 & 0.135582684 & 0.102679051 \\ 
        GFN2-xTB & 0 & 0 & 11.8196830253 & 0.152755742 & 0.088788801 \\ 
        DFTBML DFT 20000 & 0 & 0 & 2.844527111 & 0.034611572 & 0.056926236 \\ 
        DFTBML DFT 10000 & 0 & 0 & 2.780649563 & 0.033498379 & 0.05520389 \\ 
        DFTBML DFT 5000 & 0 & 0 & 2.942230868 & 0.03718845 & 0.059594985 \\ 
        DFTBML DFT 2500 & 0 & 0 & 2.931086849 & 0.03466201 & 0.045701002 \\ 
        DFTBML DFT 1000 & -3 & 0 & 3.257273694 & 0.043716532 & 0.062218114 \\ 
        DFTBML DFT 300 & -5 & 0 & 4.094116102 & 0.048632409 & 0.062578825 \\ \bottomrule
    \end{tabular}}
\end{table}

\begin{table}[!ht]
    \centering
    \caption{MAE in energy per atom and per heavy atom when trained against CC total energy targets}
    \label{Tab:cc_eatom}
    \sisetup{round-mode=places}
    \large
    \resizebox{\textwidth}{!}{\begin{tabular}{lS[round-precision=2]S[round-precision=2]}
    \toprule
        \textbf{Parameterization} & \textbf{MAE energy (kcal mol\boldmath{$^{-1} N_{heavy}^{-1}$})} & \textbf{MAE energy (kcal mol\boldmath{$^{-1} N_{atom}^{-1}$})} \\ \midrule
        Auorg & 1.708164528 & 0.971925778 \\ 
        MIO & 1.740547040 & 0.975882385 \\ 
        GFN1-xTB & 1.826831025 & 1.053434836 \\ 
        GFN2-xTB & 2.233148843 & 1.330365437 \\ 
        DFTBML CC 20000 & 0.438858399 & 0.248618750 \\ 
        DFTBML CC 10000 & 0.449219866 & 0.258354643  \\ 
        DFTBML CC 5000 & 0.466576385 & 0.264609215  \\ 
        DFTBML CC 2500 & 0.488295067 & 0.276970893  \\ 
        DFTBML CC 1000 & 0.559219334 & 0.321913653  \\ 
        DFTBML CC 300 & 0.711151181 & 0.40359237  \\ \bottomrule
    \end{tabular}}
\end{table}

\begin{table}[!ht]
    \centering
    \caption{MAE in energy per atom and per heavy atom when trained against DFT total energy targets}
    \label{Tab:dft_eatom}
    \sisetup{round-mode=places}
    \large
    \resizebox{\textwidth}{!}{\begin{tabular}{lS[round-precision=2]S[round-precision=2]}
    \toprule
        \textbf{Parameterization} & \textbf{MAE energy (kcal mol\boldmath{$^{-1} N_{heavy}^{-1}$})} & \textbf{MAE energy (kcal mol\boldmath{$^{-1} N_{atom}^{-1}$})} \\ \midrule
        Auorg & 1.923890536 & 1.098888961 \\ 
        MIO & 1.880101281 & 1.069081415 \\ 
        GFN1-xTB & 1.706850090 & 0.971032201 \\ 
        GFN2-xTB & 2.053740821 & 1.210437811 \\ 
        DFTBML DFT 20000 & 0.475096489 & 0.270624295 \\ 
        DFTBML DFT 10000 & 0.463771676 & 0.261961818 \\ 
        DFTBML DFT 5000 & 0.491323492 & 0.277491494 \\ 
        DFTBML DFT 2500 & 0.492500263 & 0.277799826 \\ 
        DFTBML DFT 1000 & 0.571089061 & 0.331627576 \\ 
        DFTBML DFT 300 & 0.711228816 & 0.402026949 \\ \bottomrule
    \end{tabular}}
\end{table}

\begin{figure}[ht!]
    \centering
    \includegraphics[width=\columnwidth]{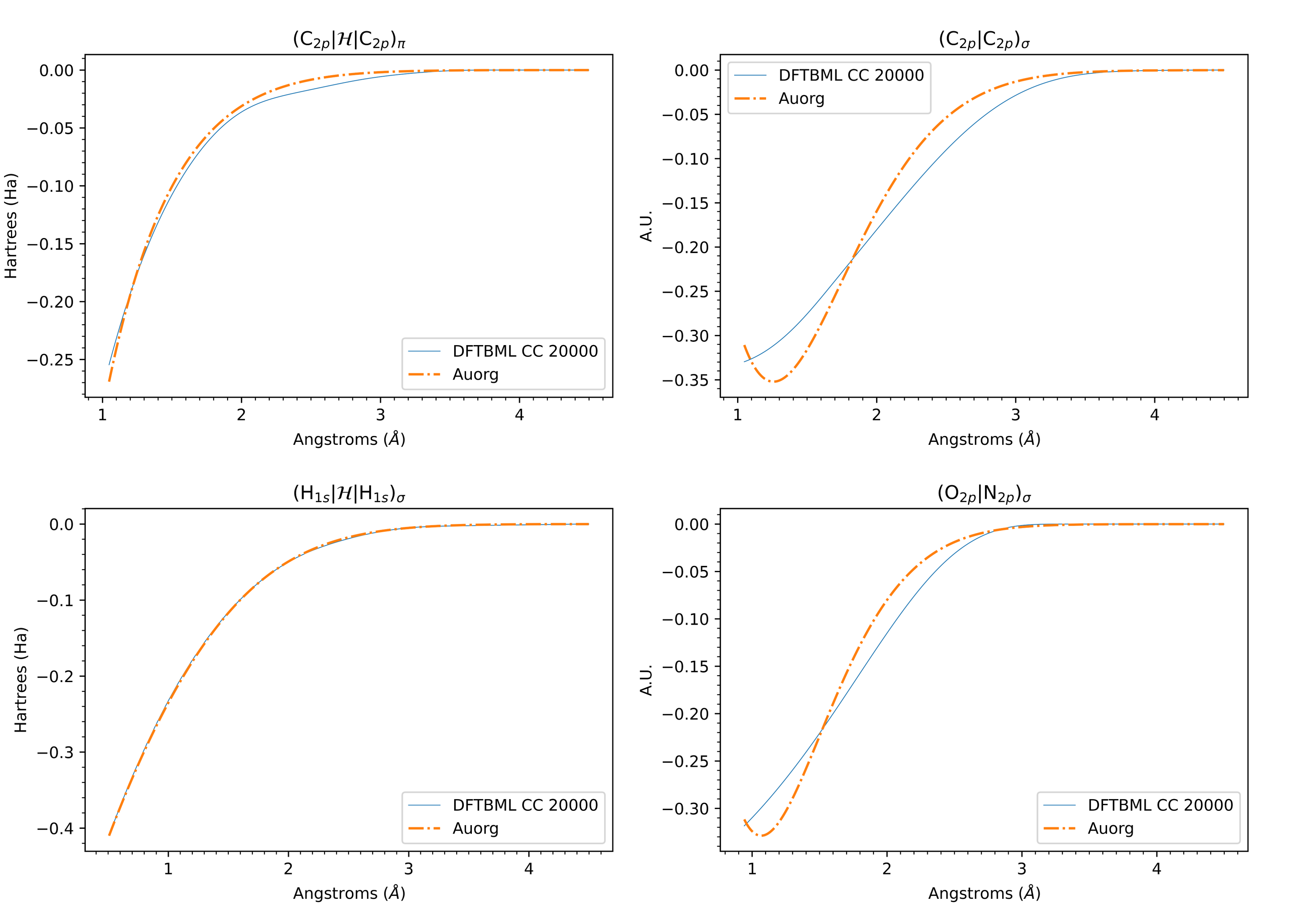}
    \caption{Representative splines generated from DFTBML CC 20000 for a few of the matrix elements involved in computing properties for organic molecules. The two plots on the left hand column involve Hamiltonian matrix elements and the two plots on the right hand column detail overlap matrix elements. $\sigma$ and $\pi$ are used to indicate the orientation of the interaction. Hamiltonian matrix elements are given in units of Hartrees and the overlap matrix elements have arbitrary units (A.U.). The potential functions from the Auorg reference set are overlaid for comparison.}
    \label{fig:representative_splines}
\end{figure}

\clearpage

\subsection{Model transferability and reproducibility}

\begin{figure}[ht!]
    \centering
    \includegraphics[width=\columnwidth]{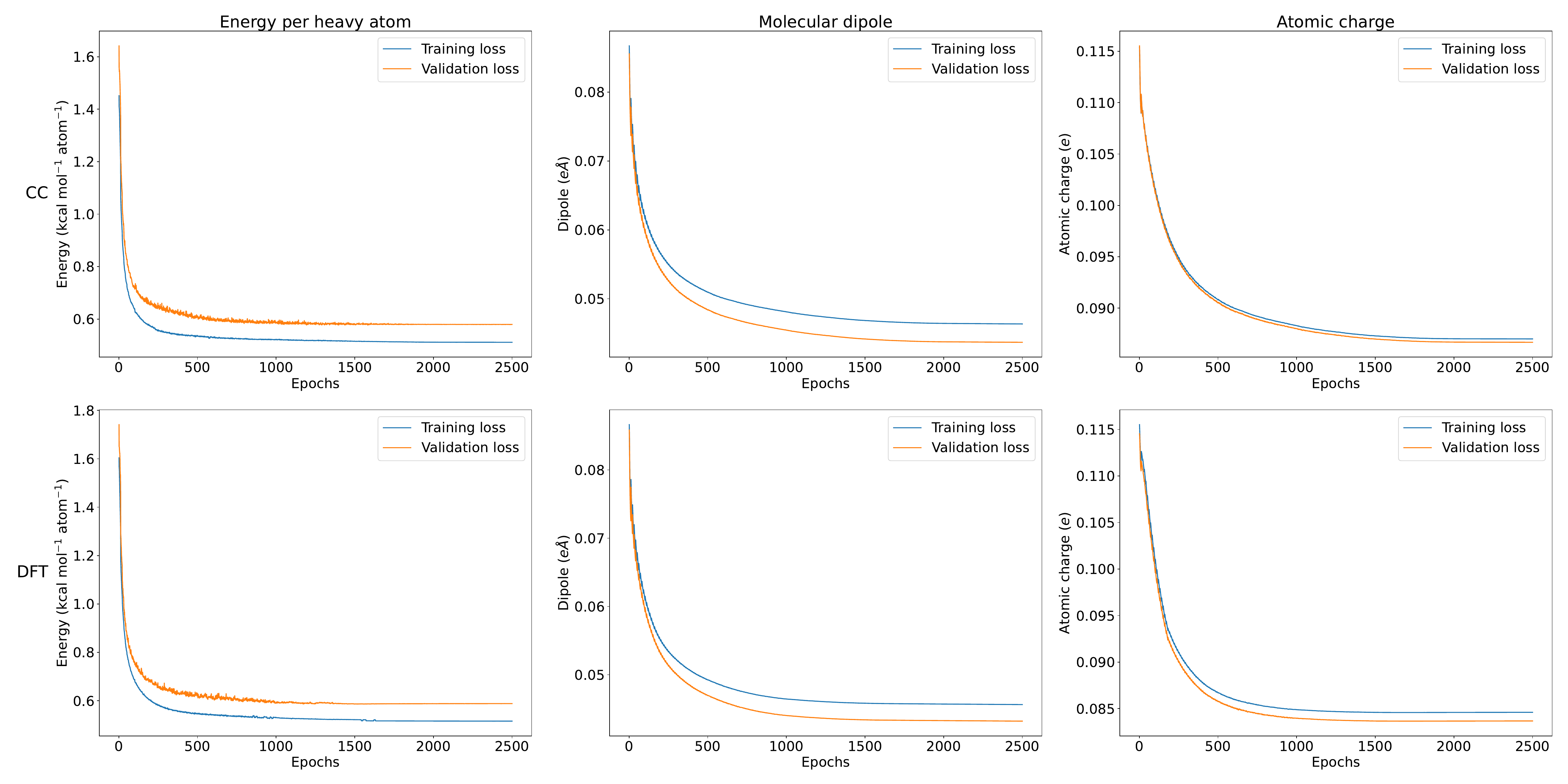}
    \caption{Learning curves for total energy per heavy atom, molecular dipole, and atomic charge for far-transfer training. The top row is for training to the CC total energy target and the bottom row is for training to the DFT total energy target.}
    \label{fig:transfer_learning_curves}
\end{figure}

\begin{table}[!ht]
    \centering
    \caption{MAE in energy per atom and per heavy atom in far-transfer}
    \label{Tab:eatom_far_transfer}
    \sisetup{round-mode=places}
    \large
    \resizebox{\textwidth}{!}{\begin{tabular}{lS[round-precision=2]S[round-precision=2]}
    \toprule
        \textbf{Parameterization} & \textbf{MAE energy (kcal mol\boldmath{$^{-1} N_{heavy}^{-1}$})} & \textbf{MAE energy (kcal mol\boldmath{$^{-1} N_{atom}^{-1}$})} \\ \midrule
        Auorg CC & 1.644885647 & 0.970457777 \\ 
        MIO CC & 1.651700437 & 0.966793710 \\ 
        Auorg DFT & 1.846739579 & 1.096933306 \\ 
        MIO DFT & 1.815826239 & 1.074810495 \\ 
        GFN1-xTB CC & 1.515169543 & 0.909799076 \\ 
        GFN2-xTB CC & 1.883735044 & 1.155594396  \\ 
        GFN1-xTB DFT & 1.412688788 & 0.831117175  \\ 
        GFN2-xTB DFT & 1.705860984 & 1.030303543  \\ 
        Transfer CC 2500 & 0.661599906 & 0.407295490  \\ 
        Transfer DFT 2500 & 0.656844974 & 0.411345279  \\ \bottomrule 
    \end{tabular}}
\end{table}

\begin{figure}[ht!]
    \centering
    \includegraphics[width=\columnwidth]{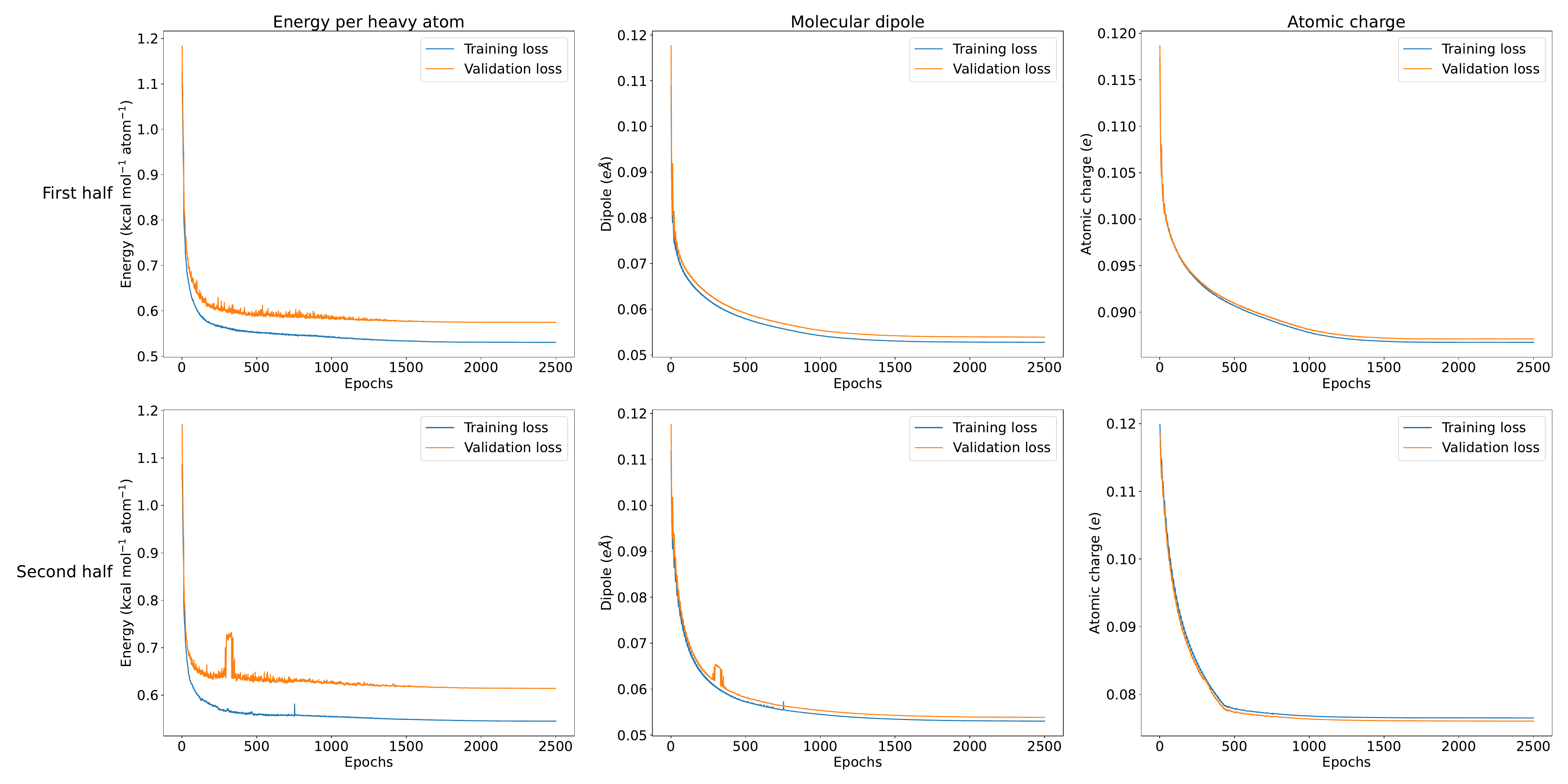}
    \caption{Learning curves for total energy per heavy atom, molecular dipole, and atomic charge from assessing model reproducibility from two disjoint training sets.}
    \label{fig:reprod_learning_curves}
\end{figure}

\begin{table}[!ht] 
    \centering
    \caption{MAE in energy per atom and per heavy atom trained on two disjoint training sets}
    \label{Tab:eatom_reproducibility}
    \sisetup{round-mode=places}
    \large
    \resizebox{\textwidth}{!}{\begin{tabular}{lS[round-precision=2]S[round-precision=2]}
    \toprule
        \textbf{Parameterization} & \textbf{MAE energy (kcal mol\boldmath{$^{-1} N_{heavy}^{-1}$})} & \textbf{MAE energy (kcal mol\boldmath{$^{-1} N_{atom}^{-1}$})} \\ \midrule
        Auorg & 1.708164528 & 0.971925778 \\ 
        MIO & 1.740547040 & 0.975882385 \\ 
        GFN1-xTB & 1.826831025 & 1.053434836 \\ 
        GFN2-xTB & 2.233148843 & 1.330365437 \\ 
        DFTBML CC 5000 First Half & 0.463624168 & 0.264851041 \\ 
        DFTBML CC 5000 Second Half & 0.488735369 & 0.277856475 \\ \bottomrule
    \end{tabular}}
\end{table}

\clearpage

\subsection{DFTBML performance on COMP6 benchmark}\label{sec:SI_COMP6_results}

\begin{table*}[!ht]
    \centering
    \caption{MAE total energy in kcal/mol for DFTBML and various models on COMP6 benchmark suite}
    \label{Tab:SI_COMP6_tot_ener}
    \resizebox{\textwidth}{!}{\begin{tabular}{lcccccccccc}
    \toprule
        \textbf{Test set} & \textbf{Auorg} & \textbf{MIO} & \textbf{GFN1-xTB} & \textbf{GFN2-xTB} & \textbf{DFTBML CC} & \textbf{Repulsive CC} & \textbf{DFTBML DFT} & \textbf{Repulsive DFT} & \textbf{Transfer CC} & \textbf{Transfer DFT} \\ \midrule
        Ani MD & 6.23 & 5.78 & 4.80 & 7.94 & 5.13 & 6.56 & 4.48 & 5.69 & 5.30 & \textbf{4.17} \\ 
        Drugbank & 13.39 & 12.66 & 11.40 & 10.97 & 8.14 & 10.78 & \textbf{6.51} & 9.57 & 8.74 & 6.71 \\ 
        GDB 7 & 9.33 & 9.02 & 8.40 & 7.64 & 3.84 & 6.26 & \textbf{3.17} & 5.68 & 4.32 & 3.43 \\ 
        GDB 8 & 10.20 & 9.89 & 8.99 & 8.09 & 4.06 & 6.43 & \textbf{3.29} & 5.92 & 4.56 & 3.53 \\ 
        GDB 9 & 10.99 & 10.66 & 8.96 & 8.62 & 4.38 & 7.37 & \textbf{3.44} & 6.59 & 4.99 & 3.72 \\ 
        GDB 10 & 10.70 & 10.45 & 9.79 & 9.18 & 4.58 & 7.17 & \textbf{3.67} & 6.47 & 5.29 & 3.96 \\ 
        GDB 11 & 12.69 & 12.51 & 10.32 & 10.66 & 5.83 & 8.83 & \textbf{4.11} & 7.92 & 6.66 & 4.49 \\ 
        GDB 12 & 13.70 & 13.53 & 10.59 & 11.03 & 6.18 & 9.65 & \textbf{4.21} & 8.44 & 6.71 & 4.46 \\ 
        GDB 13 & 14.10 & 13.91 & 11.37 & 11.44 & 6.52 & 10.09 & \textbf{4.38} & 8.85 & 7.25 & 4.72 \\ 
        S66x8 & 4.28 & 3.65 & 4.68 & 4.65 & 2.64 & 4.52 & \textbf{2.64} & 4.54 & 3.90 & 2.94 \\ 
        Tripeptide & 9.11 & 8.96 & 6.25 & 7.36 & 5.04 & 7.74 & \textbf{4.02} & 6.23 & 6.11 & 4.83 \\ \bottomrule
    \end{tabular}}
\end{table*}

\begin{table*}[!ht]
    \centering
    \caption{MAE dipole in e\AA{} for DFTBML and various models on COMP6 benchmark suite}
    \label{Tab:SI_COMP6_dipole}
    \resizebox{\textwidth}{!}{\begin{tabular}{lcccccccccc}
    \toprule
        \textbf{Test set} & \textbf{Auorg} & \textbf{MIO} & \textbf{GFN1-xTB} & \textbf{GFN2-xTB} & \textbf{DFTBML CC} & \textbf{Repulsive CC} & \textbf{DFTBML DFT} & \textbf{Repulsive DFT} & \textbf{Transfer CC} & \textbf{Transfer DFT} \\ \midrule
        Ani MD & 0.167 & 0.168 & 0.170 & 0.205 & \textbf{0.104} & 0.167 & 0.106 & 0.167 & 0.119 & 0.115 \\ 
        Drugbank & 0.111 & 0.111 & 0.207 & 0.242 & \textbf{0.053} & 0.111 & 0.055 & 0.111 & 0.066 & 0.064 \\ 
        GDB 7 & 0.088 & 0.088 & 0.138 & 0.157 & \textbf{0.031} & 0.088 & 0.033 & 0.088 & 0.045 & 0.043 \\ 
        GDB 8 & 0.094 & 0.094 & 0.144 & 0.163 & \textbf{0.034} & 0.094 & 0.036 & 0.094 & 0.050 & 0.049 \\ 
        GDB 9 & 0.095 & 0.095 & 0.155 & 0.174 & \textbf{0.034} & 0.095 & 0.036 & 0.095 & 0.049 & 0.047 \\ 
        GDB 10 & 0.099 & 0.099 & 0.165 & 0.185 & \textbf{0.038} & 0.099 & 0.040 & 0.099 & 0.055 & 0.053 \\ 
        GDB 11 & 0.116 & 0.116 & 0.169 & 0.192 & \textbf{0.049} & 0.116 & 0.051 & 0.116 & 0.065 & 0.063 \\ 
        GDB 12 & 0.117 & 0.117 & 0.172 & 0.195 & \textbf{0.048} & 0.117 & 0.050 & 0.117 & 0.065 & 0.062 \\ 
        GDB 13 & 0.122 & 0.122 & 0.182 & 0.207 & \textbf{0.051} & 0.122 & 0.053 & 0.122 & 0.068 & 0.065 \\ 
        S66x8 & 0.062 & 0.062 & 0.129 & 0.146 & \textbf{0.031} & 0.062 & 0.033 & 0.062 & 0.043 & 0.040 \\ 
        Tripeptide & 0.128 & 0.128 & 0.311 & 0.371 & 0.073 & 0.128 & \textbf{0.070} & 0.128 & 0.079 & 0.074 \\ \bottomrule
    \end{tabular}}
\end{table*}

\begin{table*}[!ht]
    \centering
    \caption{MAE charge in e for DFTBML and various models on COMP6 benchmark suite}
    \label{Tab:SI_COMP6_charge}
    \resizebox{\textwidth}{!}{\begin{tabular}{lcccccccccc}
    \toprule
        \textbf{Test set} & \textbf{Auorg} & \textbf{MIO} & \textbf{GFN1-xTB} & \textbf{GFN2-xTB} & \textbf{DFTBML CC} & \textbf{Repulsive CC} & \textbf{DFTBML DFT} & \textbf{Repulsive DFT} & \textbf{Transfer CC} & \textbf{Transfer DFT} \\ \midrule
        Ani MD & 0.071 & 0.071 & 0.090 & 0.076 &\textbf{0.053} & 0.071 & 0.057 & 0.071 & 0.058 & 0.057 \\ 
        Drugbank & 0.065 & 0.065 & 0.091 & 0.074 & \textbf{0.049} & 0.065 & 0.053 & 0.065 & 0.052 & 0.051 \\ 
        GDB 7 & 0.075 & 0.075 & 0.101 & 0.089 & \textbf{0.046} & 0.075 & 0.050 & 0.075 & 0.054 & 0.052 \\ 
        GDB 8 & 0.074 & 0.074 & 0.100 & 0.088 & \textbf{0.047} & 0.074 & 0.052 & 0.074 & 0.055 & 0.054 \\ 
        GDB 9 & 0.073 & 0.074 & 0.097 & 0.087 & \textbf{0.046} & 0.073 & 0.050 & 0.073 & 0.053 & 0.052 \\ 
        GDB 10 & 0.073 & 0.074 & 0.098 & 0.085 & \textbf{0.048} & 0.073 & 0.052 & 0.073 & 0.055 & 0.054 \\ 
        GDB 11 & 0.071 & 0.071 & 0.095 & 0.083 & \textbf{0.050} & 0.071 & 0.055 & 0.071 & 0.056 & 0.055 \\ 
        GDB 12 & 0.068 & 0.068 & 0.093 & 0.080 & \textbf{0.047} & 0.068 & 0.051 & 0.068 & 0.052 & 0.051 \\ 
        GDB 13 & 0.068 & 0.068 & 0.093 & 0.081 & \textbf{0.047} & 0.068 & 0.051 & 0.068 & 0.053 & 0.051 \\
        S66x8 & 0.065 & 0.065 & 0.094 & 0.084 & \textbf{0.045} & 0.065 & 0.048 & 0.065 & 0.049 & 0.047 \\ 
        Tripeptide & 0.085 & 0.085 & 0.100 & 0.087 & \textbf{0.056} & 0.085 & 0.060 & 0.085 & 0.064 & 0.062 \\ \bottomrule
    \end{tabular}}
\end{table*}

\begin{table*}[!ht]
    \centering
    \caption{MAE total energy per heavy atom in kcal/mol for DFTBML and various models on the COMP6 benchmark suite}
    \label{Tab:COMP6_tot_ener_per_heavy}
    \resizebox{\textwidth}{!}{\begin{tabular}{lcccccccccc}
    \toprule
        \textbf{Test set} & \textbf{Auorg} & \textbf{MIO} & \textbf{GFN1-xTB} & \textbf{GFN2-xTB} & \textbf{DFTBML CC} & \textbf{Repulsive CC} & \textbf{DFTBML DFT} & \textbf{Repulsive DFT} & \textbf{Transfer CC} & \textbf{Transfer DFT} \\ \midrule
        Ani MD & 0.27 & 0.24 & 0.24 & 0.43 & 0.24 & 0.31 & 0.20 & 0.26 & 0.21 & \textbf{0.16} \\ 
        Drugbank & 0.68 & 0.64 & 0.59 & 0.57 & 0.39 & 0.54 & \textbf{0.31} & 0.48 & 0.41 & 0.32 \\ 
        GDB 7 & 1.33 & 1.29 & 1.20 & 1.09 & 0.55 & 0.89 & \textbf{0.45} & 0.81 & 0.62 & 0.49 \\ 
        GDB 8 & 1.28 & 1.24 & 1.12 & 1.01 & 0.51 & 0.80 & \textbf{0.41} & 0.74 & 0.57 & 0.44 \\ 
        GDB 9 & 1.22 & 1.18 & 1.00 & 0.96 & 0.49 & 0.82 & \textbf{0.38} & 0.73 & 0.55 & 0.41 \\ 
        GDB 10 & 1.07 & 1.04 & 0.98 & 0.92 & 0.46 & 0.72 & \textbf{0.37} & 0.65 & 0.53 & 0.40 \\ 
        GDB 11 & 1.15 & 1.14 & 0.94 & 0.97 & 0.53 & 0.80 & \textbf{0.37} & 0.72 & 0.61 & 0.41 \\ 
        GDB 12 & 1.14 & 1.13 & 0.88 & 0.92 & 0.51 & 0.80 & \textbf{0.35} & 0.70 & 0.56 & 0.37 \\ 
        GDB 13 & 1.08 & 1.07 & 0.87 & 0.88 & 0.50 & 0.78 & \textbf{0.34} & 0.68 & 0.56 & 0.36 \\ 
        S66x8 & 0.70 & 0.58 & 0.64 & 0.65 & 0.41 & 0.74 & \textbf{0.41} & 0.71 & 0.58 & 0.44 \\ 
        Tripeptide & 0.35 & 0.35 & 0.24 & 0.28 & 0.19 & 0.30 & \textbf{0.15} & 0.24 & 0.24 & 0.18 \\ \bottomrule
    \end{tabular}}
\end{table*}

\begin{table*}[!ht]
    \centering
    \caption{MAE total energy per atom in kcal/mol for DFTBML and various models on the COMP6 benchmark suite}
    \label{Tab:COMP6_tot_ener_per_atom}
    \resizebox{\textwidth}{!}{\begin{tabular}{lcccccccccc}
    \toprule
        \textbf{Test set} & \textbf{Auorg} & \textbf{MIO} & \textbf{GFN1-xTB} & \textbf{GFN2-xTB} & \textbf{DFTBML CC} & \textbf{Repulsive CC} & \textbf{DFTBML DFT} & \textbf{Repulsive DFT} & \textbf{Transfer CC} & \textbf{Transfer DFT} \\ \midrule
        Ani MD & 0.13 & 0.12 & 0.12 & 0.21 & 0.11 & 0.15 & 0.09 & 0.12 & 0.10 & \textbf{0.07} \\ 
        Drugbank & 0.35 & 0.33 & 0.30 & 0.29 & 0.20 & 0.27 & \textbf{0.16} & 0.24 & 0.21 & 0.16 \\ 
        GDB 7 & 0.66 & 0.64 & 0.59 & 0.56 & 0.27 & 0.43 & \textbf{0.22} & 0.39 & 0.30 & 0.25 \\ 
        GDB 8 & 0.64 & 0.62 & 0.56 & 0.52 & 0.26 & 0.39 & \textbf{0.21} & 0.36 & 0.29 & 0.22 \\ 
        GDB 9 & 0.60 & 0.59 & 0.49 & 0.49 & 0.24 & 0.40 & \textbf{0.19} & 0.36 & 0.28 & 0.21 \\ 
        GDB 10 & 0.54 & 0.52 & 0.49 & 0.47 & 0.23 & 0.36 & \textbf{0.19} & 0.33 & 0.27 & 0.20 \\ 
        GDB 11 & 0.56 & 0.55 & 0.46 & 0.48 & 0.26 & 0.39 & \textbf{0.18} & 0.35 & 0.30 & 0.20 \\ 
        GDB 12 & 0.56 & 0.55 & 0.43 & 0.45 & 0.25 & 0.39 & \textbf{0.17} & 0.34 & 0.27 & 0.18 \\ 
        GDB 13 & 0.52 & 0.52 & 0.42 & 0.42 & 0.24 & 0.37 & \textbf{0.16} & 0.33 & 0.27 & 0.18 \\ 
        S66x8 & 0.28 & 0.23 & 0.27 & 0.27 & \textbf{0.16} & 0.30 & 0.17 & 0.29 & 0.23 & 0.18 \\ 
        Tripeptide & 0.18 & 0.18 & 0.12 & 0.14 & 0.10 & 0.15 & \textbf{0.08} & 0.12 & 0.12 & 0.09 \\ \bottomrule
    \end{tabular}}
\end{table*}

\clearpage

\bibliography{references.bib}